\documentclass[twocolumn, twocolappendix, floatfix]{aastex631}

\newcommand{\rowname}[1]

\shorttitle{Multiphase galactic outflows in dwarfs}
\shortauthors{Steinwandel et al.}

\graphicspath{{./}{figures/}}

\begin{document}

\title{The structure and composition of multiphase galactic winds in a Large Magellanic Cloud mass simulated galaxy}

\author[0000-0002-0786-7307]{Ulrich P. Steinwandel}
\affiliation{Center for Computational Astrophysics, Flatiron Institute, 162 5th Avenue, New York, NY 10010, USA}
\email{usteinwandel@flatironinstitute.org}

\author[0000-0003-2896-3725]{Chang-Goo Kim}
\affiliation{Department of Astrophysical Sciences, Princeton University, Princeton, NJ 08544, USA}

\author[0000-0003-2630-9228]{Greg L. Bryan}
\affiliation{Department of Astronomy, Columbia University, 550 West 120th Street, New York, NY 10027, USA}

\author[0000-0002-0509-9113]{Eve C. Ostriker}
\affiliation{Department of Astrophysical Sciences, Princeton University, Princeton, NJ 08544, USA}

\author[0000-0003-2835-8533]{Rachel S. Somerville}
\affiliation{Center for Computational Astrophysics, Flatiron Institute, 162 5th Avenue, New York, NY 10010, USA}

\author[0000-0003-3806-8548]{Drummond B. Fielding}
\affiliation{Center for Computational Astrophysics, Flatiron Institute, 162 5th Avenue, New York, NY 10010, USA}

\begin{abstract}

We present the first results from a high resolution simulation with a focus on galactic wind driving for an isolated galaxy with a halo mass of $\sim 10^{11}$ M$_{\odot}$ (similar to the Large Magellanic Cloud) and a total gas mass of  $\sim 6 \times 10^{8}$ M$_{\odot}$, resulting in $\sim 10^{8}$ gas cells at $\sim 4$ M$_{\odot}$ mass resolution. We adopt a resolved stellar feedback model with non-equilibrium cooling and heating, including photoelectric heating and photo-ionizing radiation, as well as supernovae (SNe), coupled to the second order meshless finite mass (MFM) method for hydrodynamics. These features make this the largest resolved-ISM galaxy model run to date. We find mean star formation rates around $0.05$ M$_{\odot}$ yr$^{-1}$ and evaluate typical time averaged loading factors for mass ($\eta_\mathrm{M}$ $\sim$ 1.0, in good agreement with recent observations) and energy ($\eta_\mathrm{E}$ $\sim$ 0.01).
The bulk of the mass of the wind is transported by the warm ($T < 5 \times 10^5$K) phase, while there is a similar amount of energy transported in the warm and the hot phases ($T > 5 \times 10^5$K). We find an average opening angle of 30 degrees for the wind, decreasing with higher altitude above the midplane. The wind mass loading is decreasing (flat) for the warm (hot) phase as a function of the star formation surface rate density $\Sigma_{\rm SFR}$, while the energy loading shows inverted trends with $\Sigma_{\rm SFR}$, decreasing for the warm wind and increasing for the hot wind, although with very shallow slopes. These scalings are in good agreement with previous simulations of resolved wind driving in the multi-phase ISM. 

\end{abstract}

\keywords{Galactic winds (572) --- Galaxy evolution (594) --- Hydrodynamical simulations (767) --- Stellar feedback (1602) --- Interstellar medium (847)}

\section{Introduction} \label{sec:intro}

Galactic winds are common across galaxy masses and sizes, and are generally believed to play a crucial role in regulating star formation and the galactic baryon budget, as well as the growth of supermassive black holes (SMBH). The role of galactic winds is twofold as they actively \textit{eject} gas from the ISM, transporting mass, energy and metals into the circumgalactic medium (CGM) where they can then \textit{prevent} the hot X-ray emitting gas in the CGM from falling victim to radiative cooling processes that would lead to over-cooling of the hot CGM gas. Thus, it has become apparent that galactic winds are important contributors to the assembly and formation of galaxies and represent one of the key aspects that needs to be understood to develop a detailed theory of galaxy formation and evolution.

The signatures of galactic winds are observed in the local universe \citep[e.g.][]{Lehnert1996, Martin1999, McQuinn2019, Xu2022} as well as at high redshift \citep[][]{Pettini2001, Shapely2003, FS2019}. Some observational work suggests that  there is a rather weak correlation between outflow velocity and star formation rate (SFR) \citep[][]{Chisholm2015}, while other groups suggest that there is a stronger correlation between outflow velocity and SFR \citep[see e.g.][]{Martin2005, Heckman2015}. 

While the trend in the literature on the correlation between outflow velocity and galaxy mass is still somewhat under debate one can find more consensus in the literature on the mass loading factor $\eta_\mathrm{m} = \dot{M}_\mathrm{out}/\mathrm{SFR}$, which tends to decrease with increasing galaxy mass which is not only established by observations at low redshift \citep[e.g.][]{Newmann2012, Chisolm2017}, but also appears to be robust against trends found at higher redshift \citep[e.g.][]{FS2014, FS2019}. Similarly, one can find evidence that $\eta_{m}$ is positively correlated with star formation rate surface density ($\Sigma_\mathrm{sfr}$), which is again a trend that manifests at lower redshift \citep[e.g.][]{McQuinn2019} and in high redshift systems \citep[e.g.][]{Davis2019}. 

However, despite the important findings on scaling relations of wind proprieties,  galactic winds are much more complex than simple mass transport at a given outflow velocity, due to the observed multiphase nature of the winds. This fact makes observations of galactic winds extremely complicated since the cold, warm, and hot gas phases each require different observational techniques and/or instruments. The cold gas can be traced by neutral hydrogen \citep[e.g.][]{Walter2002, Martini2018} or via CO \citep[e.g.][]{Bolatto2013, Cicone2014, Leroy2015} and metals such as NaD, OI or MgI \citep[e.g.][]{Heckman2000,Rupke2005, Chen2010}, and is characterized by temperatures between 10 and a few 100 K with rather low outflow velocities. The warm phase of the outflow, with temperatures around 10$^{4}$ K, can be studied by utilizing either H$\alpha$ \citep[e.g.][]{Keith1995, Westermoquette2009} or metal lines of photo-ionized species such as MgII, OIII and SiIV \citep[e.g.][]{Martin2009, Rubin2011, Nielsen2015, Chisolm2017}. Even hotter winds ($10^5-10^6$ K) can be traced by the higher ionization states of different metal species \citep[e.g][]{Steidel2010, Kacprzak2015, Chisholm2018, Ashley2020}, with the hottest phase ($>10^7$) being accessible in the X-ray emitting gas \cite[e.g.][]{Ptak1997, Stricklan2009, Lopez2020, Hodges2020}. 

While there are local starburst systems such as M82 which show evidence of the presence of \textit{all} of these different phases simultaneously, the situation is more complicated in local dwarf galaxies due to the transient nature of galactic winds and the intrinsically low surface brightness of dwarf galaxy systems \citep[see][and references therein for a detailed discussion]{McQuinn2019}. Specifically, the presence of a hot wind in nearby dwarfs remains uncertain with only a small fraction of systems showing a hot outflow component \citep[e.g][]{Heckman1995, Summers2003, Summers2004, Hartwell2004, Ott2005, McQuinn2018}, while there is very good observational evidence for warm and cold winds in nearby dwarf galaxy systems \citep[e.g.][]{Meurer2004, Meurer2006, Barnes2011, Kobulnicky2008, Lelli2014}. 
  
Numerical simulations of large cosmological volumes such as Illustris \citep[e.g.][]{Vogelsberger2014}, Magneticum \citep[e.g.][]{Hirschmann2014}, EAGLE \citep[e.g.][]{Schaye2015}, IllustrisTNG \citep[e.g.][]{Pillepich2018} and SIMBA \citep[e.g.][]{Dave2019}, are not able to directly resolve the processes that drive galactic winds, and therefore rely on phenomenological ``sub-grid'' prescriptions to implement these winds \citep[see][for reviews]{Somerville2015, Naab2017}. These prescriptions typically involve depositing momentum or energy into the ISM, in a manner that is controlled by adjustable parameters that are tuned to match observations such as the $z=0$ stellar mass function \citep[e.g.][]{Baldry2008, Baldry2012, Bernardi2013, Dsouza2015}, the mass-metallicity relation \citep[e.g.][]{Tremonti2004}, and other observables. While this approach has overall been very successful at reproducing many key observations, it is interesting to note that the \emph{required} mass loadings are considerably higher than those derived from observations such as the ones referred to earlier \citep[see also][]{Marasco2022,McQuinn2019, Xu2022}. However, it is worth noting that the mass loadings derived from observations are highly uncertain, and a true ``apples to apples'' comparison with simulations (comparing gas outflows with the same phase, measured at the same spatial scale) has not yet been carried out. 

It has recently become possible to carry out simulations of sub-galactic portions of the ISM, or even more recently, entire idealized galaxies that can capture the momentum and hot-phase build-up in a resolved Sedov-Taylor phase of individual supernova (SN) remnants \citep[e.g.][]{Kim2015, Martizzi2015, Hu2017, Emerick2019, Ohlin2019, Steinwandel2020, Fujii2021, Hirai2021, Grudic2021}. 

In these simulations, phenomenological tuning is no longer adopted, and the properties of large scale multiphase winds can be considered emergent predictions following from the physical processes operating on parsec scales. The solar neighborhood environment has been probed extensively in so called ``tall box'' simulations, which have explored outflow launching mechanisms by supernovae (SNe) either in simulations with fixed SN-rate \citep[e.g.][]{deAvillez2001, Joung2009, Hill2012, Creasey2013, Fielding2018} or simulations where the SN-rate follows the star formation rate with a characteristic time-delay \citep[e.g.][]{Girichidis2016, Girichidis2018, Kim2018, Kannan2019, Li2020, Kim2020, Kim2020_smaug, Vijayan2020, Rathjen2021, Hu2021, Hu2022, Kim2022, Rathjen2022}.

In addition, there have been a number of simulations of wind driving in global galaxies with resolved stellar feedback, but up until recently these were limited to very low mass and low metallicity, isolated dwarf galaxies that resemble the Wolf-Lundmark-Moyette (WLM) dwarf galaxy \citep[e.g.][]{Emerick2019, Fielding2017, Hu2016, Hu2017, Hu2019, Smith2021, Gutcke2021, Hislop2022, Steinwandel2022}, with typical star formation rates between $10^{-5}$ to $10^{-3}$ M$_{\odot}$ yr$^{-1}$ and star formation rate surface densities between $10^{-9}$ and $10^{-6}$ M$_{\odot}$ kpc$^{-2}$ yr$^{-1}$. While investigations of the outflow driving by SN-feedback have been performed in these kinds of simulations, the fairly low star formation rates under WLM-like conditions and the small system size allow for little variation locally within the galaxy and studies of radial and/or vertical analysis of outflow rates and loading factors are mostly absent from the literature \citep[but see][for an analysis of radially averaged loading factors in WLM-like systems]{Hu2019}. In addition, the low potential depth of such systems makes it hard to generalize to larger galaxies. 

Finally, it is worthwhile to note the immense progress made over the past decade in the regime of ``semi-resolved'' simulations that adopt schemes for resolved SN-injection if the cooling radius is resolved and otherwise input the terminal momentum produced by SN-remnants with a given boost factor, often following the previous results of isolated and turbulent blast wave evolution \citet{Cioffi1988, Geen2016, Kim2015, Martizzi2015, Thornton1998, Walch2015}. Notable examples of these type of simulations are the FIRE-1 \citep[e.g.][]{Hopkins2014}, FIRE-2 \citep[e.g.][]{Hopkins2018} and FIRE-3 \citep[e.g.][]{Hopkins_fire3}, the simulations of \citet{Kimm2015}, the SMUGGLE model \citep[e.g.][]{Marinacci2019} as well as the very recent SATIN model \citep[][]{Bieri2022}. In this context it is interesting to point out that these simulations tend to reproduce the rather high mass loading factors in large-scale cosmological simulations. 

In this paper we introduce the first results from a simulation of a galaxy that resembles more massive local dwarfs, such as the large Magellanic cloud (LMC) or M33, with resolved stellar feedback, and discuss the outflow launching and driving process in phase space and as a function of radius and vertical height.  
The structure of this paper is as follows. In Section \ref{sec:methods} we will briefly discuss the numerical methods used and describe the initial conditions adopted. In Section \ref{sec:results} we will present the main findings of our study for the time evolution of outflow rates and loading factors, their spatial properties (radial and vertical profiles), as well as the key aspect of this work: the joint PDFs of outflow velocity and sound speed measured at differing heights and radii. In Section \ref{sec:discussion} we will discuss our results and compare to previous work where appropriate. In Section \ref{sec:conclusions} we will summarize our findings and discuss pathways for future improvements.

\section{Numerical Methods and Simulations} \label{sec:methods}

\subsection{Gravity and Hydrodynamics}
The simulation discussed in this paper has been carried out with the code \textsc{p-gadget3} \citep{springel05} with major improvements to the routines for hydrodynamics and stellar feedback included  \citep[see][for details]{Hu2014, Hu2016, Hu2017, Hu2019, Steinwandel2020}. The code is a multi-method framework for solving the equations of hydrodynamics and has the option to solve for the fluid flow either with the ``pressure-energy'' Smoothed Particle Hydrodynamics \citep[SPH][]{Hopkins2013, Hu2014} method or the second order meshless finite mass (MFM) method \citep[e.g][]{Lanson2008, Gaburov2011, Hopkins2015}. We use the MFM method as implemented in \citet{Steinwandel2020} which is itself based on the multi-method code \textsc{gizmo} as presented in \citet{Hopkins2015}. MFM is realized as a second order Godunov scheme and we solve for fluid fluxes with a Harten-Lax-van-Leer (HLL) Riemann solver with an appropriate reconstruction of the rarefaction wave (HLLC) between tracer particles. This is combined with a second order ``kick-drift-kick (KDK)'' leap-frog scheme for the time integration. The gravitational forces are computed via the tree particle mesh method \citep[e.g.][]{Barnes1986, springel05, Hopkins2015} and the appropriate corrections for the coupling between MFM and Gadget's gravity solver are applied as outlined in the appendix of \citet{Hopkins2015}.  

\subsection{Star formation and stellar feedback}

Star formation is implemented in the simulation by sampling the stellar Initial Mass Function (IMF) following the methodology first described in \citet[][]{Hu2016} and \citet[][]{Hu2017} based on a Schmidt-type star formation scaling with a fixed star formation efficiency of 2 per cent per free fall time. In addition, we compute the Jeans mass for all gas cells and convert any gas cell with less than 0.5 M$_\mathrm{J}$ in the MFM-kernel radius instantly to a star particle in the next time step. The Jeans mass is given via: 
\begin{equation}
    M_{\mathrm{J},i} = \frac{\pi^{5/2}{c_{\mathrm{s},i}^3}}{6G^{3/2}\rho_{i}^{1/2}}.
\end{equation}
Additionally, we note that the kernel mass is computed on the average number of neighbors and not the true number of neighbors as computed by the code, to avoid an additional SPH-neighbor search.   
All star particles above the resolution limit of 4 M$_{\odot}$ are treated as single stars.
We follow several feedback channels for the stars with a mass larger than 4 M$_{\odot}$ formed within the simulation, including photoelectric heating and photo-ionising radiation, as well as the metal enrichment from AGB stellar winds (as a single kernel weighted metal dump at the end of their lifetimes). The core of the stellar feedback scheme is SN-feedback as implemented in \citet{Hu2017} for the SPH version of the code  and \citet{Steinwandel2020} for the MFM version of the code, which at the target resolution of our simulation of 4 M$_{\odot}$ allows for a self consistent build-up of the hot phase and the momentum of individual SN-events in a resolved Sedov-Taylor phase \citep[see][]{Hu2021, Hu2022} with the deposition of the canonical SN-energy of $10^{51}$ erg into the ambient ISM. Moreover we adopt lifetimes of massive stars based on \citet{Georgy2013}, allowing us to self-consistently trace the locations in the ISM where SNe explode. Thus, all the outflow properties in these simulations can be interpreted as being self-consistently launched from the ISM.

Our simulation utilizes a treatment of the far-UV radiation from single stars (above 4 M$_{\odot}$) following the implementation of \citet{Hu2017} where we update the local radiation field $G_{0}$ during the gravity-tree walk (optically thin approximation). We  follow the lifetime functions and effective surface temperatures of \citet{Georgy2013}. UV luminosities for each star is taken from the BaSel \citep{Lejeune1997, Lejeune1998, Westera2002} stellar library, in the (single bin) energy band $6 - 13.6$ eV, which dominates the photo-electric heating rate in the ISM. 
We then follow photoelectric heating rates of\citet{Bakes1994, Wolfire2003} and \citet{Bergin2004} with the rate:
\begin{equation}
    \Gamma_\mathrm{pe} = 1.3 \cdot 10^{-24} \epsilon D G_\mathrm{eff} n \  \mathrm{erg} \ \mathrm{s}^{-1} \mathrm{cm}^{-3},
\end{equation}
by multiplying the local radiation field $G_{0}$, the effective attenuation radiation field $G_\mathrm{eff} = G_{0} \exp(-1.33 \cdot 10^{-21} {\rm cm^2} D N_\mathrm{H, tot})$, where $n$ denotes the hydrogen number density, $D$ is the dust-to-gas mass ratio, $N_\mathrm{H, tot}$ is the total (local) hydrogen column density and $\epsilon$ is the photoelectric heating rate given via:
\begin{equation}
    \epsilon = \frac{0.049}{1 + 0.004 \psi^{0.73}} + \frac{0.037 (T/10000)^{0.7}}{1+2 \cdot10^{-4}\psi}
\end{equation}
with $\psi = G_\mathrm{eff} \sqrt{T} /n_\mathrm{e}$, with the electron number density $n_\mathrm{e}$ and the gas temperature $T$. 
We include a treatment for photo-ionizing radiation since it can have a significant impact on ISM properties and star formation rates \citep[e.g.][]{Hu2017, Smith2021, Hislop2022}. Ideally, we would utilize a proper radiative transfer scheme for this purpose, but this is prohibitively computationally expensive. Thus we model the photo-ionizing feedback via a Str\"omgren approximation locally around each photo-ionizing source in the simulation, similar to the approach used in \citet{Hopkins2012} and later improved by \citet{Hu2017}. 

Every star particle with mass m$_\mathrm{IMF} > 8$ M$_{\odot}$ is treated as a photo-ionizing source in the presented simulation. All the neighboring gas particles within the Str\"omgren radius R$_\mathrm{S}$ of the star: 
\begin{equation}
    R_\mathrm{S} = \left(\frac{3}{4 \pi} \frac{S_{\star}}{n^2 \beta}\right)^{1/3}
\end{equation}
are labeled as photo-ionized, with $\beta$ as the (electron) recombination coefficient, $n$ the number density and S$_{\star}$ the photo-ionizing flux. For the gas cells in the Str\"omgren radius R$_\mathrm{S}$ we adjust the chemical abundance, destroy all the molecular and neutral hydrogen and transfer it to H$^{+}$. Furthermore, we heat the surrounding gas within R$_\mathrm{S}$ to 10$^4$ K (if it is already warmer than $10^{4}$ K we do not update its temperature), which is the temperature in \ion{H}{2}-regions, by providing the necessary energy input from the ionizing source. 
This procedure is rather simple if R$_{S}$ is a known quantity. In practice, the density structure if the ISM is much more complicated than a constant density background for which R$_\mathrm{S}$ is well defined and we compute $R_{S}$ in an iterative fashion, by iterating until our ionising photon budget is depleted. Given this procedure, it is possible that two gas particles are flagged as photo-ionized by two star particles in which case we only adopt the flux of the closer star particle. The current treatment of this procedure is isotropic which leads to a strong mass bias in a scheme. Since we are aware of this we set the maximum of R$_{s}$ to a distance of 50 pc from the host star. This scheme can be improved in the future by healpix tessellating the the ionisation sphere as implemented in \citet{Smith2021}.

Our simulation includes a resolved treatment of core collapse supernovae (CCSN) for which the key is ultimately the high mass and spatial resolution with which we run the simulation as pointed out by \citet{Steinwandel2020}for Lagrangian codes one needs a spatial resolution of between 1 and 10 $M_{\odot}$ to resolve the momentum and hot-phase build-up during the Sedov-Taylor phase. Hence our scheme is rather simple and we distribute $10^{51}$ erg with every SNe into the nearest 32 neighbors. We adopt  metal enrichment following the yields provided by\citet{Chieffi2004}, who provide data for stellar yields of stars from zero to solar metallicity for progenitors mass from 13 M$_{\odot}$ to 35 M$_{\odot}$. We interpolate our metal enrichment routines based on this data. When a star explodes in a CCSN the mass is added to the surrounding $32$ (one kernel) gas particles with the respective metal mixture that is expected from the above yields. The choice of adopted metal yields is important because it alters the cooling properties of the ambient gas and influences the density and phase structure of the ISM. Finally, we briefly describe how we handle the splitting of particles in our code. At the target resolution we adopt for our simulations one enrichment event can add more mass to the resolution elements that they are originally hold and thus we need to split the new (heavy particles) down to the resolution level of 4 M$_{\odot}$. Splitting particles is always tricky in Lagrangian codes because it requires some careful handling of the memory available per core to make the procedure memory efficient. 
We split the gas particles after the enrichment step if the mass exceeds the initial particle mass resolution of the simulation by a factor of two, to avoid numerical artifacts. The split particles inherit all the physical properties that were present in the parent particle including specific internal energy, kinetic energy, metallicity and chemical abundances. The new particle mass is set to one half of the parent particle mass and we assign a random position within one fifth of the parents smoothing length for the new particles position. This process is isotropic within the smoothing kernel of the parent particle. Spawning the new particles can become rather expensive if this has to be carried out repeatedly in the code because if requires a new domain decomposition alongside with a new build of the gravity tree.

Cooling and heating processes are modeled in a similar fashion as in the ``SILCC'' project \citep[e.g.][]{Walch2015, Girichidis2016, Girichidis2018, Gatto2017, Peters2017} based on a non-equilibrium cooling and heating prescription. The non-equilibrium model is based on the work of \citet{Nelson1997}, \citet{Glover2007a}, \citet{Glover2007b} and \citet{Glover2012}. However, we note that some of the rate equations have been updated following \citet{Gong2016}. The model tracks six individual species given via molecular hydrogen (H$_{2}$), ionized hydrogen (H$^{+}$), neutral hydrogen (H) as well as carbon-monoxide (CO), ionized carbon (C$^{+}$) and oxygen. Additionally, we trace free electrons and assume that all silicon in the simulation is present as Si$^{+}$. There are several chemical reactions in which these species are involved and we refer the interested reader to table 1 of \citet{Micic2012}. The network models molecular hydrogen formation on dust grains for which we solve directly for the non-equilibrium rate equations to obtain $x_{H_{2}}$ and $x_{H^{+}}$ and obtain $x_H$ based on the conservation law:
\begin{equation}
    x_\mathrm{H} = 1 - 2 x_{\mathrm{H}_2} - x_{\mathrm{H}^{+}}.
\end{equation}
In the network we track n$_e$ which we compute via:
\begin{equation}
    x_\mathrm{e} = x_{\mathrm{H}^{+}} + x_{\mathrm{C}^{+}} + x_{\mathrm{Si}^{+}}
\end{equation}
We adopt non-equilibrium cooling rates that take into account the local density, temperature and chemical abundance. The most dominant cooling processes are fine structure line cooling of atoms and ions (C$^{+}$, Si$^{+}$ and O), vibration and rotation line cooling due to molecules (H$_{2}$ and CO), Lyman-$\alpha$ cooling of hydrogen, collisional dissociation of H$_{2}$, collisional ionisation of hydrogen and recombination of H$^{+}$ (in gas phase and on dust grains). The main heating  processes are photo-electric heating from dust grains and polycyclic aromatic hydrocarbonates (PAHs), a constant cosmic ray ionisation rate of $10^{-19}$ s$^{-1}$, photo-dissociation of H$_{2}$, UV-pumping of H$_{2}$ and formation of H$_{2}$. For high temperatures (T  $> 3 \times 10^{4}$ K) we adopt the equilibrium cooling formalism first described by \citet{Wiersma2009} as implemented by \citet{Aumer2013} including the elements H, He, C, N, O, Ne, Mg, Si, S, Ca, Fe, and Zn.

\subsection{Initial Conditions}

The initial conditions for this simulation were produced with the method described in \citet[][]{springel05b}, which is typically referred to as ``MakeDiskGal'' or ``MakeGalaxy''. As input parameters we choose typical values that are in agreement with observations of the Large Magellanic Cloud (LMC) \citep[e.g.][based on kinematic arguments]{Erkal2019} and values previously adopted in numerical simulations to select an LMC-type galaxy \citep[e.g.][]{Hopkins2018}. However, we note that the system is set up to be at the lower limit of halo mass and gas fraction to reduce the computational expense. Hence, we adopt a total dark matter halo mass of $2 \times 10^{11}$ M$_{\odot}$, a stellar background potential of $2 \times 10^{9}$ M$_{\odot}$ as well as a gas disk with a total mass of $0.5 \times 10^{9}$ M$_{\odot}$ for which we adopt a uniform mass resolution of 4 M$_{\odot}$, resulting in $\sim 10^8$ gas particles ($512^3$) and a mean particle spacing of $\sim$ 1.0 pc (mean effective spatial resolution) for the gaseous body of the galaxy. For the dark matter halo we adopt a concentration parameter of $c=9$ modeled as a Hernquist-profile. We adopt a disc-scale length of 1 kpc and a disc-scale height of 250 pc with an initial metallicity of 0.1 Z$_{\odot}$. The gravitational softening parameter of the simulations is set to 0.5 pc for gas and stars throughout the total run time of the simulation of 500 Myrs.

\section{Results} \label{sec:results}

\begin{figure*}
    \centering
    \includegraphics[width=0.99\textwidth]{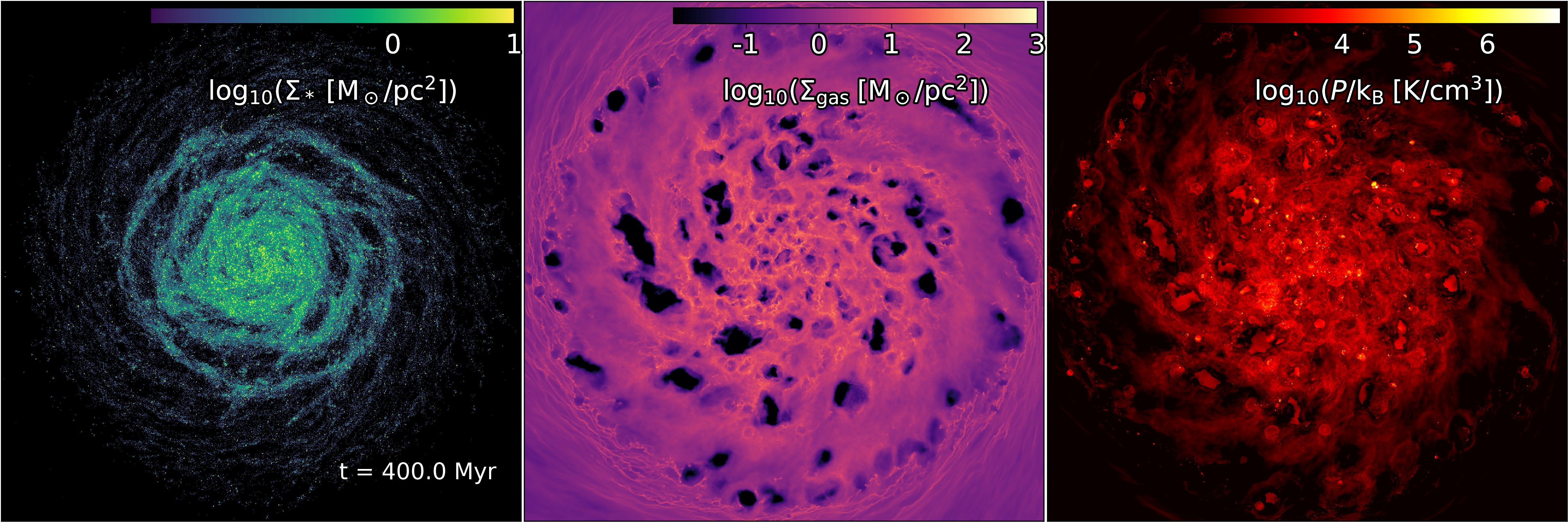}
    \includegraphics[width=0.99\textwidth]{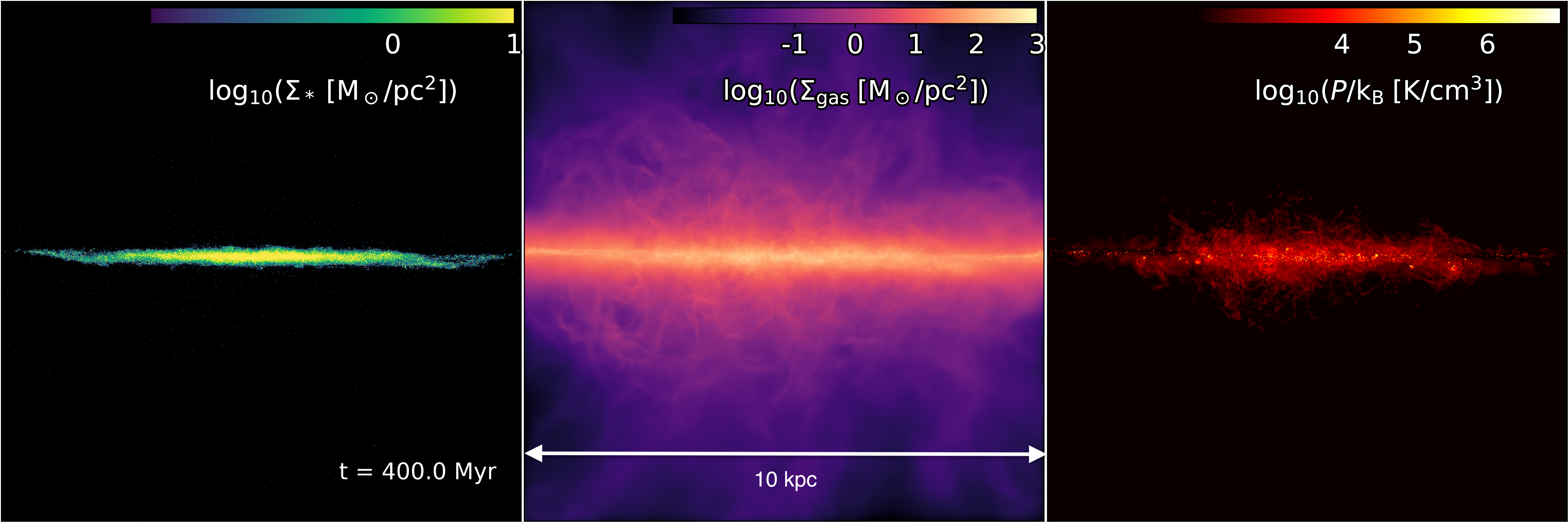}
    \caption{Face-on (top) and edge-on (bottom) projections at t=400 Myr for stellar surface density (left), gas surface density (center) as well as a slice within 100 pc of the thermal pressure (right). The top center panel clearly shows the extended SN-bubbles all across the galaxy. The edge-on view reveals a large scale outflow, easily reaching out to a height of 5 kpc with volume filling gas of density around 0.1 cm$^{-3}$. The face-on projection is restricted to 500 pc above and below the mid-plane. The edge-on view is a full projection over 10 kpc along the line of sight in order to visualize the low density gas in the outflows.}
    \label{fig:galaxy}
\end{figure*}

\begin{figure*}
    \centering
    \includegraphics[width=0.49\textwidth]{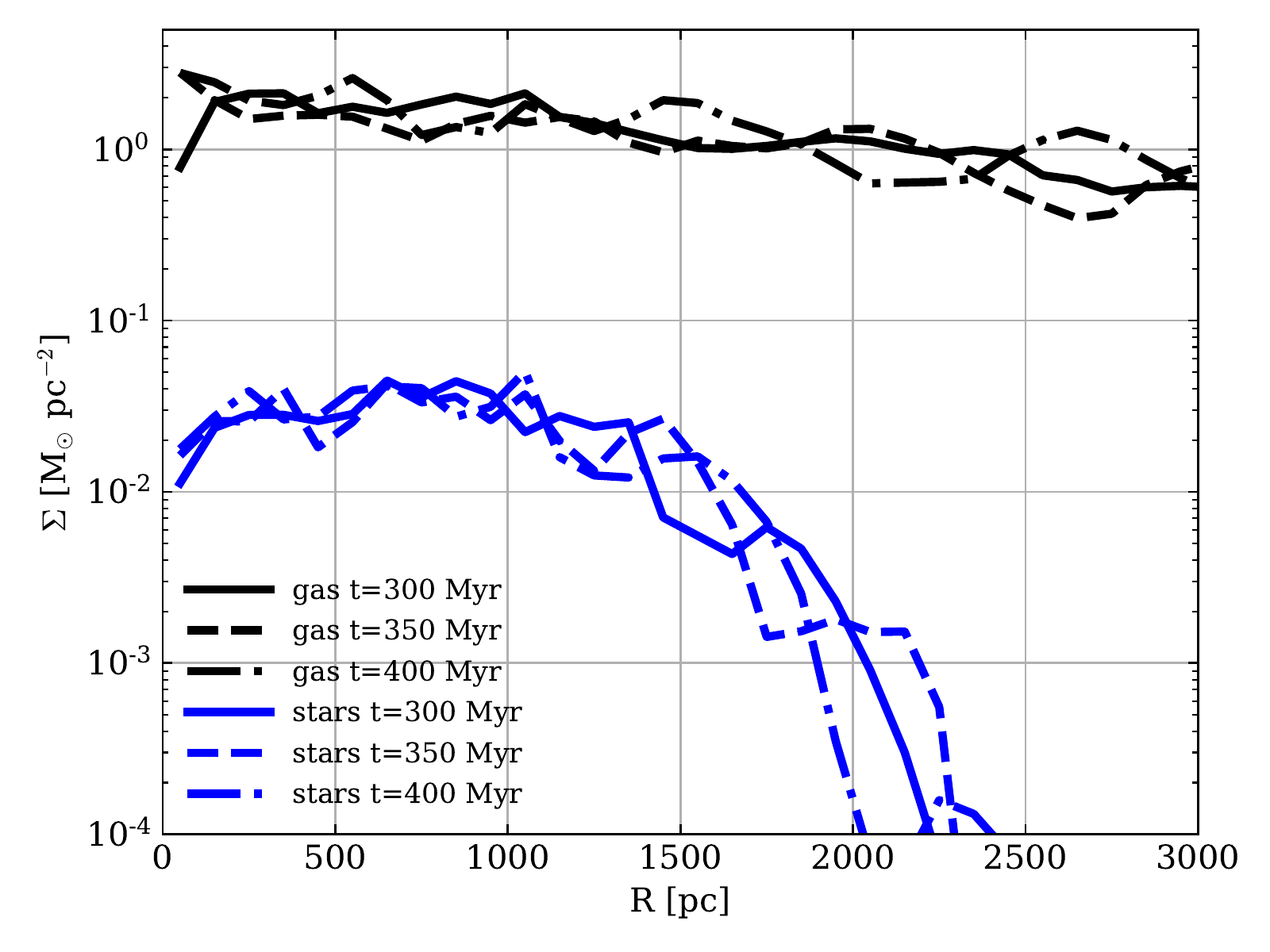}
    \includegraphics[width=0.49\textwidth]{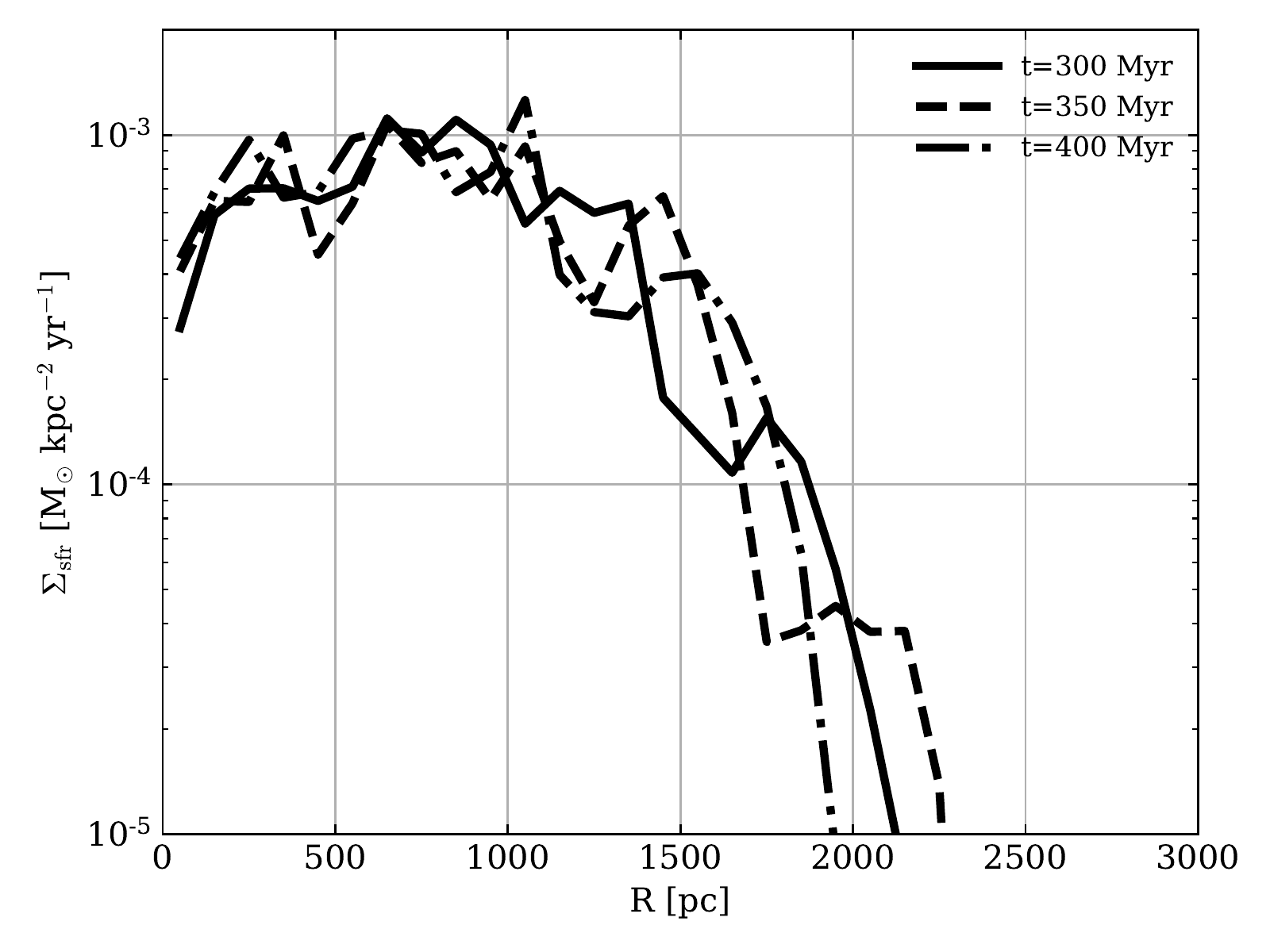}
    \caption{Radial surface density profiles of gas (black) and stars (blue) at $t=300$ Myr of evolution (left). The solid, dashed and dashed-dotted line represent the three times $t=300$, $t=350$ and $t=400$ Myr of evolution. The radial star formation rate surface density profiles is presented on the right for the same points in time.}
    \label{fig:surface_dens}
\end{figure*}

\subsection{ISM and Outflow morphology}
In Fig.~\ref{fig:galaxy} we show projections of the stellar surface density (left), the gas surface density (center) and a slice of thermal pressure within a thickness of 100 pc (right) for our LMC run in the face-on (top row) and edge-on (center row) view at an evolutionary time of t=400 Myr in a box with a side-length of 10 kpc which is indicated by a white arrow in the bottom of the lower center box. One can clearly see the turbulent structure of the ISM in the density and pressure maps as well as the complex structure of the galaxy in its stellar body. Previous studies of global galaxies with resolved ISM were thus far restricted to WLM-like systems with roughly 10-20 times lower total mass than our LMC-like system and it is worth pointing out some general differences with the much smaller WLM-like systems investigated previously \citep[e.g.][]{Hu2016, Hu2017, Fielding2017, Hu2019, Smith2021, Gutcke2021, Hislop2022, Steinwandel2022}. For instance, we note that for the first time we can observe the build-up of more complex structure in the stellar body in a resolved global ISM simulation. This is generally not seen in the lower mass WLM-like systems (or very weakly present depending on the model realization). The face-on view of the gas density reveals large holes in the galaxy, a result of the strong SN-feedback. Additionally, the edge-on view of the gas density reveals large scale outflows with volume filling gas below 0.1 cm$^{-3}$ and large scale entrained filamentary structure, that is less prominent in the lower mass WLM-runs. The pressure structure of the turbulent ISM is clearly tracing the cavities of the hot bubble-like structures that we find in our simulation and extends out to large radii.

In Fig.\ref{fig:surface_dens} we show the radial profiles of the gas and stellar surface density (left hand side) of the young stars with ages less than 40 Myr for $t=300$ Myr (solid lines), $t=300$ Myr (dashed lines), $t=300$ Myr (dash-dotted lines) as well as the star formation rate surface density (right hand side). We note that the system has a central gas surface density of around 3-4 M$_{\odot}$ pc$^{-2}$ with a stellar surface density that is two orders of magnitude lower. It is interesting to note that star formation is largely restricted to the central part within 3 kpc and is almost flat out to a radius of 2 kpc. This indicates that the wind which we will discuss in the remainder of the paper has a strong impact on the central region of the galaxy from where the wind driving process by SNe originates.

\subsection{Phase structure of the ISM}

\begin{figure*}
    \centering
    \includegraphics[width=0.49\textwidth]{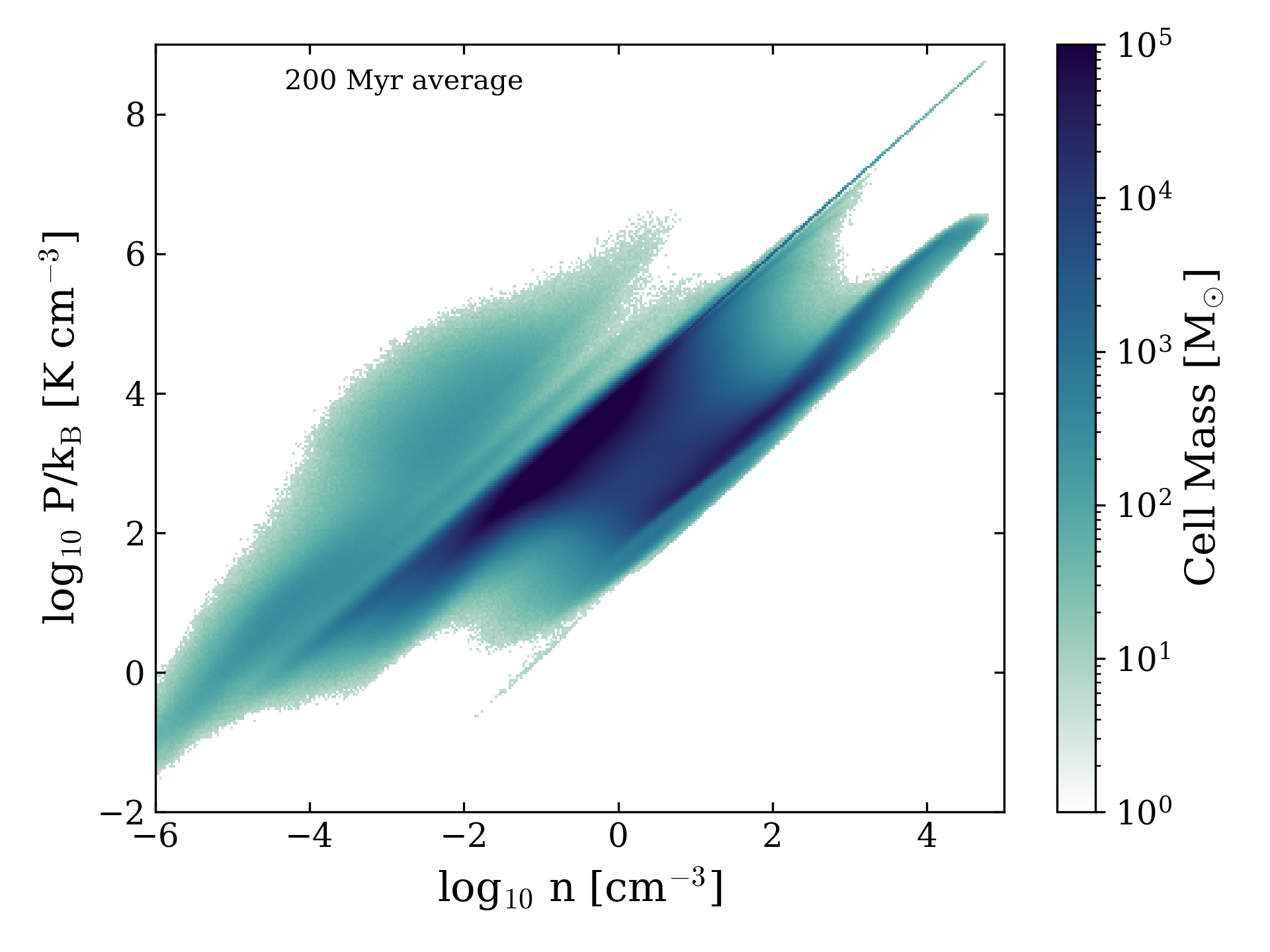}
    \includegraphics[width=0.49\textwidth]{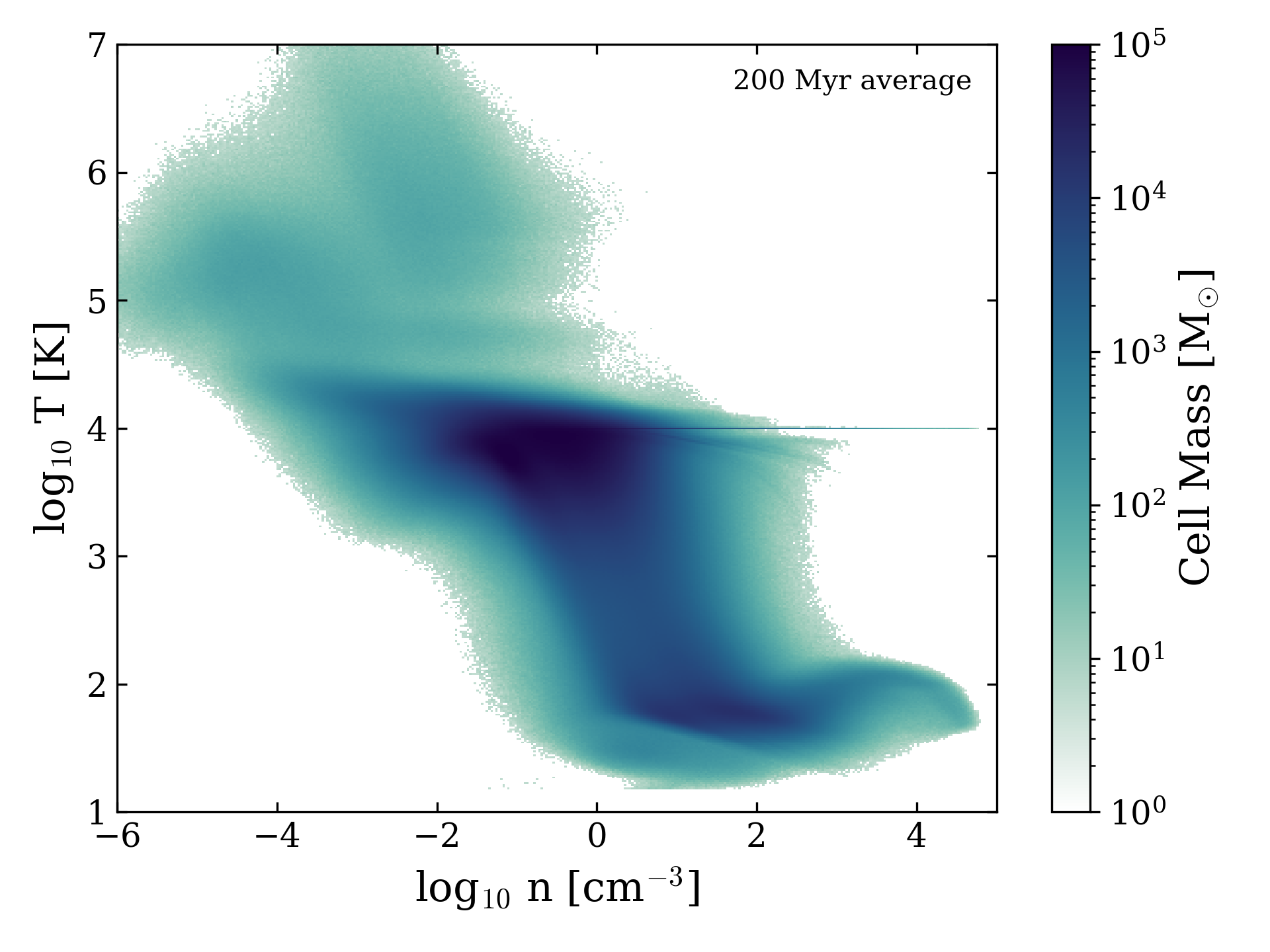}
    \caption{Mass weighted structure of our multiphase ISM in density-pressure (top) and density-temperature (bottom) phase space. The density-pressure phase space reveals the thermally unstable medium with $\partial P/\partial n < 0$ at a pressure of around $3000$ K cm$^{-3}$. In the density-temperature phase space we can see that there is clear three phase medium with a hot (SN-heated) phase, a cold (star forming phase) and a warm diffuse phase, that can be split into the warm neutral medium (WNM) and the warm ionized medium (WIM). The horizontal line at 10$^{4}$ K represents the photo-ionized gas.}
    \label{fig:phase}
\end{figure*}

\begin{figure}
    \centering
    \includegraphics[scale=0.48]{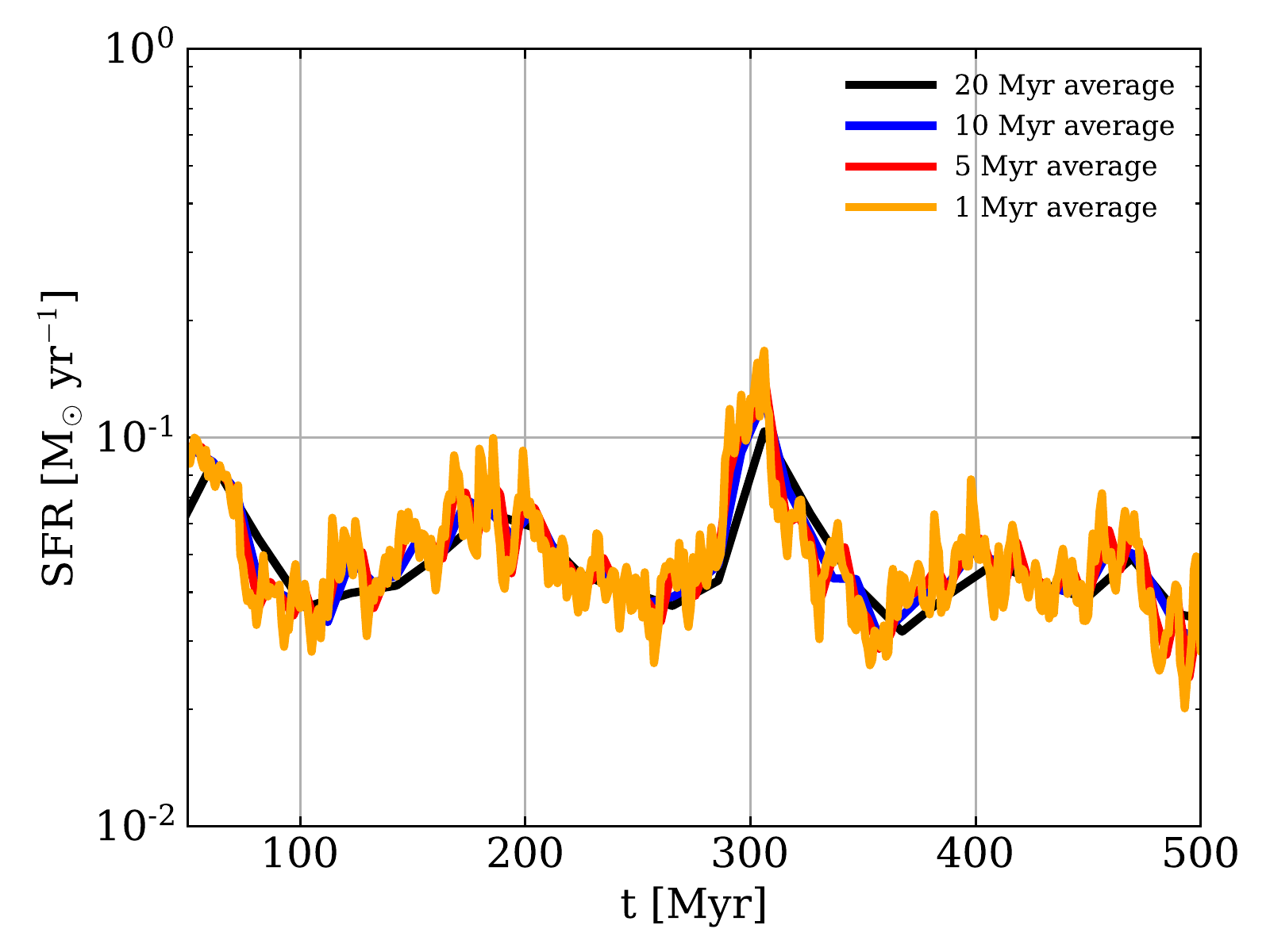}
    \includegraphics[scale=0.48]{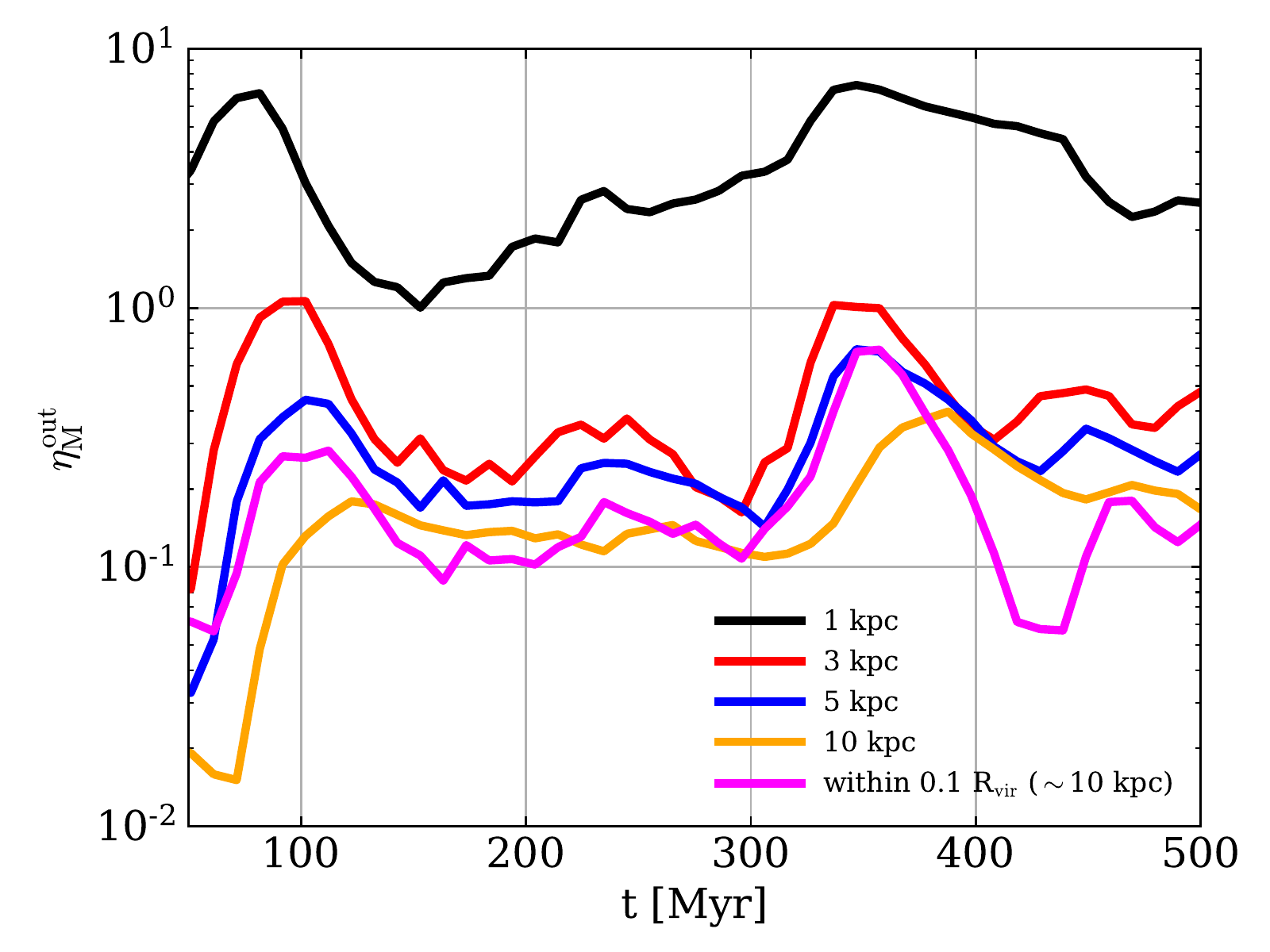}
    \includegraphics[scale=0.48]{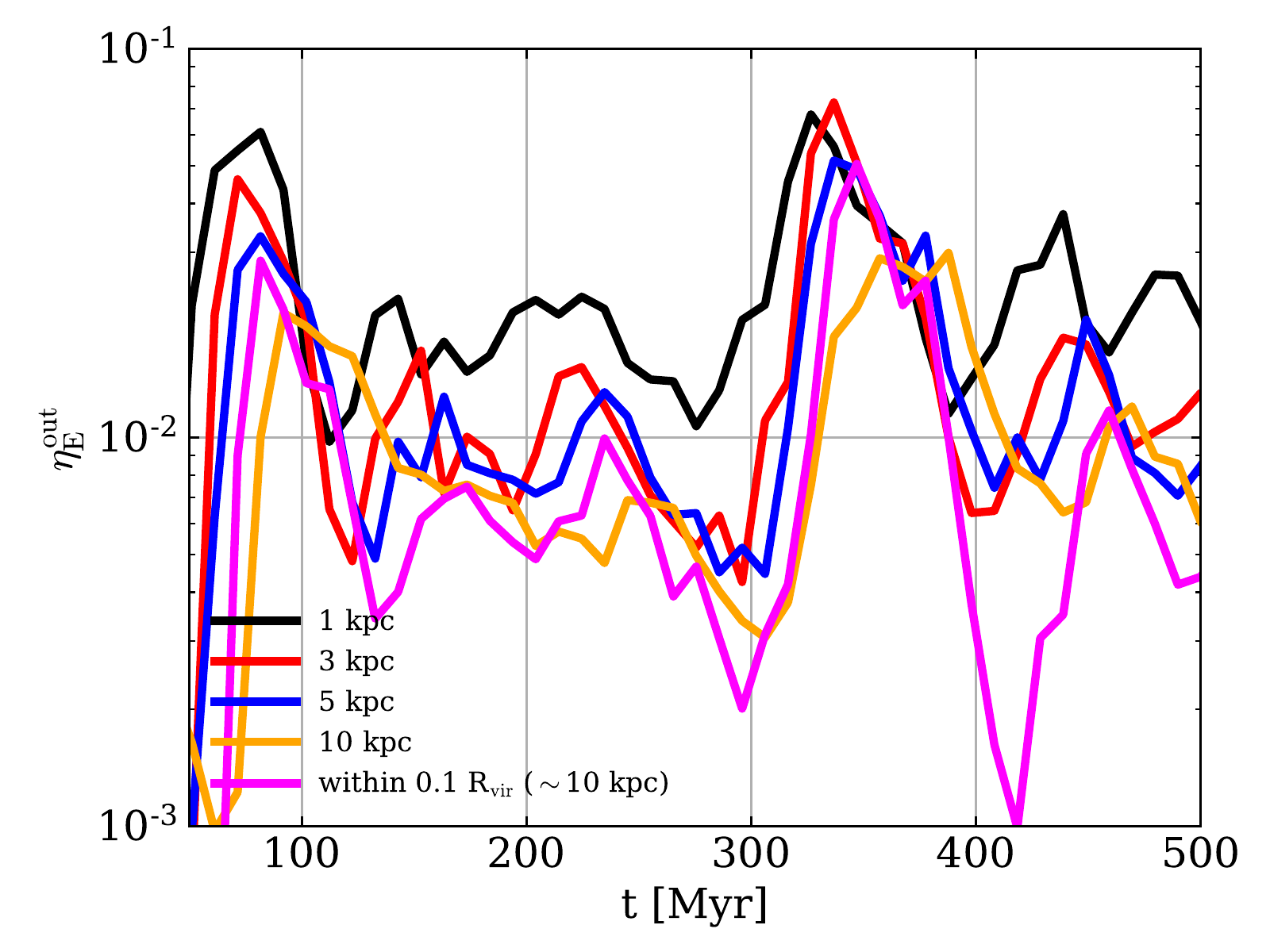}
    \caption{The top panel shows the time evolution of the star formation rate, averaged over different timescales of 20 Myr (black), 10 Myr (blue), 5 Myr (red) and 1 Myr (orange). The center panel shows the time evolution of the mass-loading factor measured at heights of 1 kpc (black), 3 kpc (red), 5 kpc (blue) and 10 kpc (orange). In the bottom panel we show the energy loading as a function of time, again measured at the same four heights above and below the mid plane. The mass loading is normalized to the mean star formation rate of $\sim0.06$ M$_{\odot}$ yr$^{-1}$ and the energy loading is normalized to the star formation rate per 100 M$_{\odot}$ times $10^{51}$ erg, as the canonical SN-energy, and is thus a measure of how much energy escapes per SN.}
    \label{fig:sfr}
\end{figure}

\begin{figure*}
    \centering
    \includegraphics[width=0.49\textwidth]{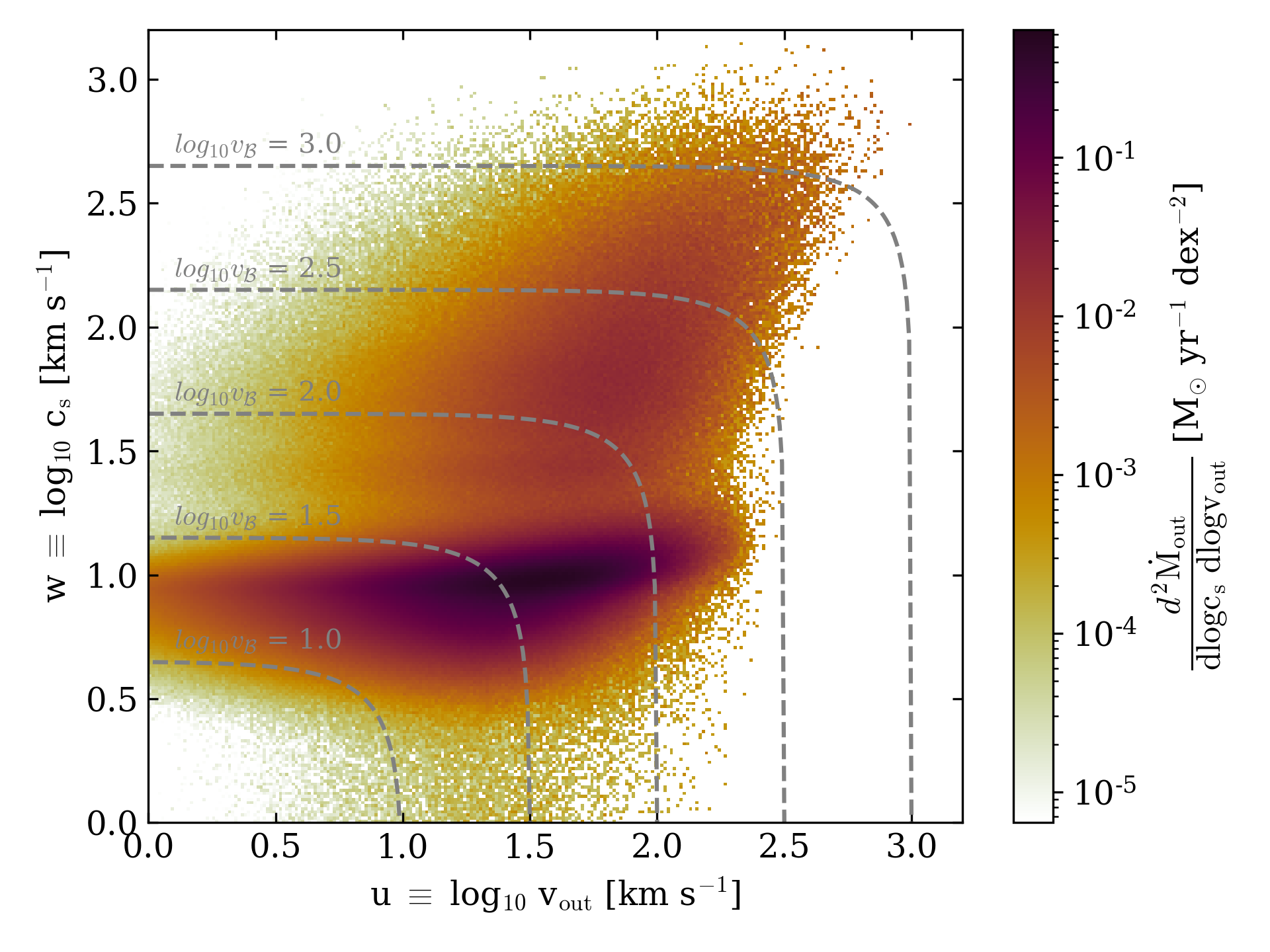}
    \includegraphics[width=0.49\textwidth]{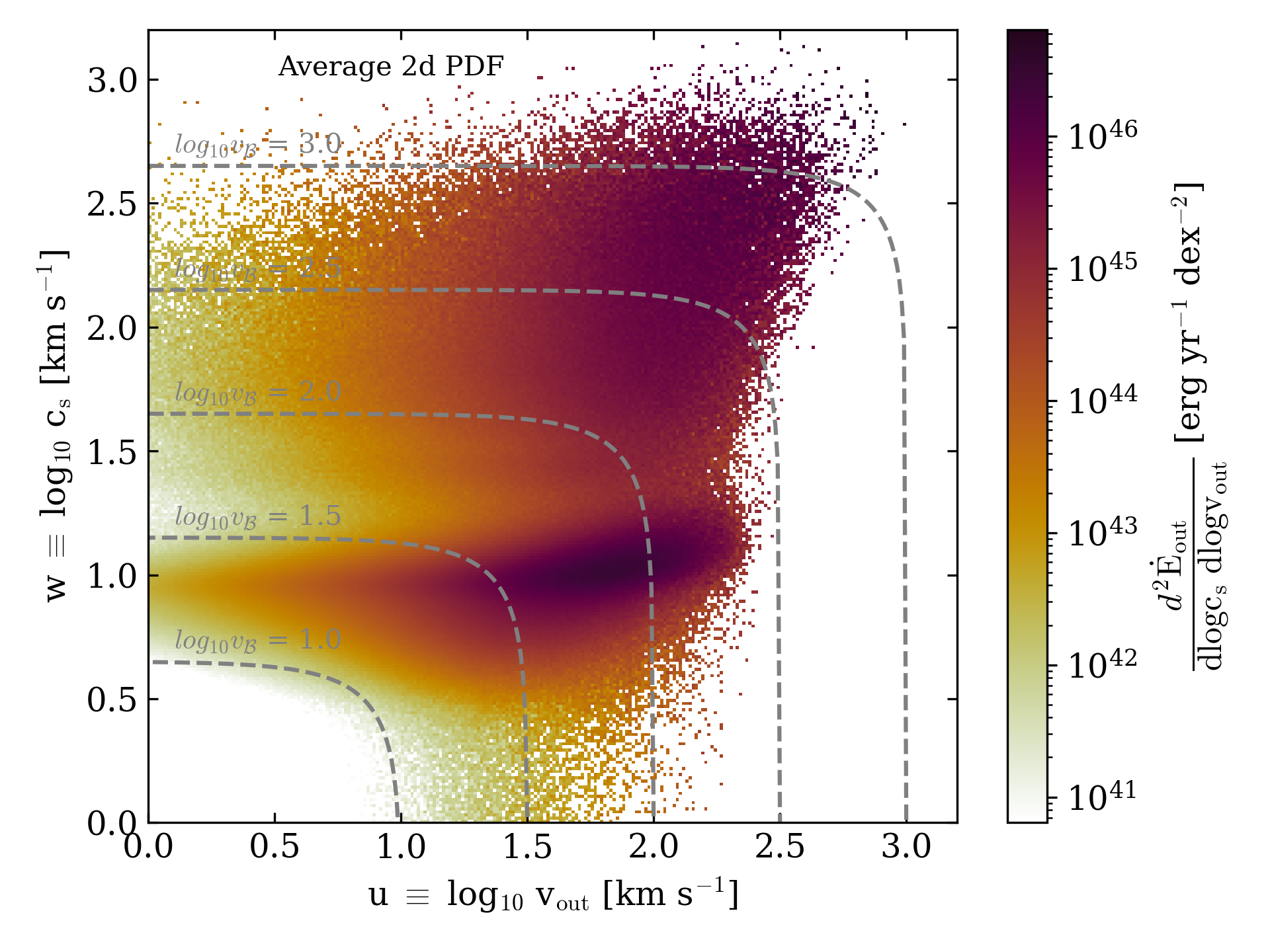}
    \caption{2d PDFs of outflow velocity and sound speed color coded by mass outflow rate (left) and energy outflow rate (right), at a height of 1 kpc for the radial region between $0.0 < R < 5.0$. We note that the left panel is the sum of the three panels of the second row in Fig.~\ref{fig:2dpdfs_mass_comp} and the right panel is the sum of the three panels of the second row in Fig.~\ref{fig:2dpdfs_energy_comp}. The grey lines in the PDFs define contours based on equal Bernoulli-velocity as defined by equation eq. \ref{eq:Bernoulli}. All the PDFs presented in this paper are time averaged between 200 and 400 Myrs of evolution.} 
    \label{fig:mass_energy_outflow}
\end{figure*}

\begin{figure*}
\begin{tabular}{ c c c }
$ \hspace{0.2cm} \mathbf{0.0 < R < 1.5}$ \textbf{kpc} & \hspace{2.65cm} $ \mathbf{1.5 < R < 3.0}$ \textbf{kpc} & \hspace{2.7cm} $ \mathbf{3.0 < R < 5.0}$ \textbf{kpc}   
\end{tabular}

    \centering
    \includegraphics[width=0.32\textwidth]{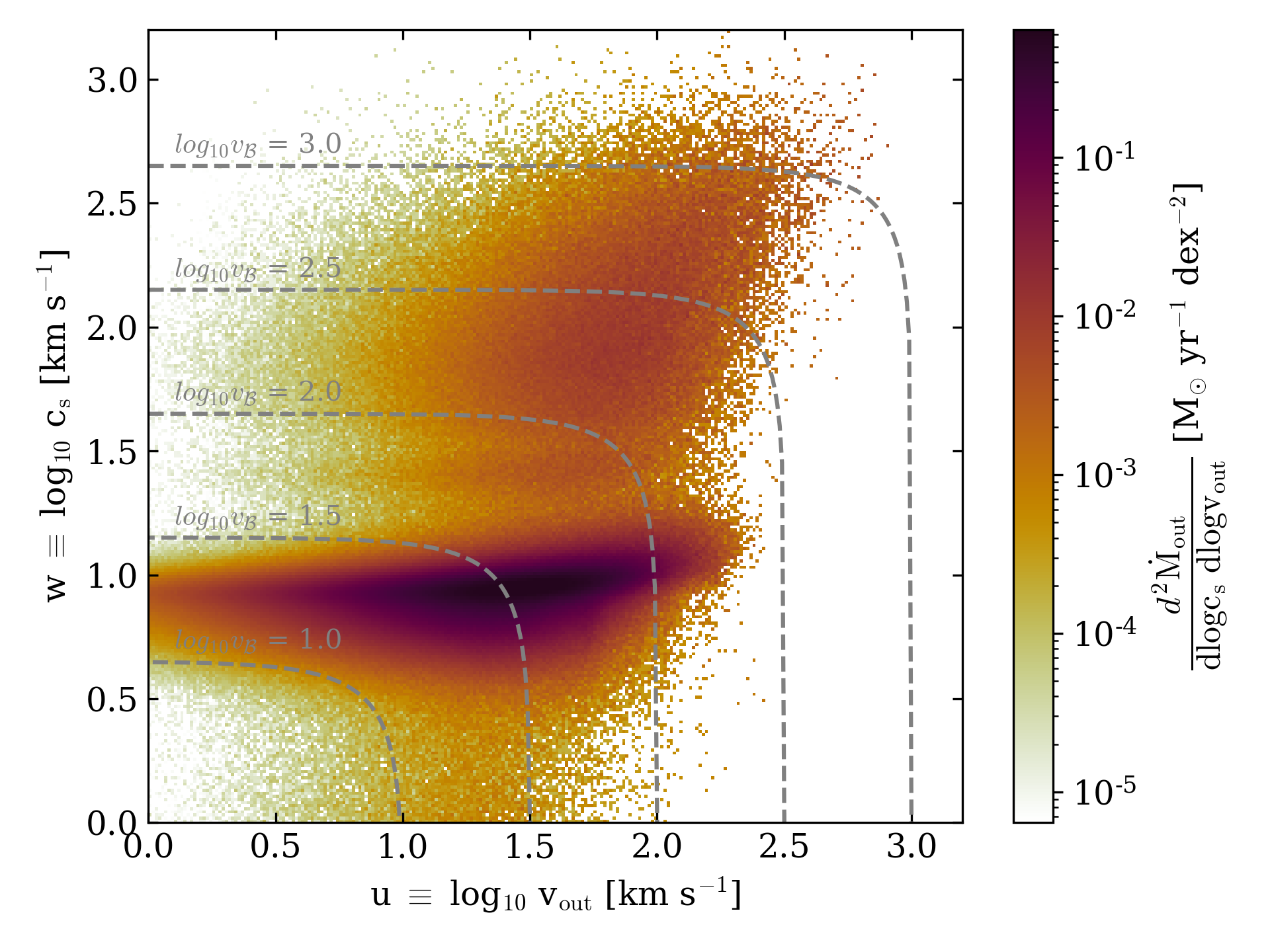}
   \includegraphics[width=0.32\textwidth]{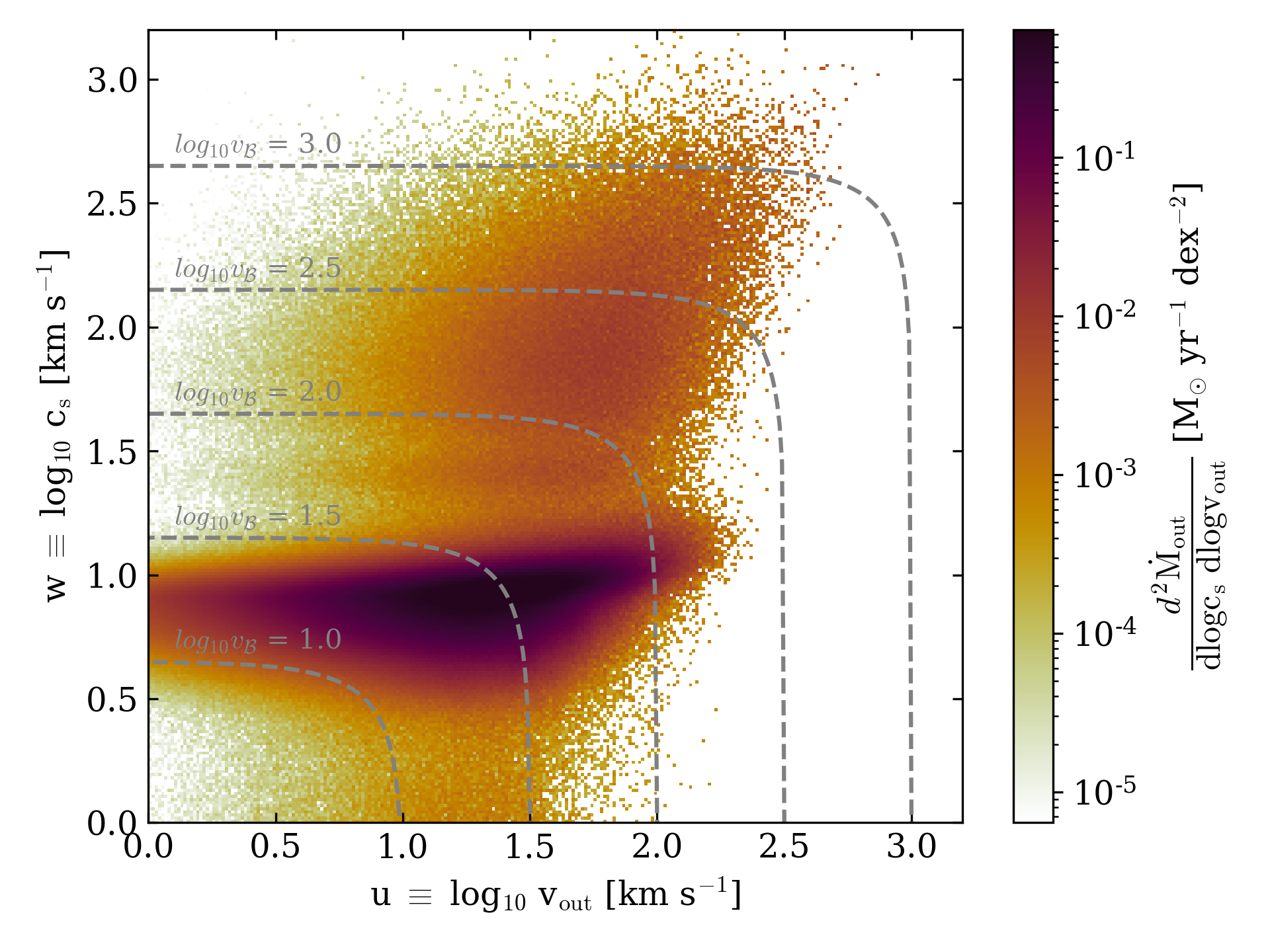}
   \includegraphics[width=0.32\textwidth]{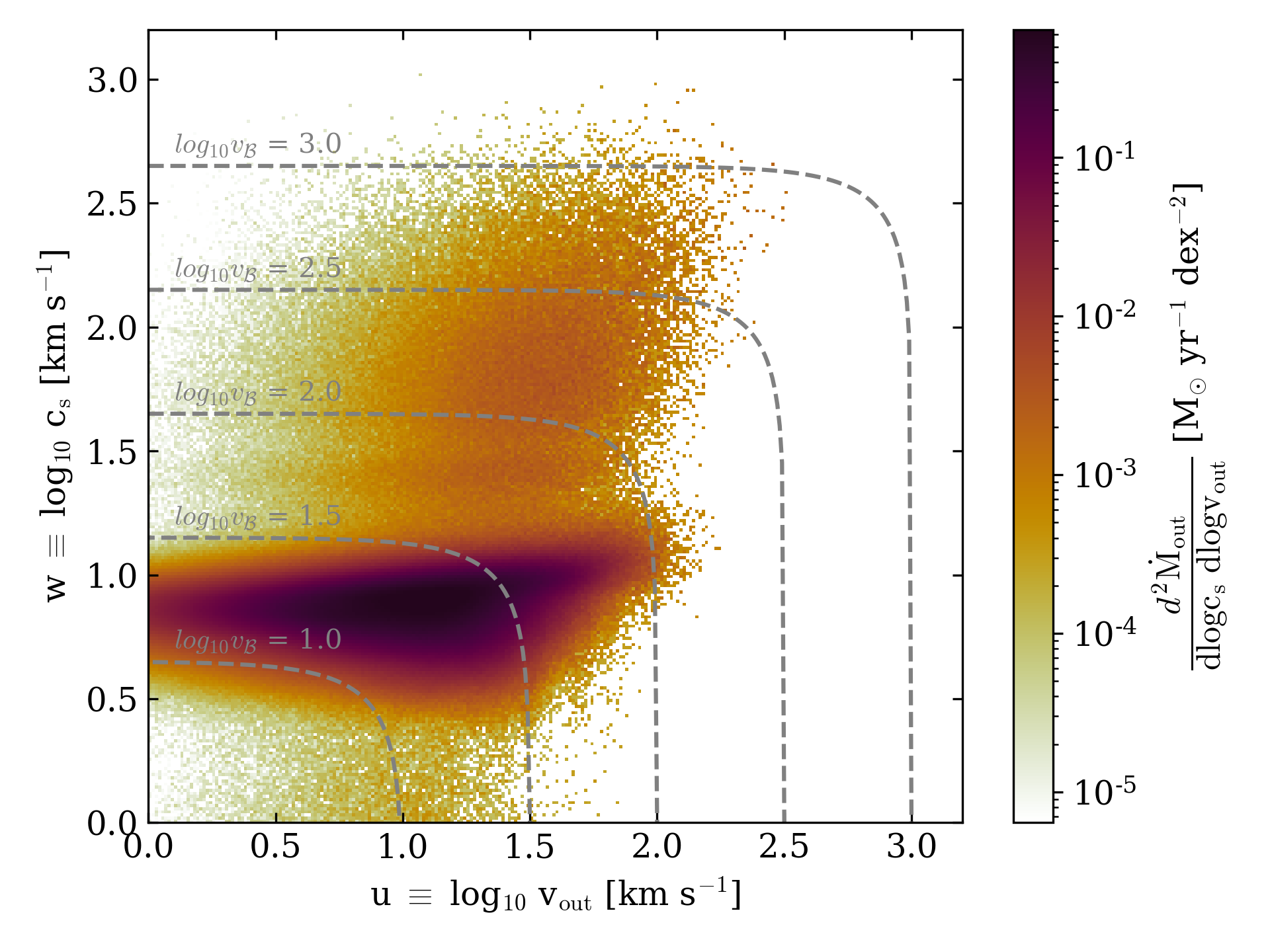}
   
    \includegraphics[width=0.32\textwidth]{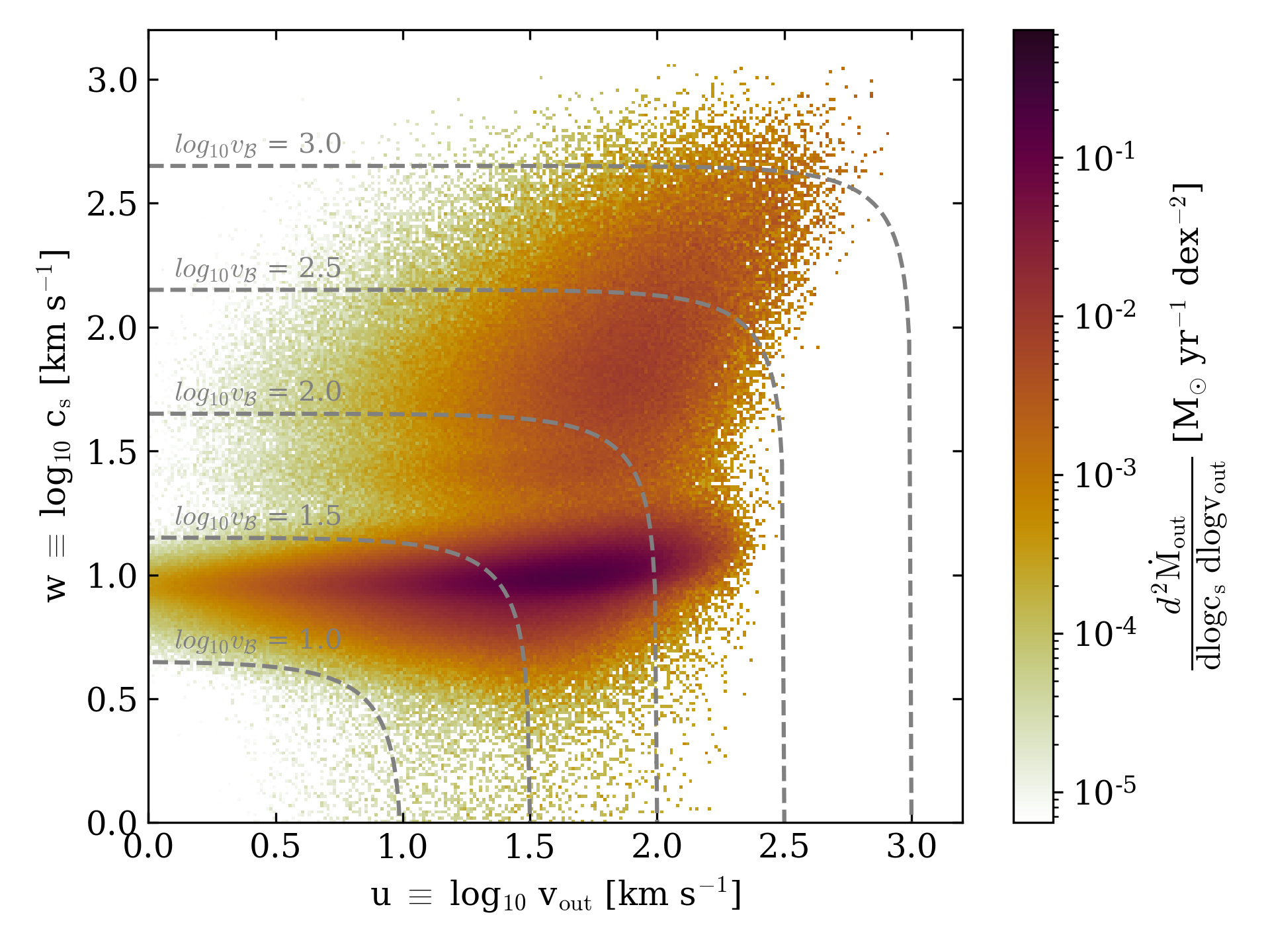}
   \includegraphics[width=0.32\textwidth]{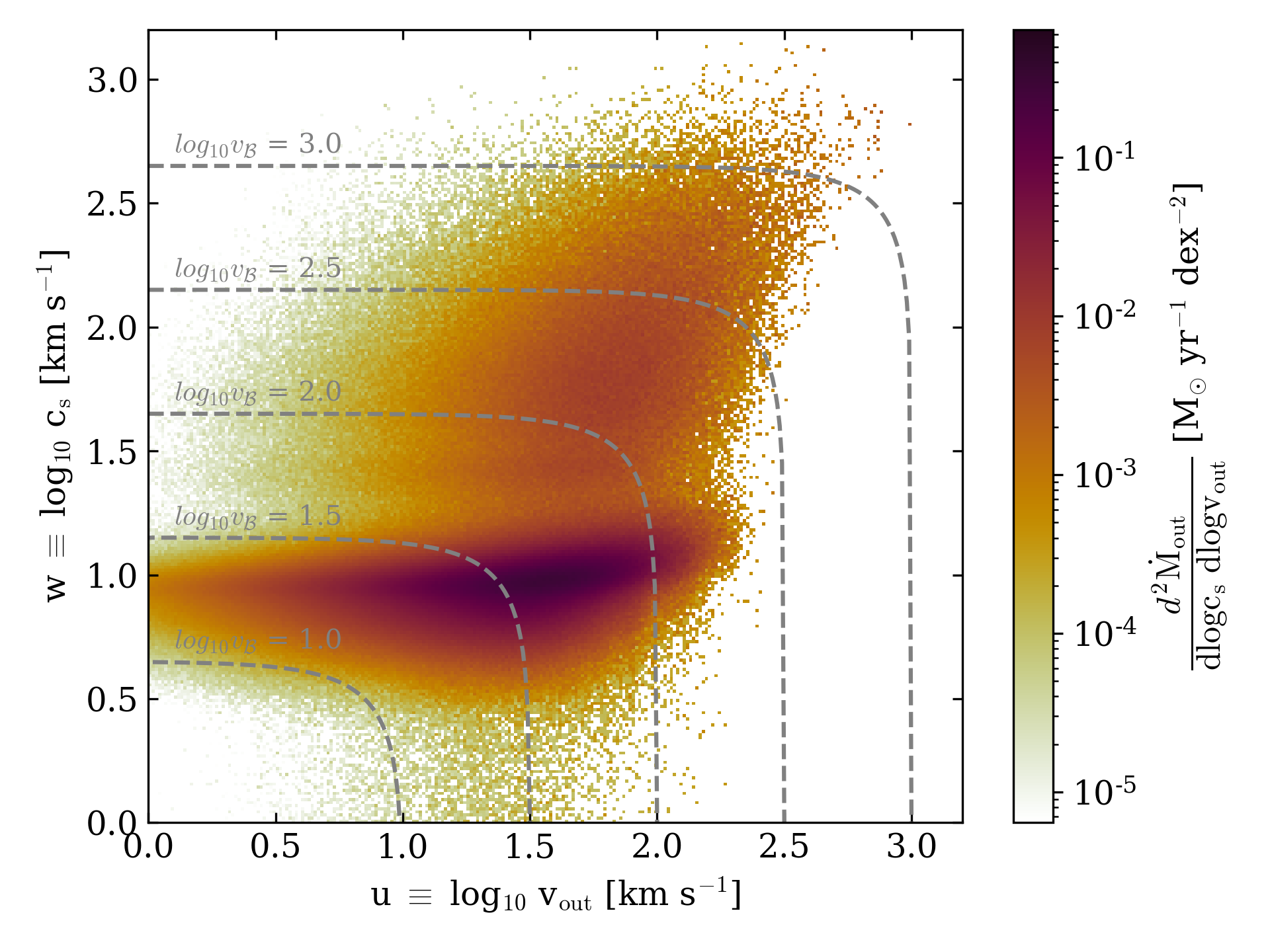}
   \includegraphics[width=0.32\textwidth]{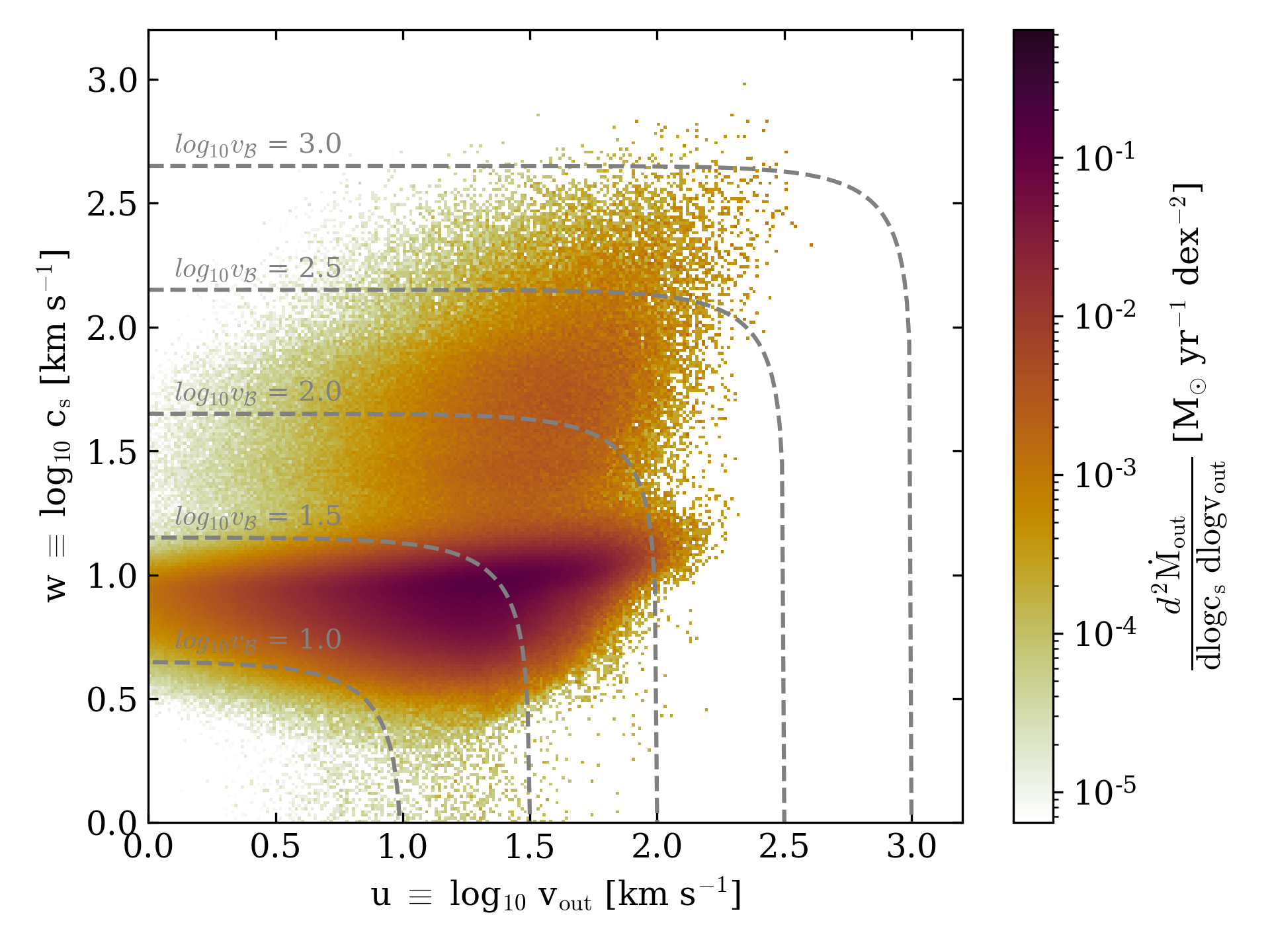}
   
   \includegraphics[width=0.32\textwidth]{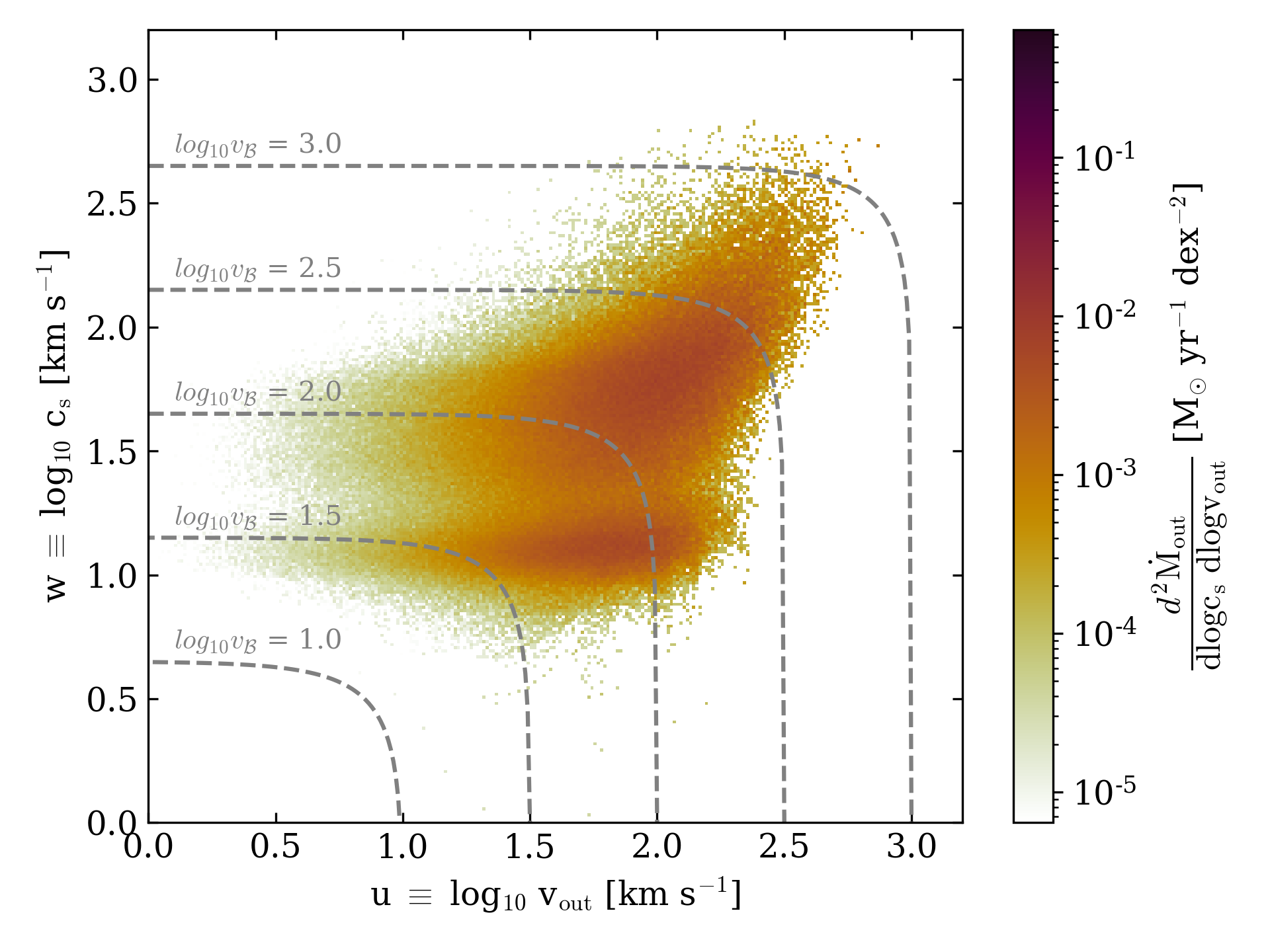}
   \includegraphics[width=0.32\textwidth]{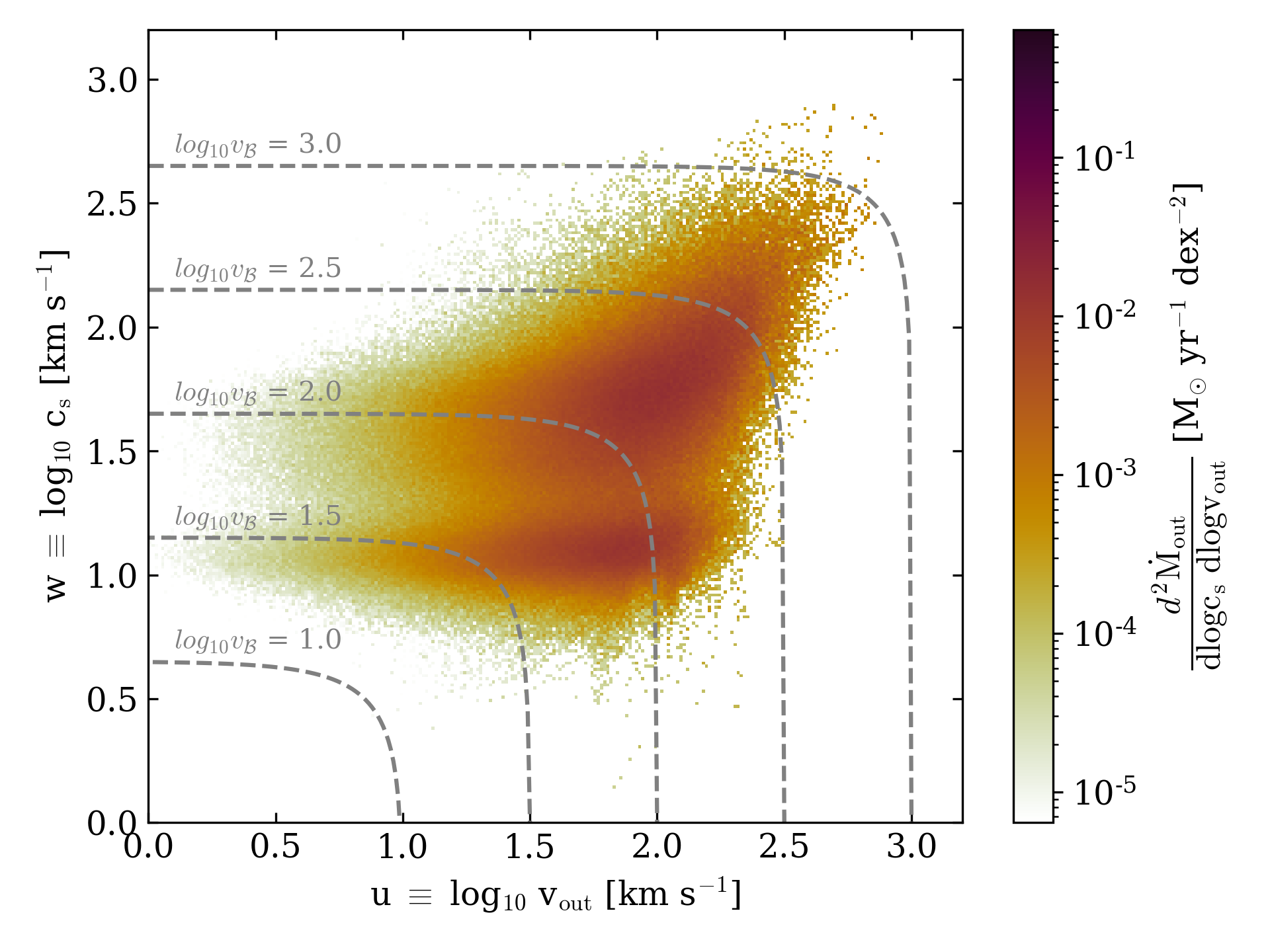}
    \includegraphics[width=0.32\textwidth]{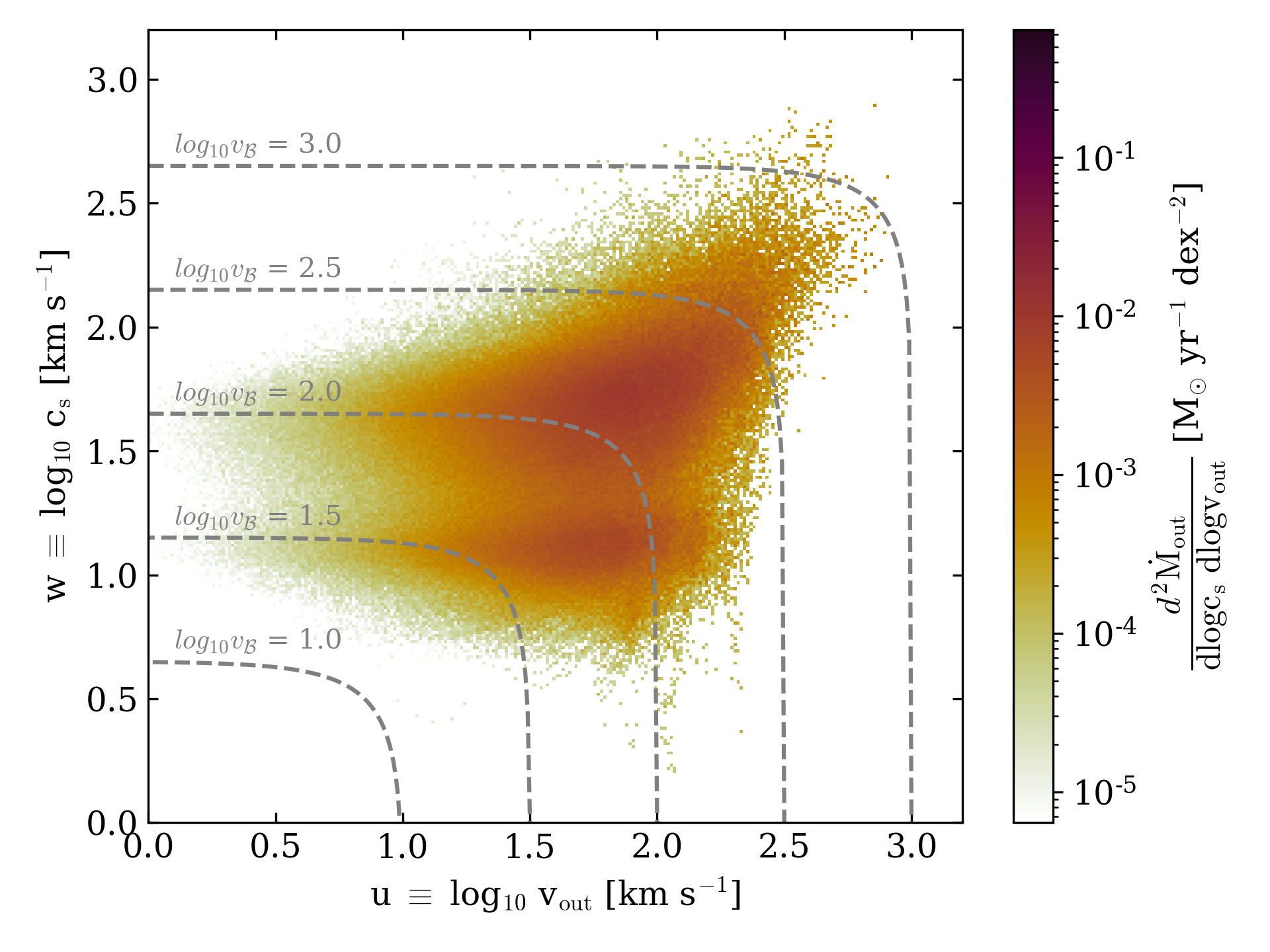}
    
    \includegraphics[width=0.32\textwidth]{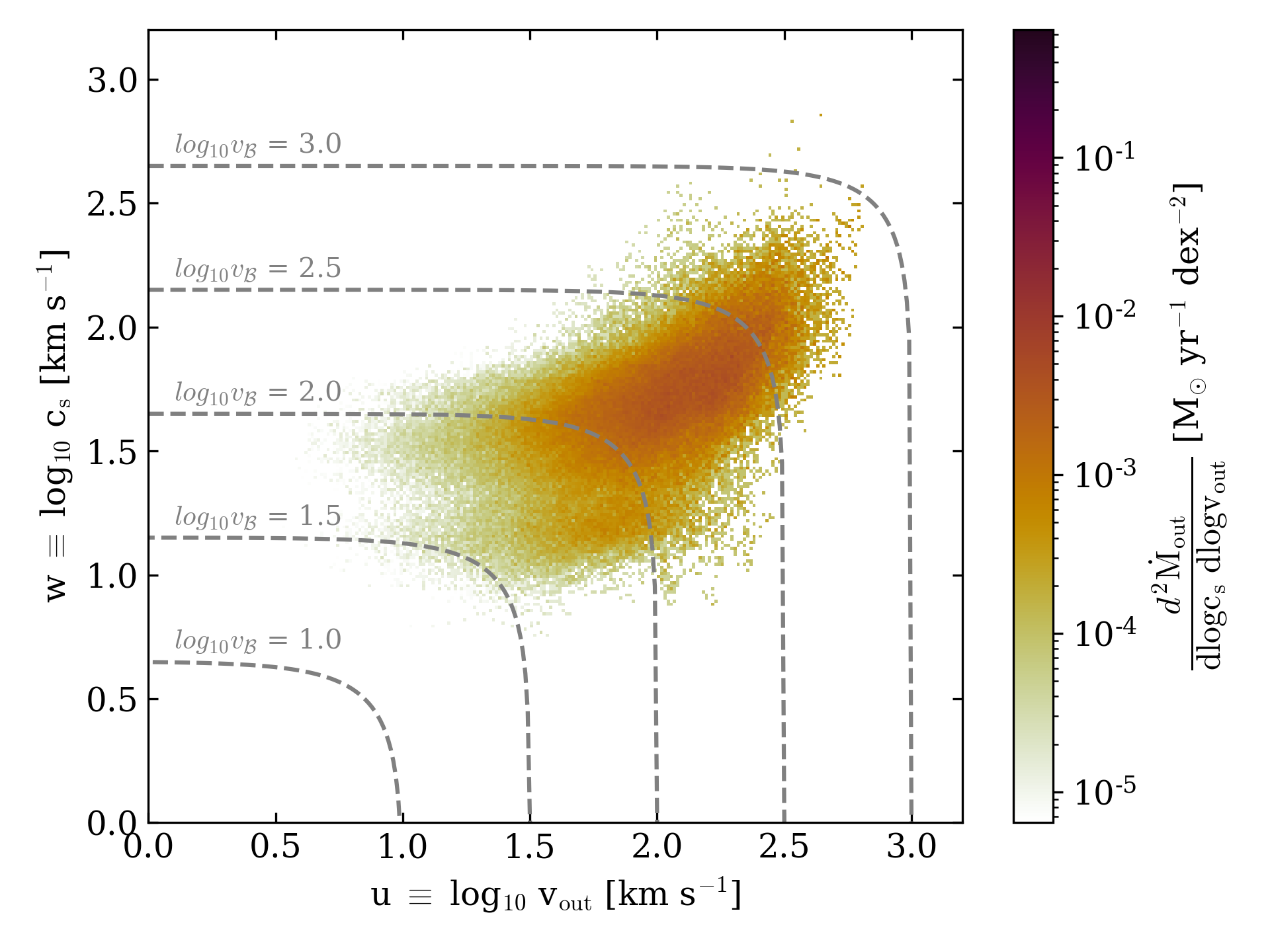}
   \includegraphics[width=0.32\textwidth]{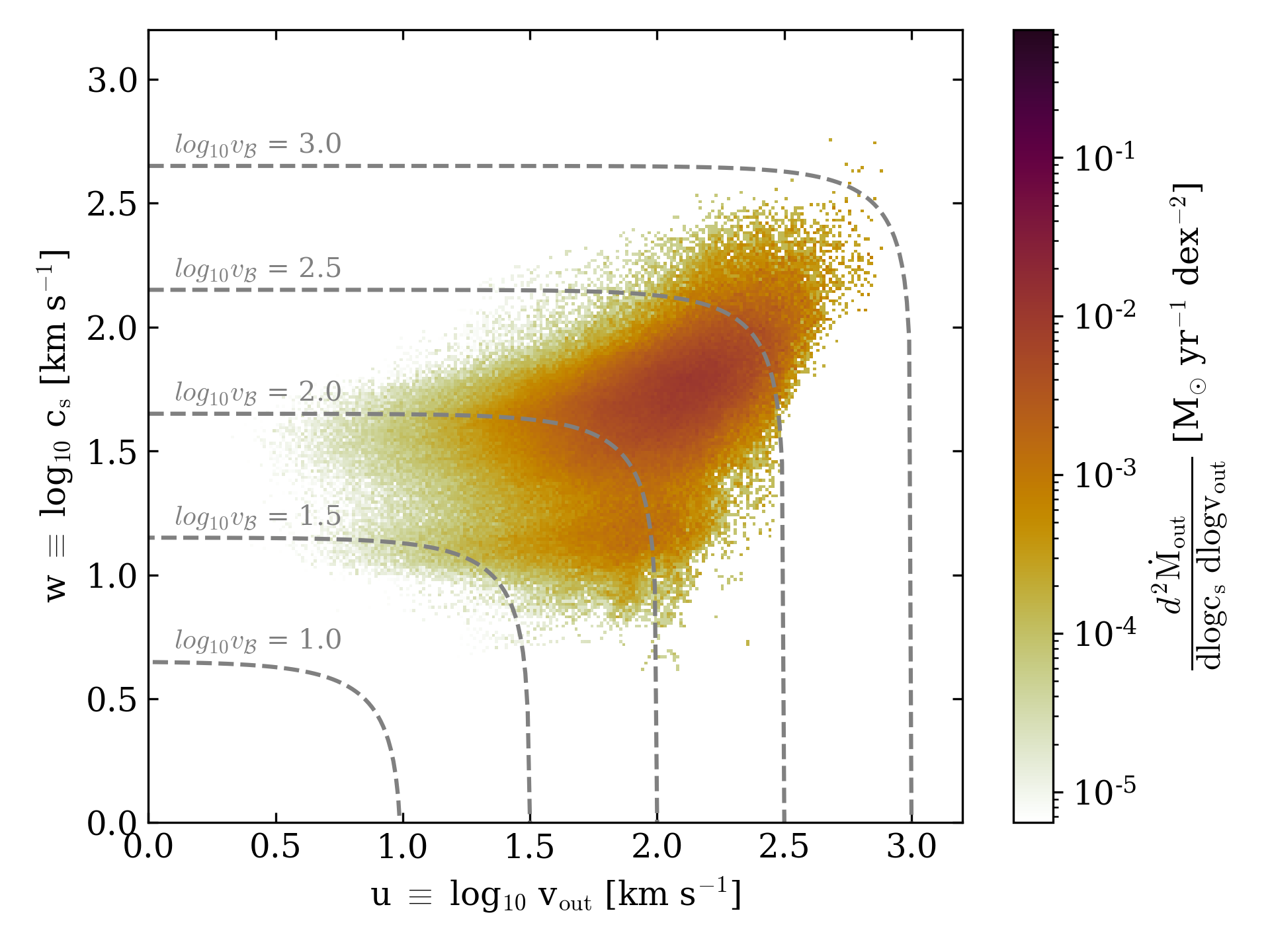}
   \includegraphics[width=0.32\textwidth]{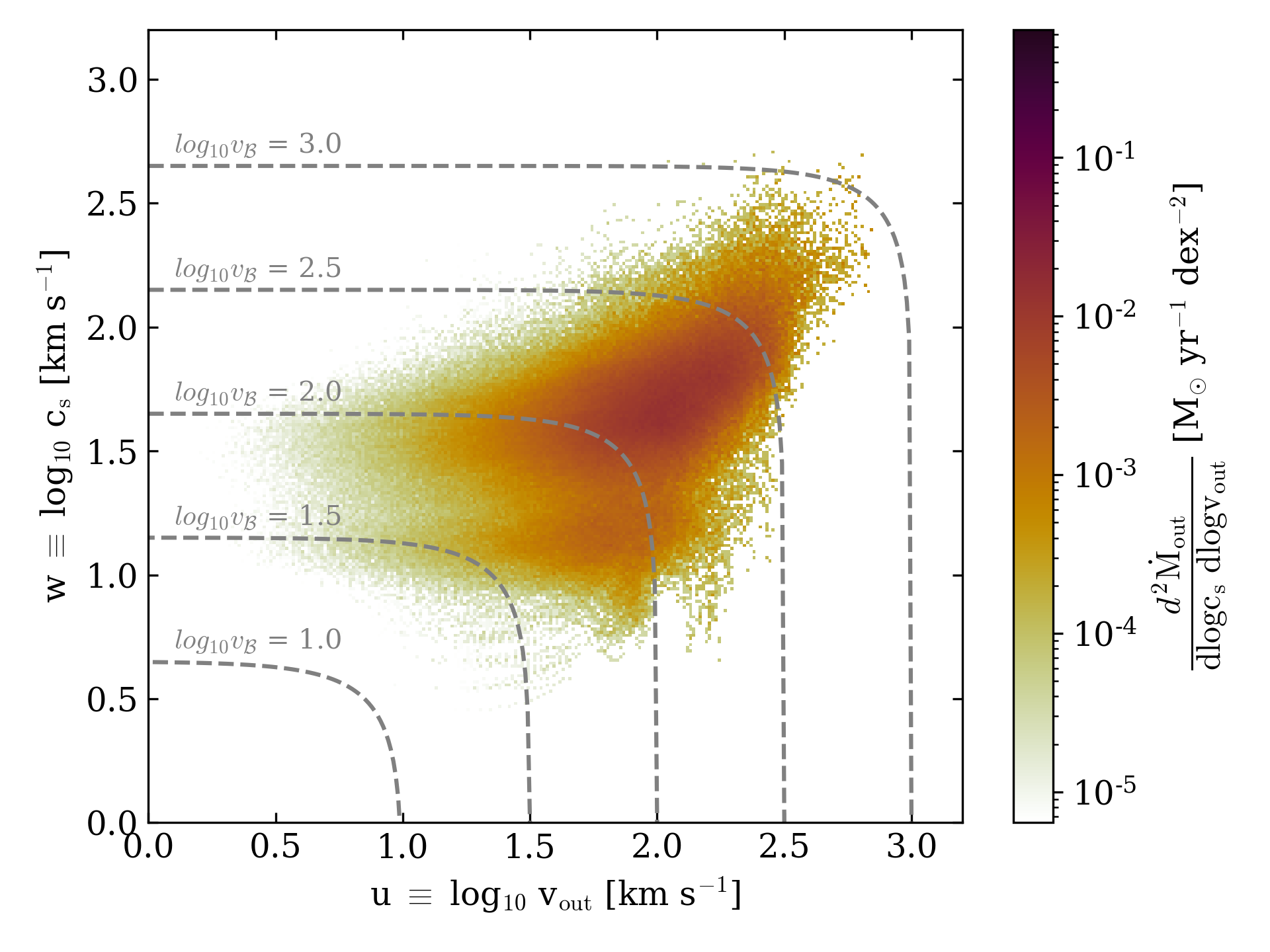}
   
    \includegraphics[width=0.32\textwidth]{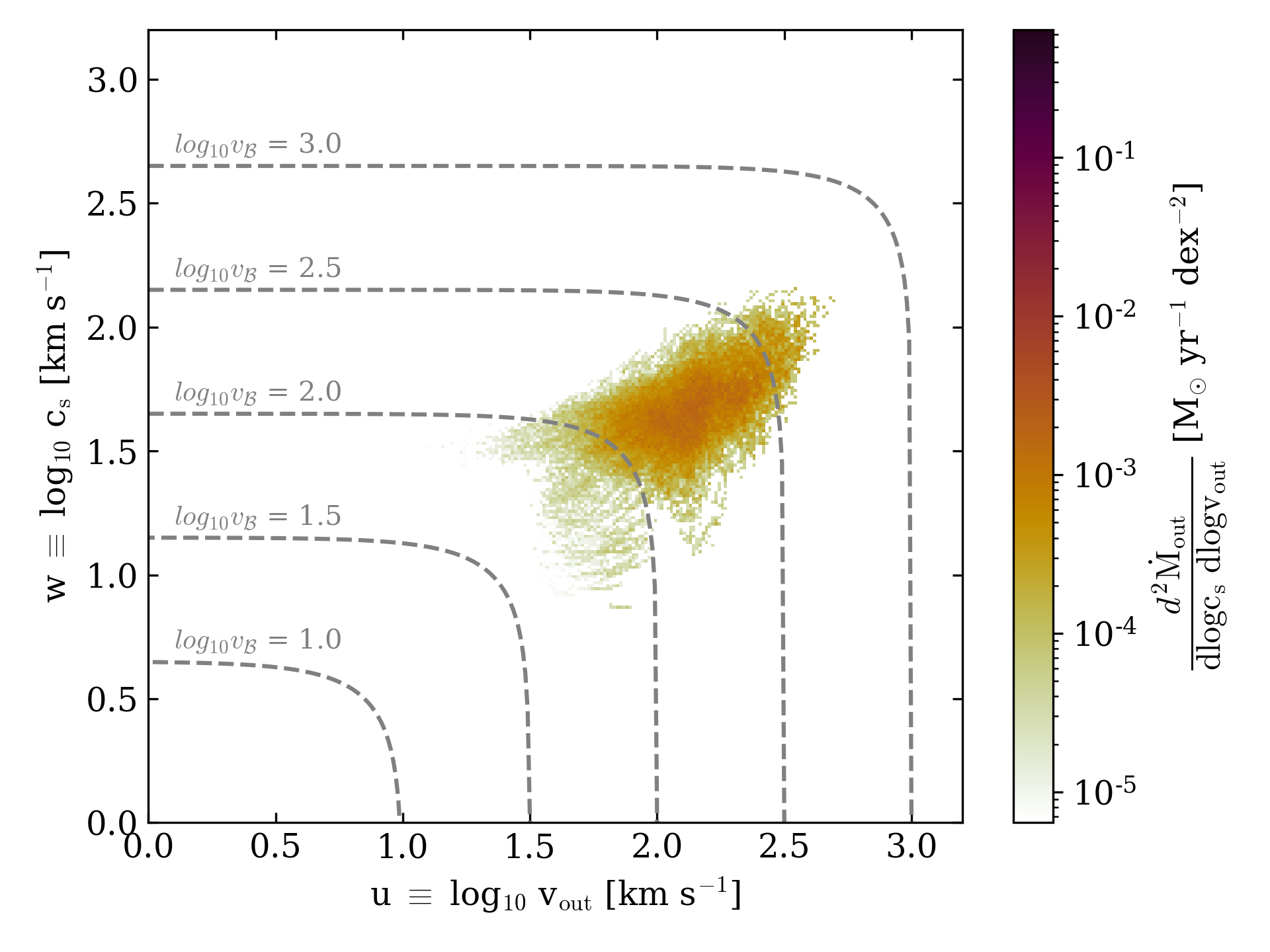}
   \includegraphics[width=0.32\textwidth]{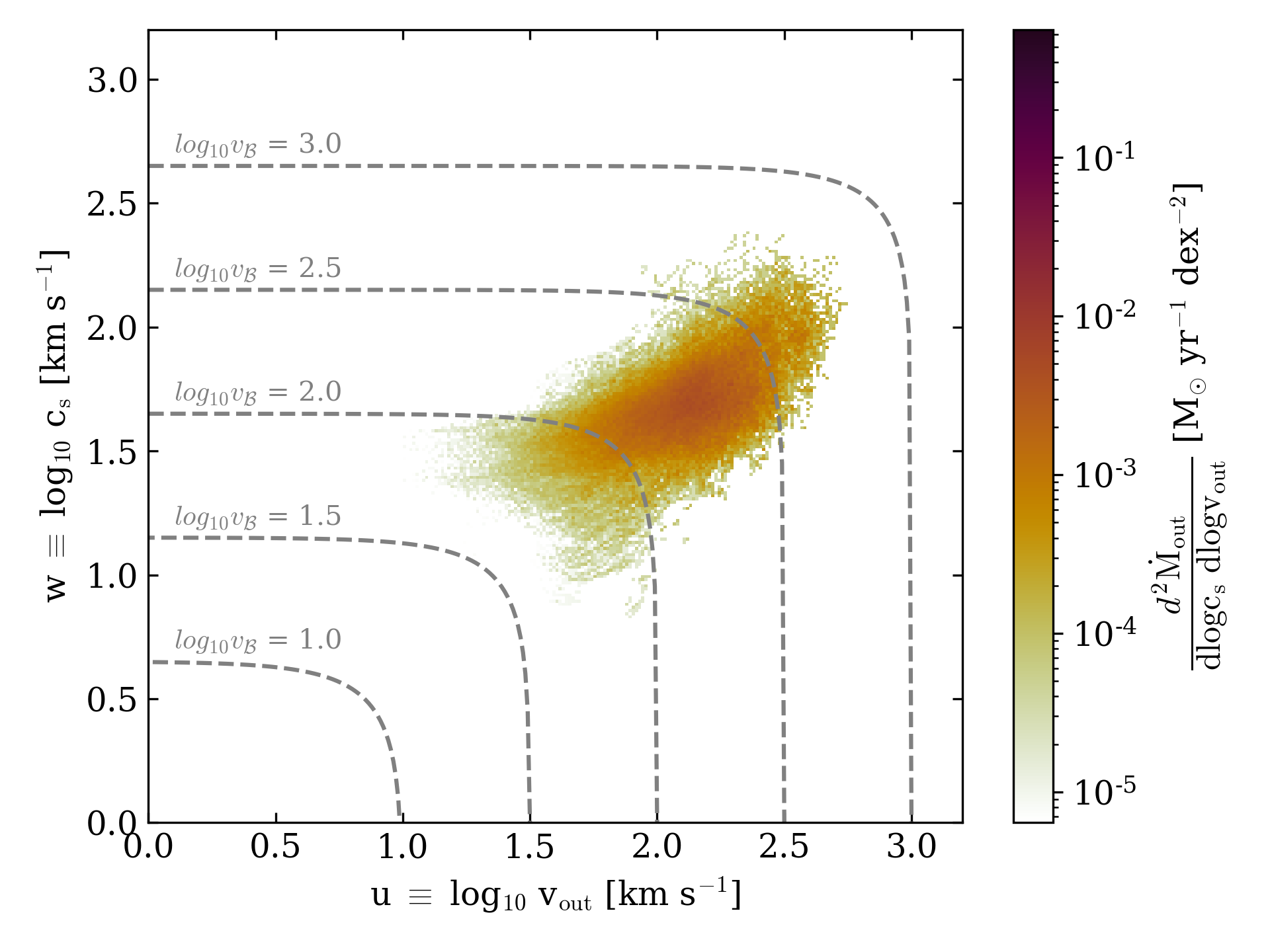}
   \includegraphics[width=0.32\textwidth]{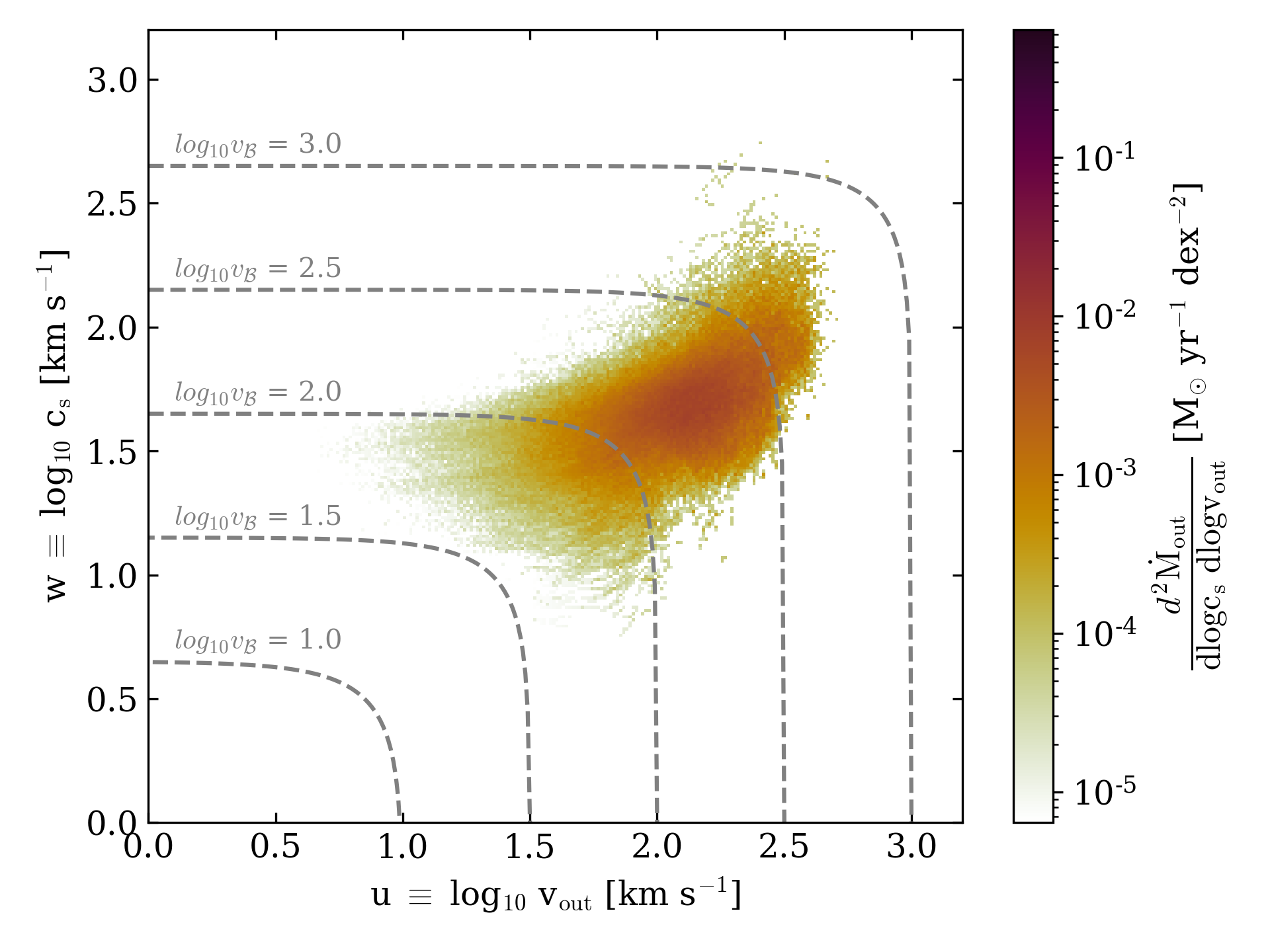}
    \caption{Compilation of the 2d joint PDFs for mass outflow rate (mass loading) for different regions in the galaxy. The columns from left to right show the PDFs in different radial regions $0.0 < R < 1.5$, $1.5 < R < 3.0$ and $3.0 < R < 5.0$. The rows from top to bottom show the joint PDFs at heights of 0.5, 1, 3, 5 and 10 kpc in slabs with a thickness of 50, 100, 300, 500 and 1000 pc, between $0.0 < R < 5.0$ kpc. The grey dashed lines represent surfaces of equal Bernoulli-velocity following eq.~\ref{eq:Bernoulli}. All PDFs are averaged between 200 and 400 Myr of evolution.}
    \label{fig:2dpdfs_mass_comp}
\end{figure*}

\begin{figure*}
\begin{tabular}{ c c c }
$ \hspace{-1.5cm} \mathbf{0.0 < R < 1.5}$ \textbf{kpc} & \hspace{2.65cm} $ \mathbf{1.5 < R < 3.0}$ \textbf{kpc} & \hspace{2.7cm} $ \mathbf{3.0 < R < 5.0}$ \textbf{kpc}   
\end{tabular}
    \centering
    
    \includegraphics[width=0.32\textwidth]{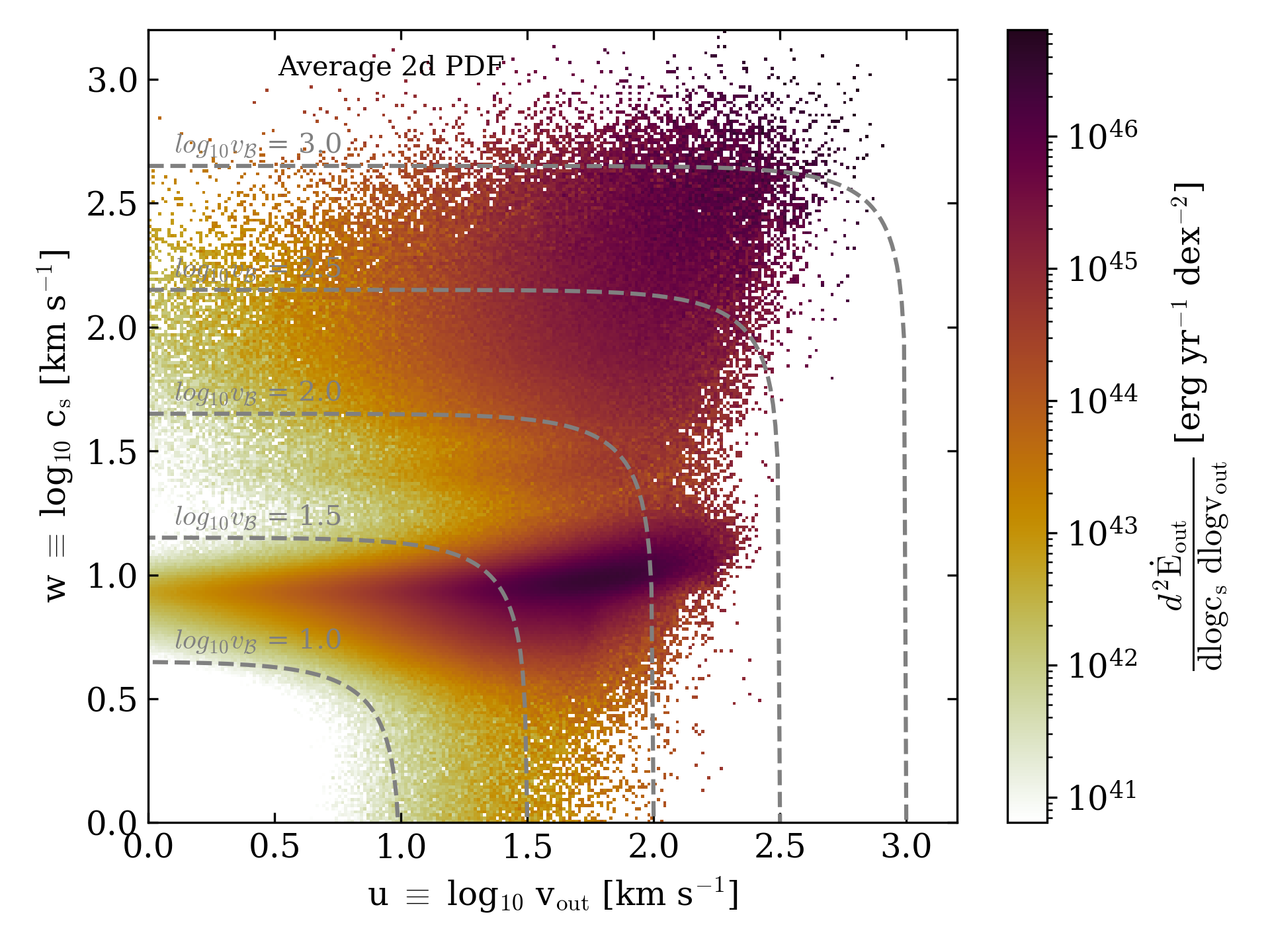}
   \includegraphics[width=0.32\textwidth]{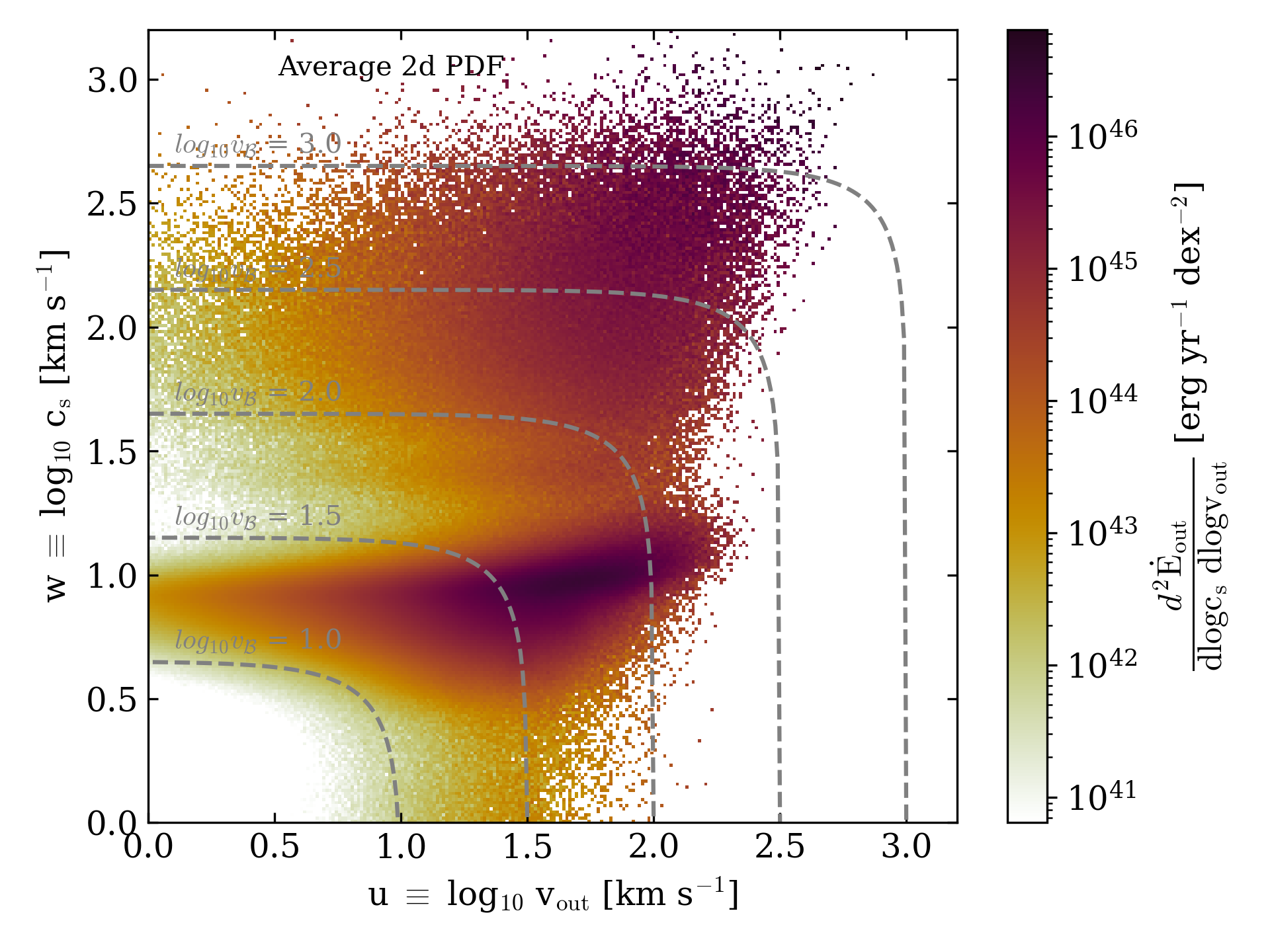}
   \includegraphics[width=0.32\textwidth]{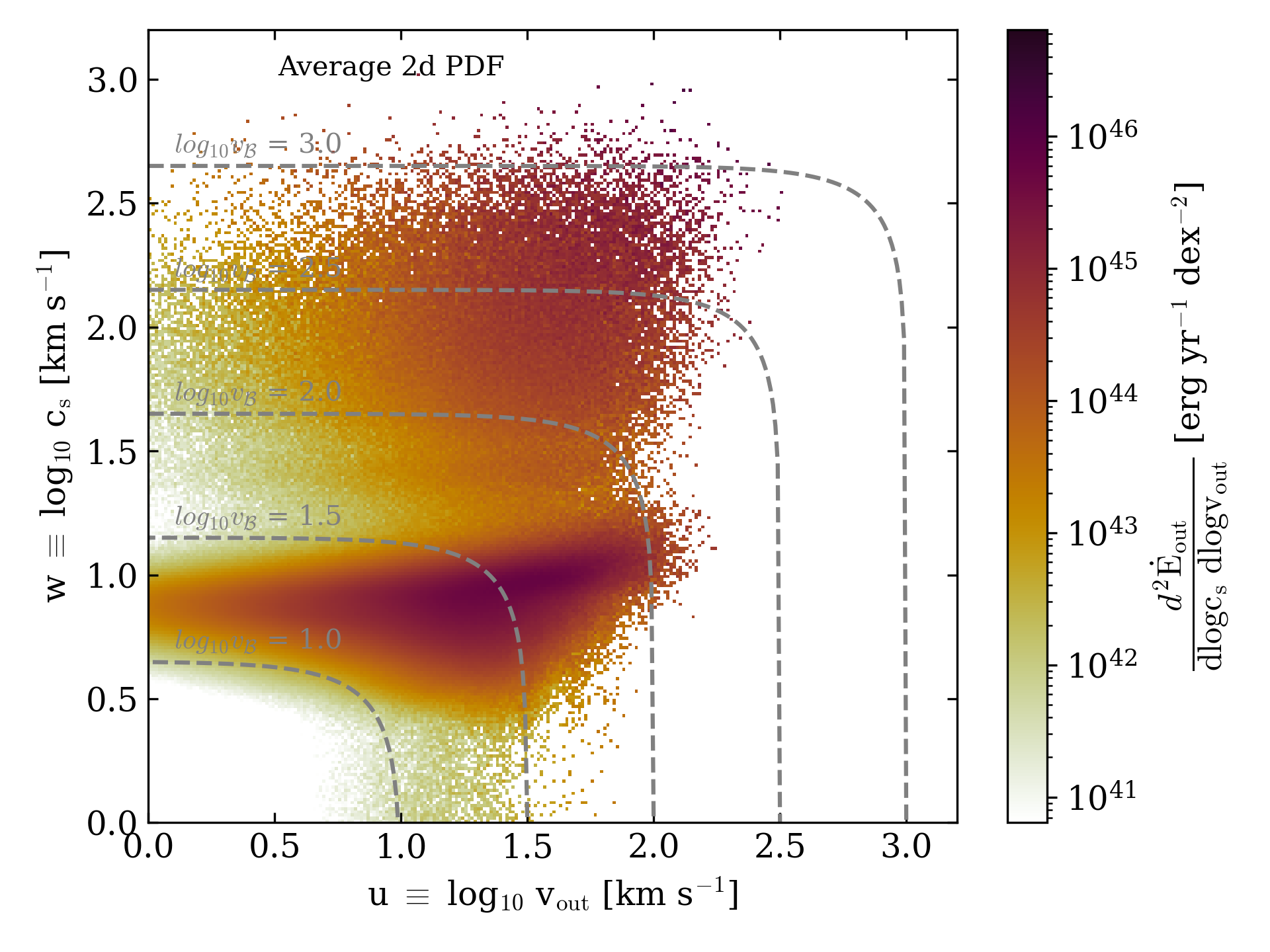}
    
    \includegraphics[width=0.32\textwidth]{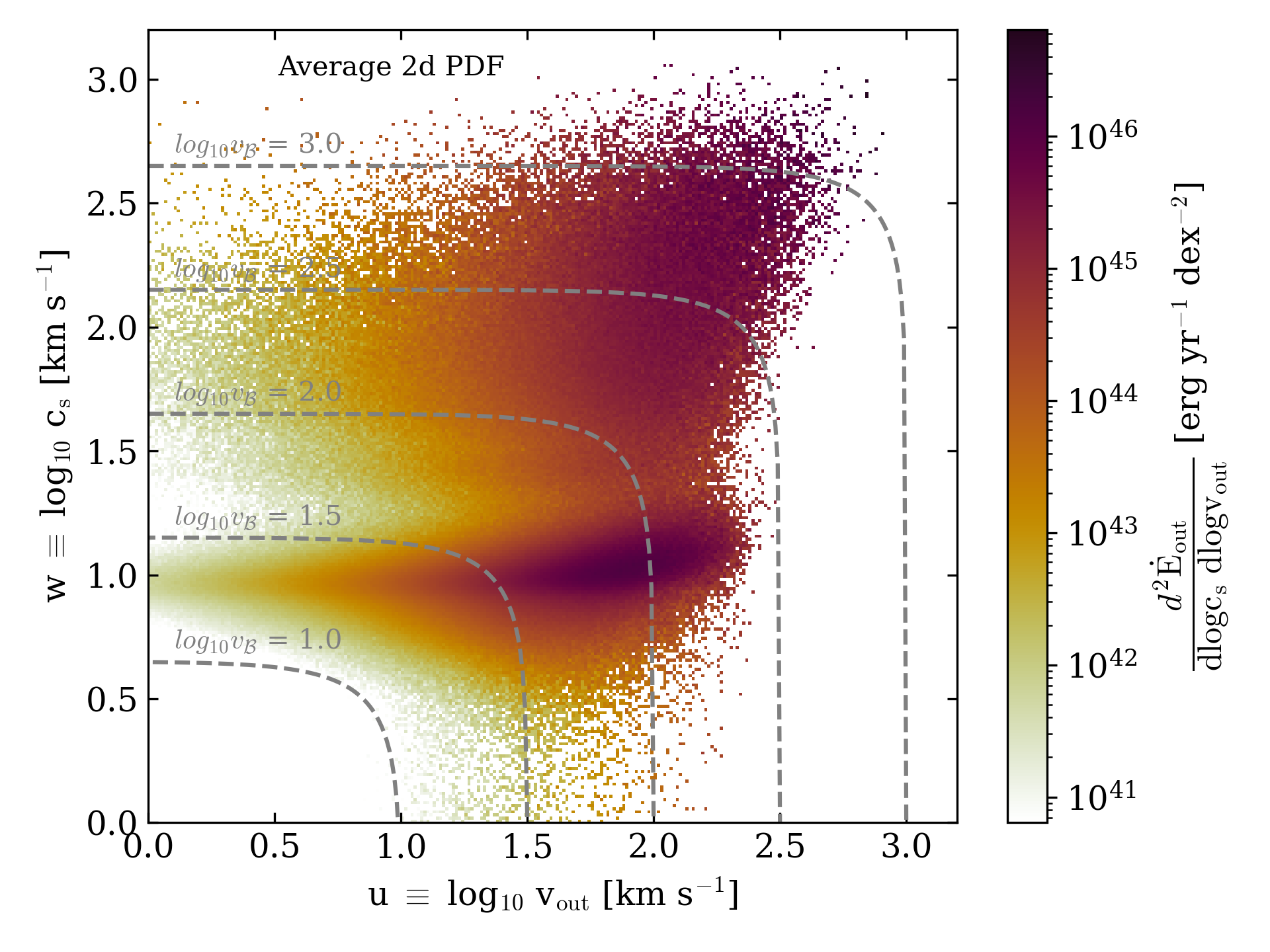}
   \includegraphics[width=0.32\textwidth]{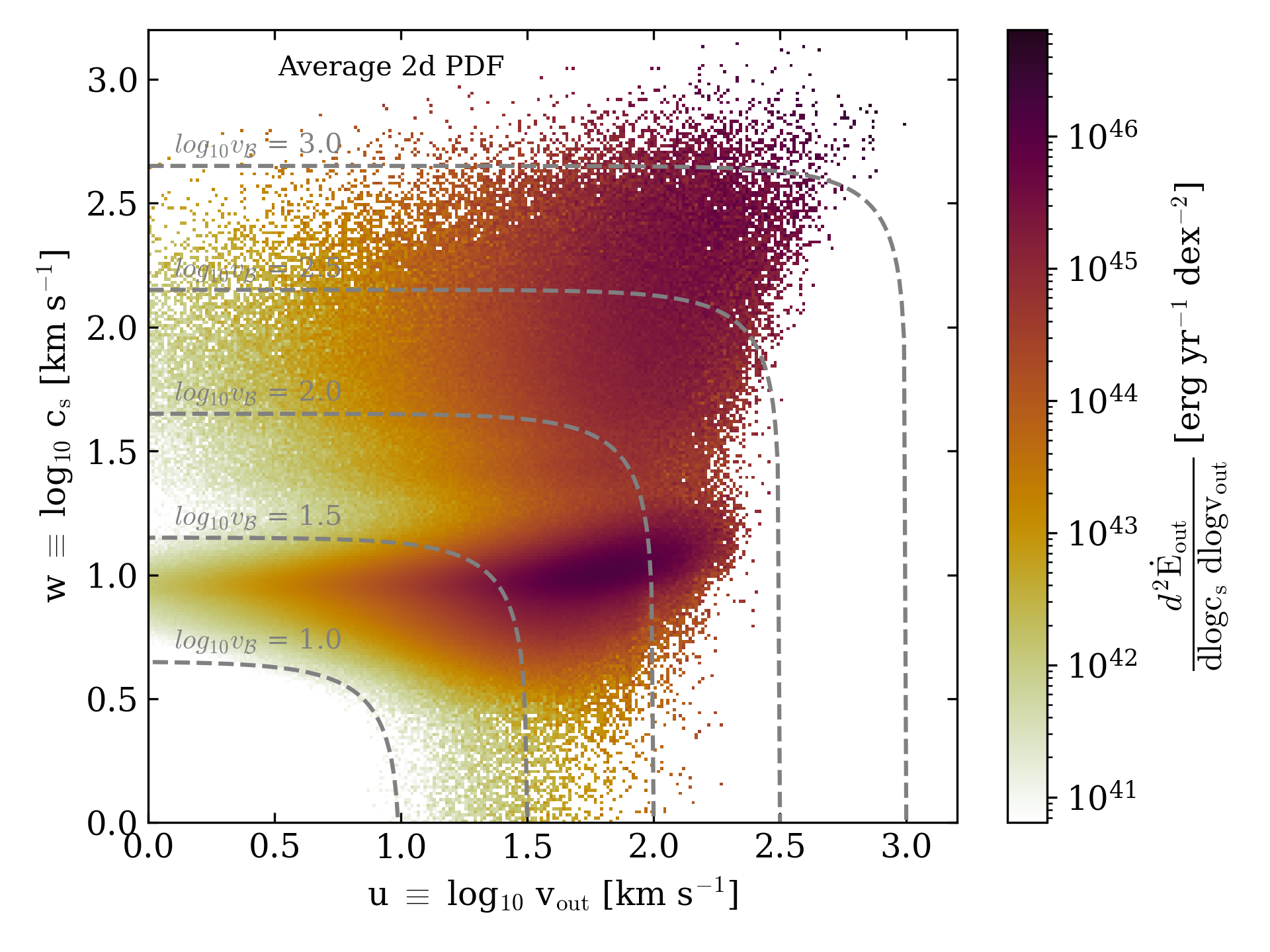}
   \includegraphics[width=0.32\textwidth]{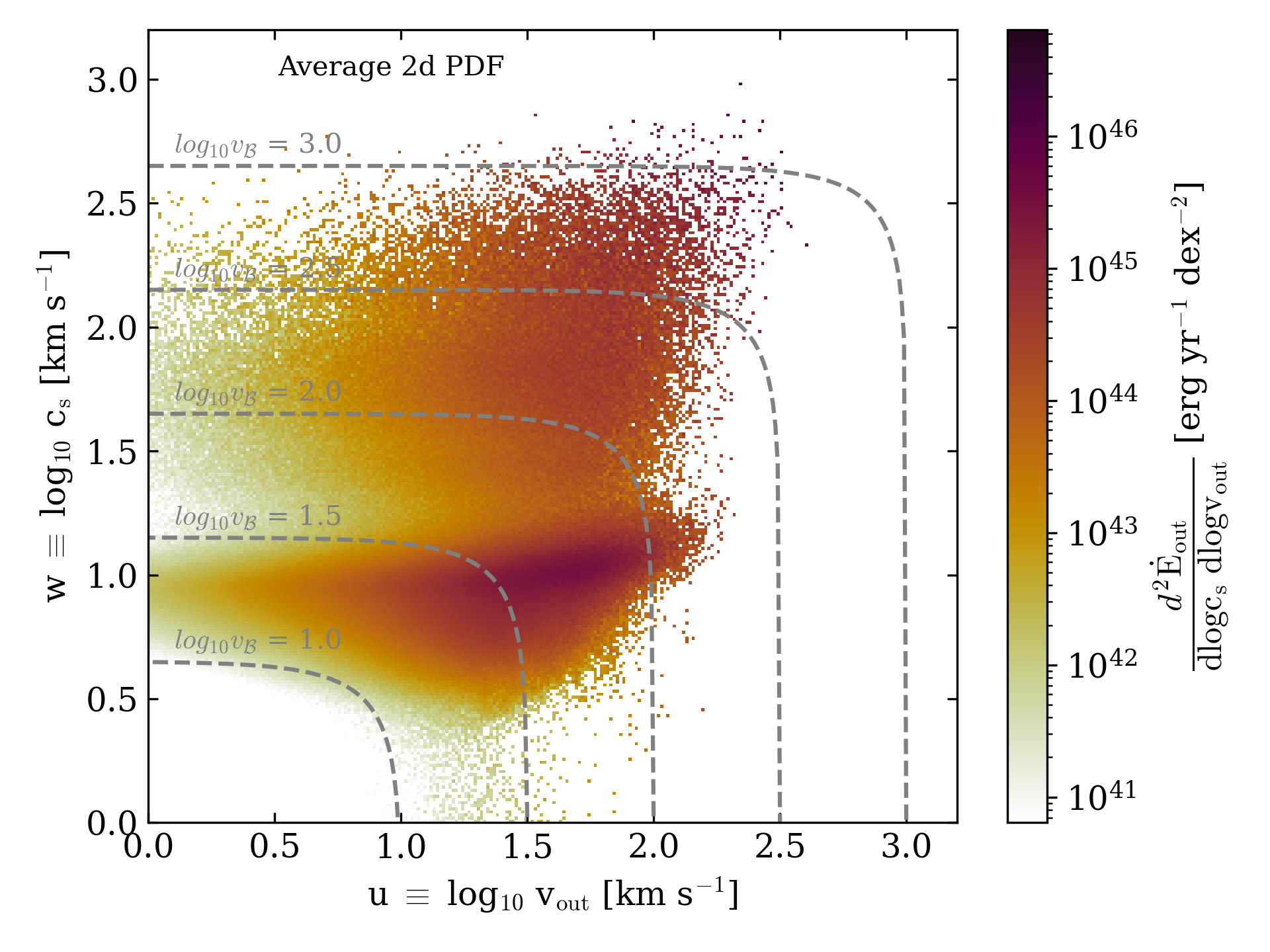}
   
    \includegraphics[width=0.32\textwidth]{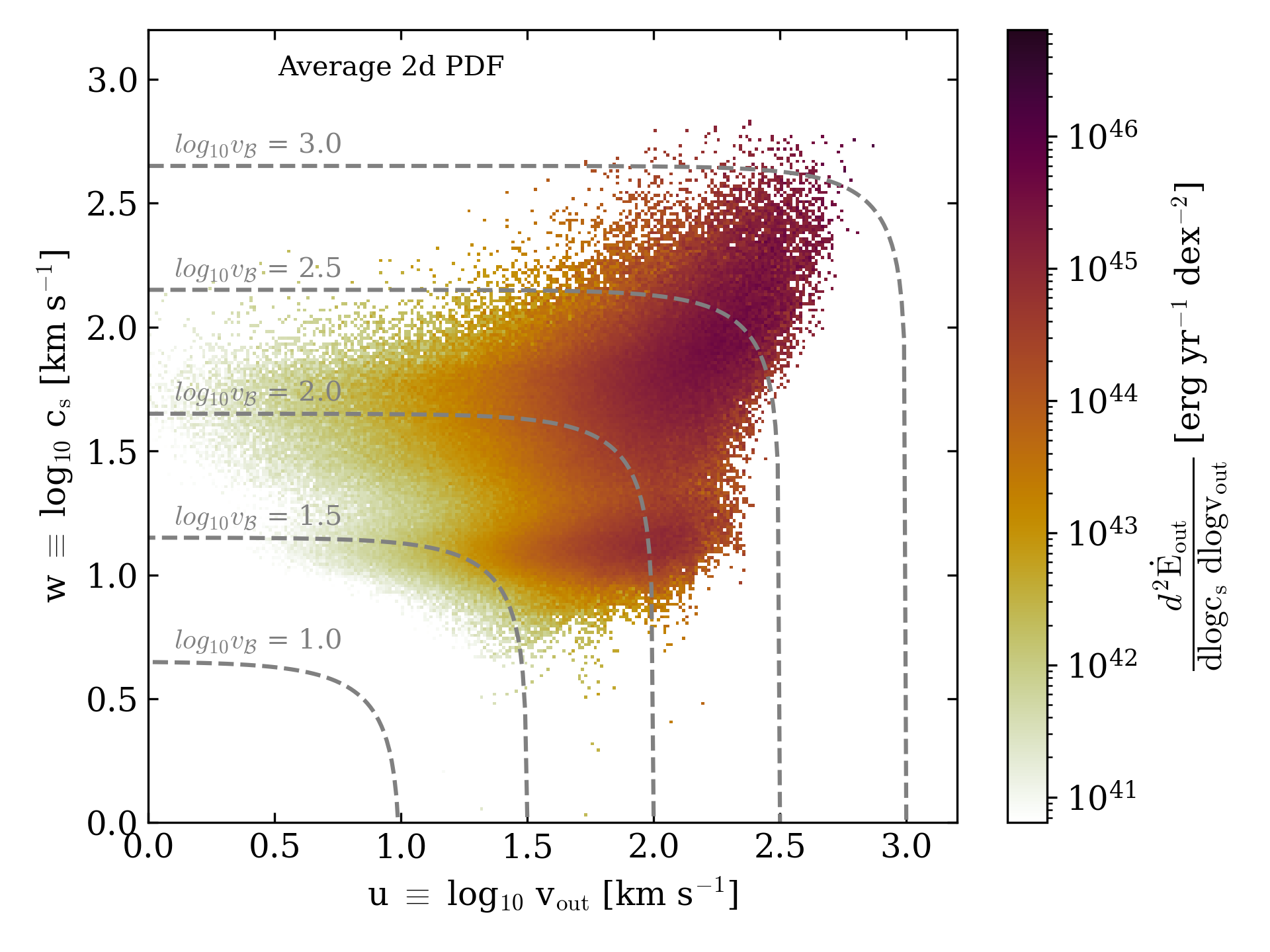}
   \includegraphics[width=0.32\textwidth]{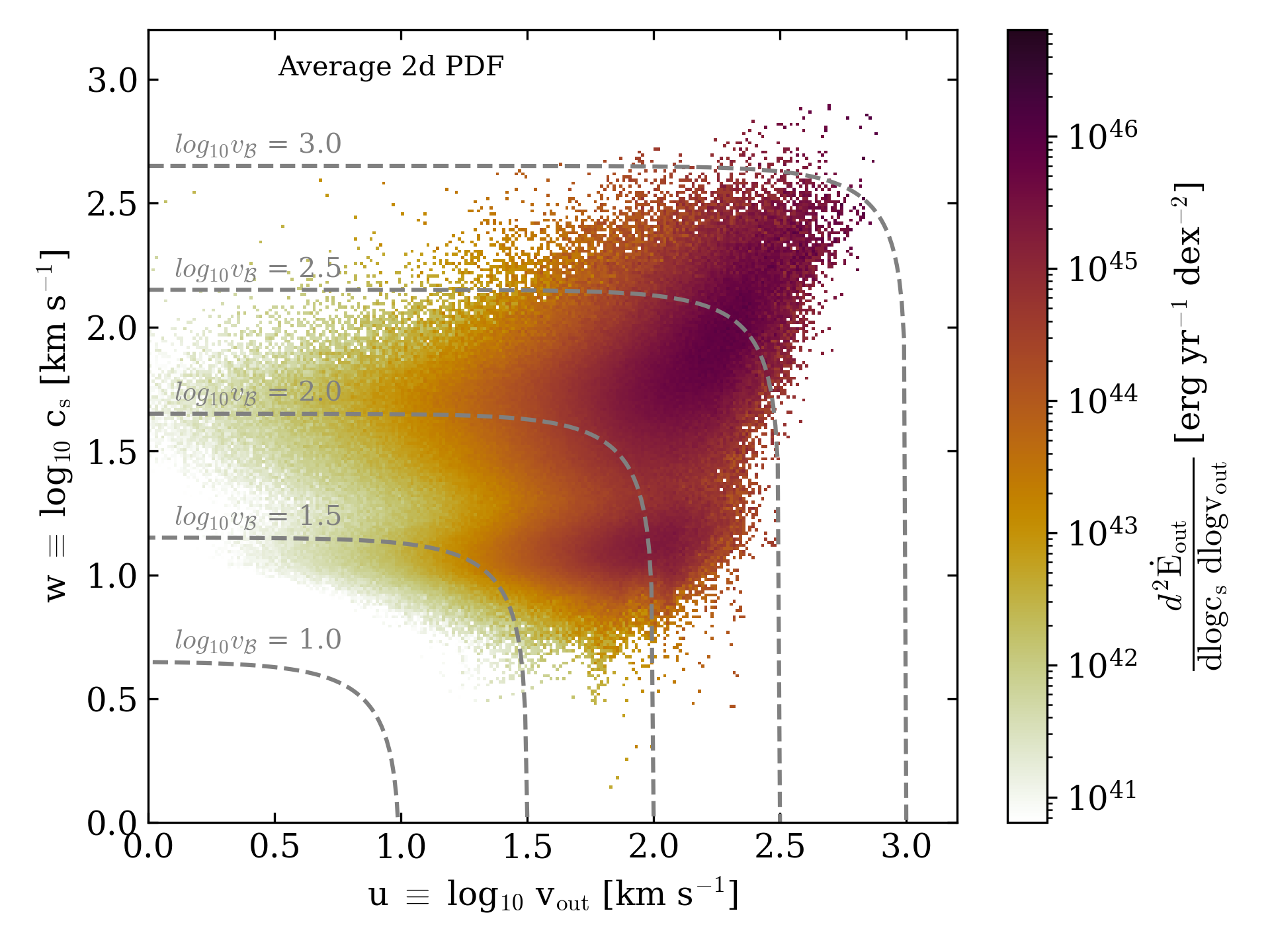}
   \includegraphics[width=0.32\textwidth]{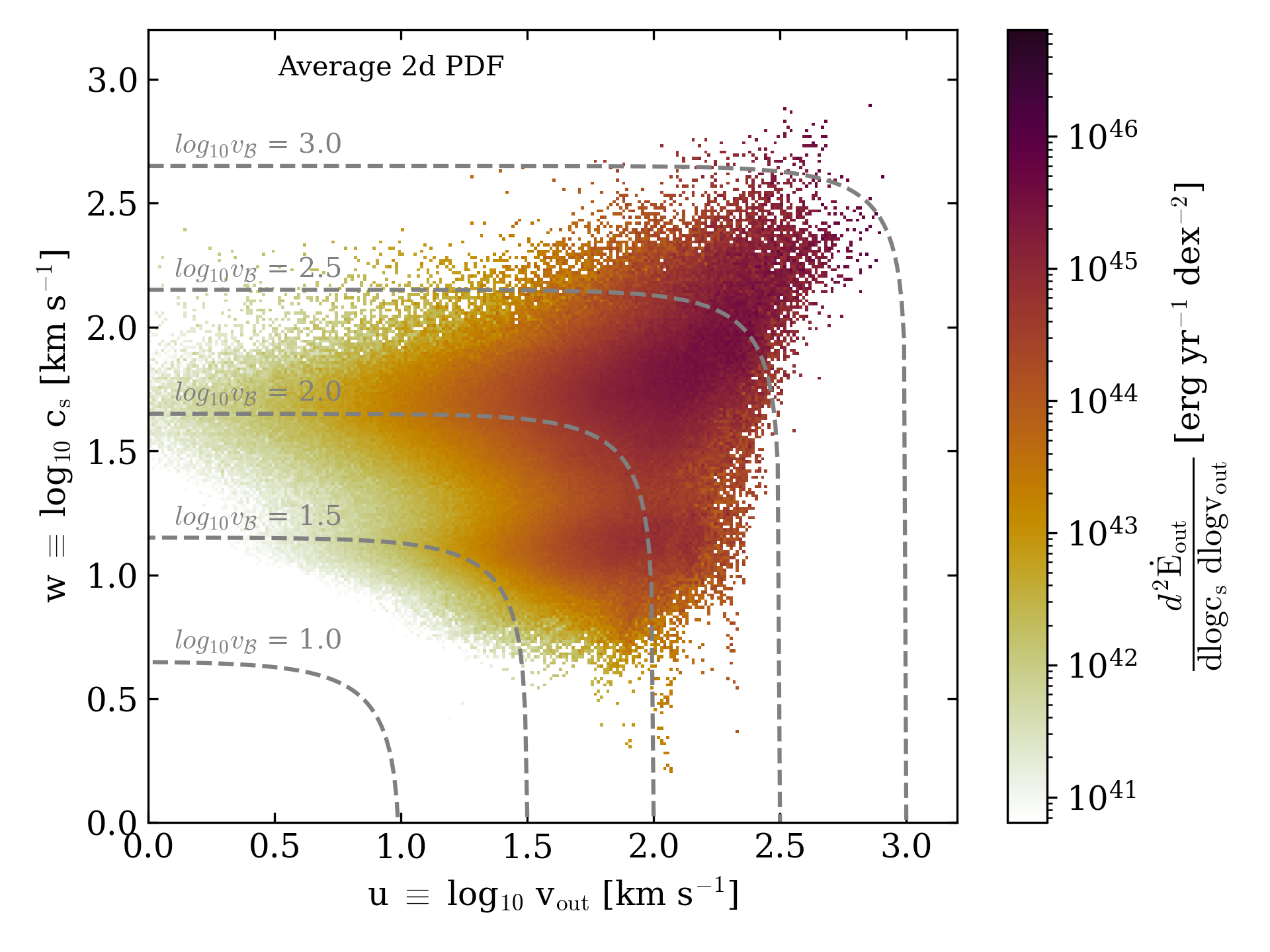} 
   
    \includegraphics[width=0.32\textwidth]{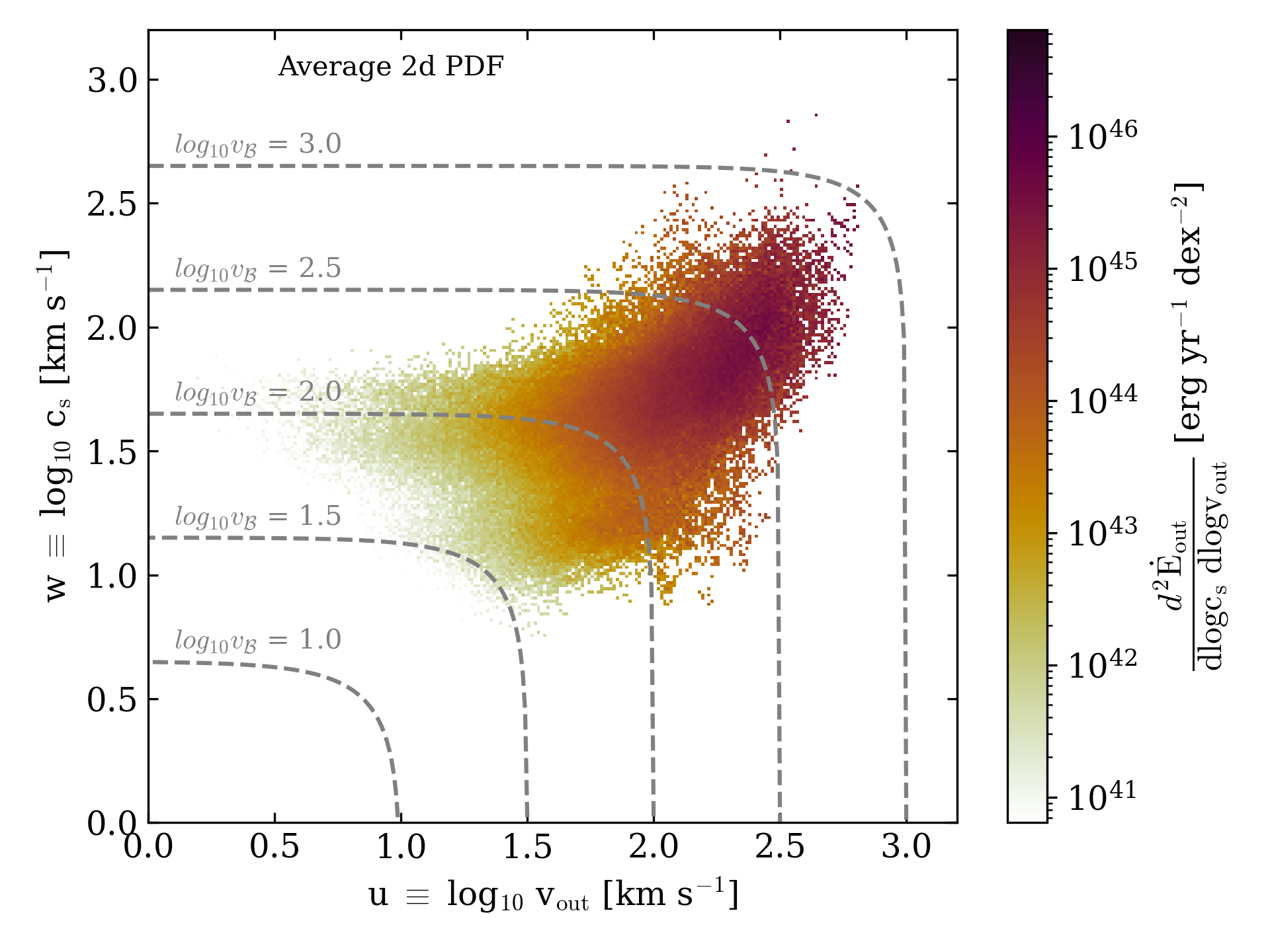}
   \includegraphics[width=0.32\textwidth]{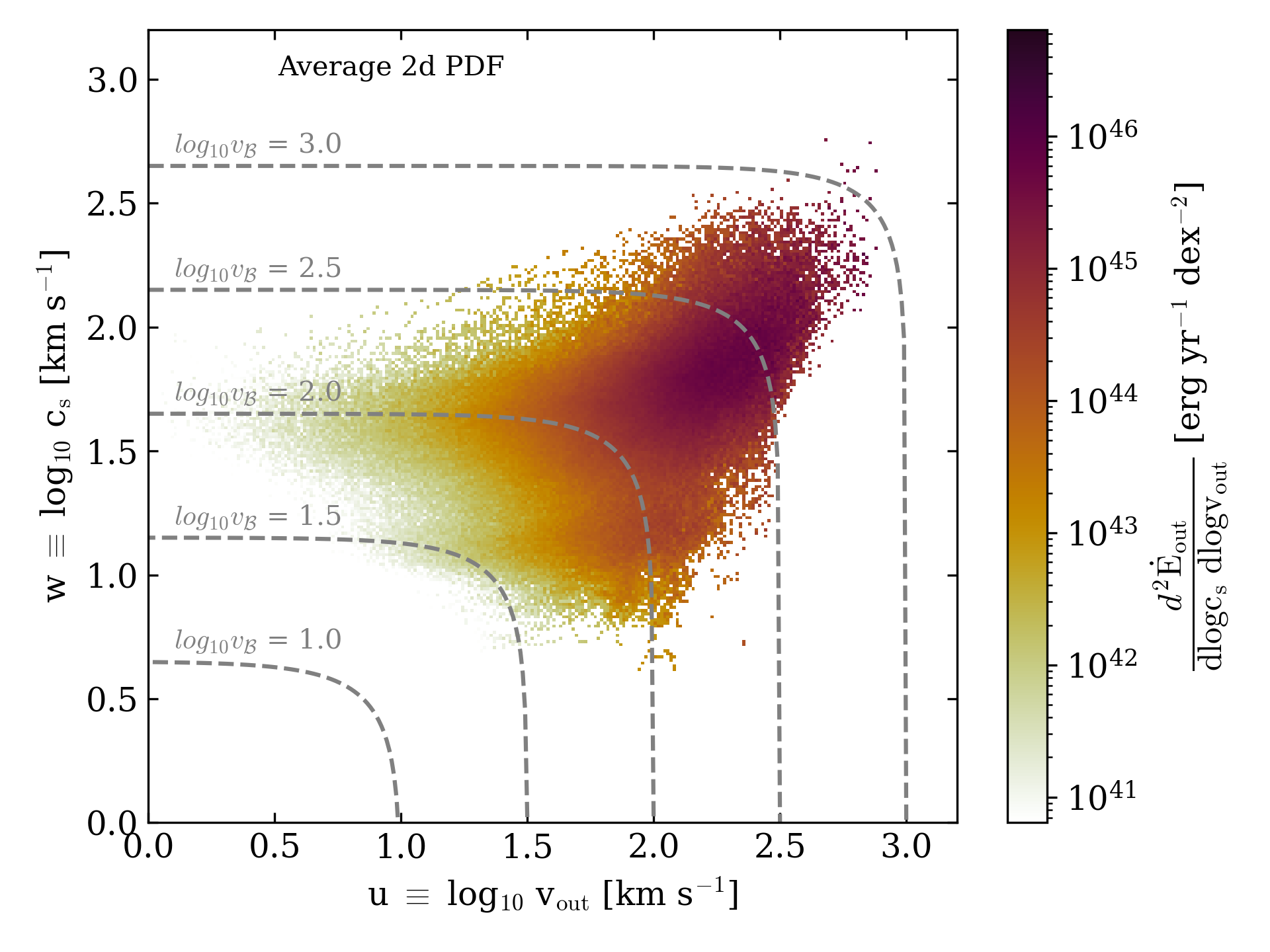}
   \includegraphics[width=0.32\textwidth]{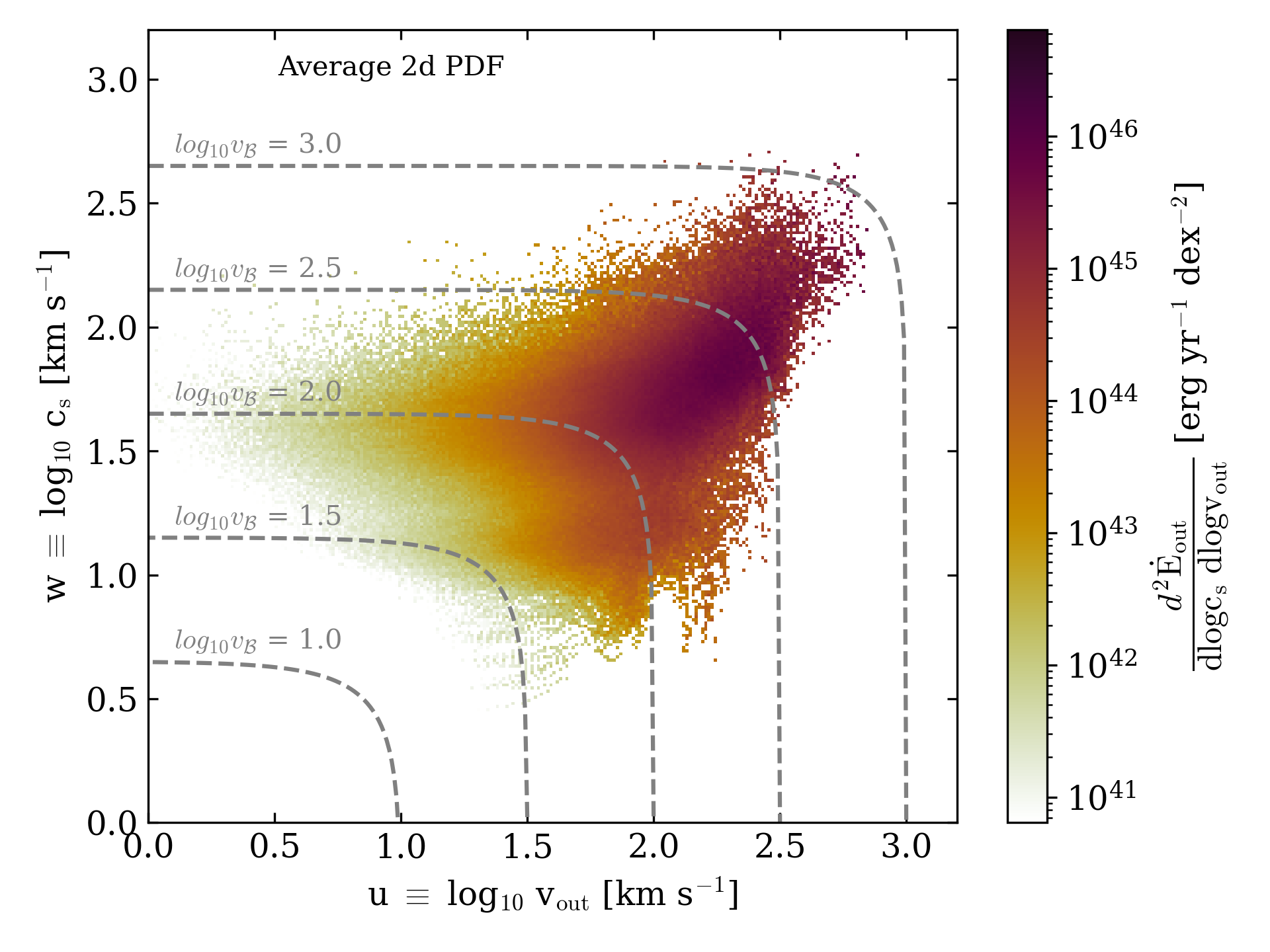}
   
    \includegraphics[width=0.32\textwidth]{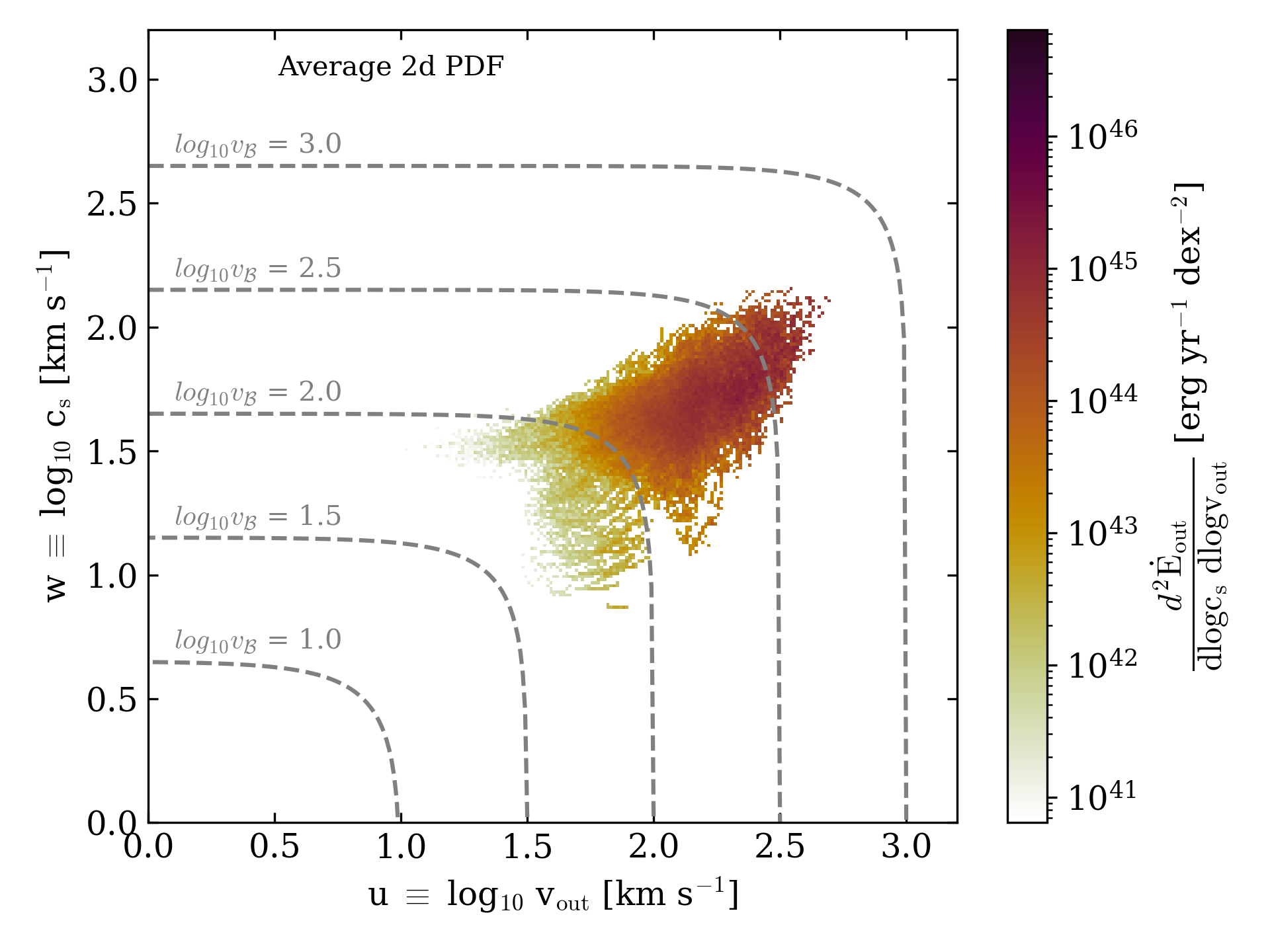}
   \includegraphics[width=0.32\textwidth]{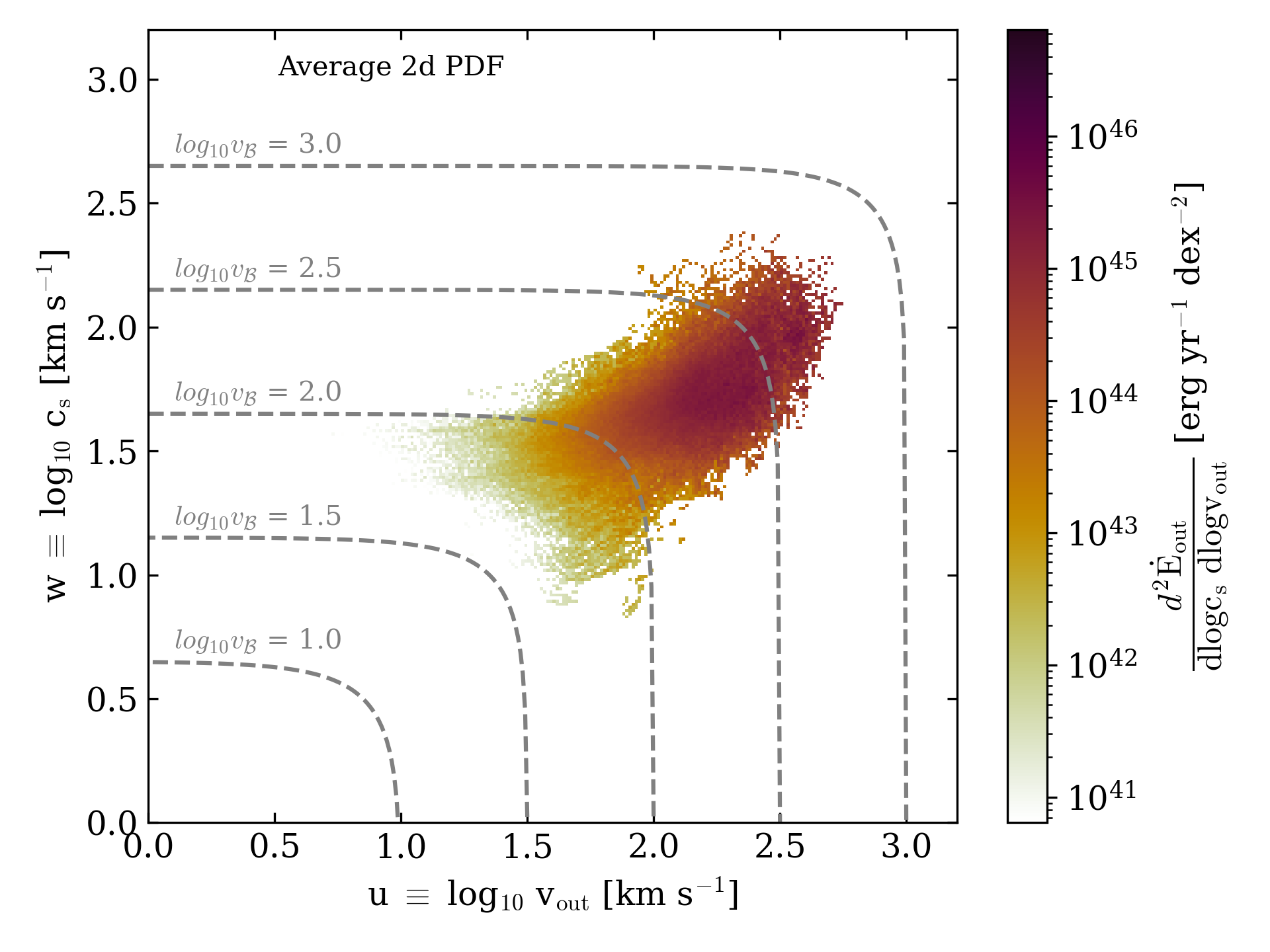}
   \includegraphics[width=0.32\textwidth]{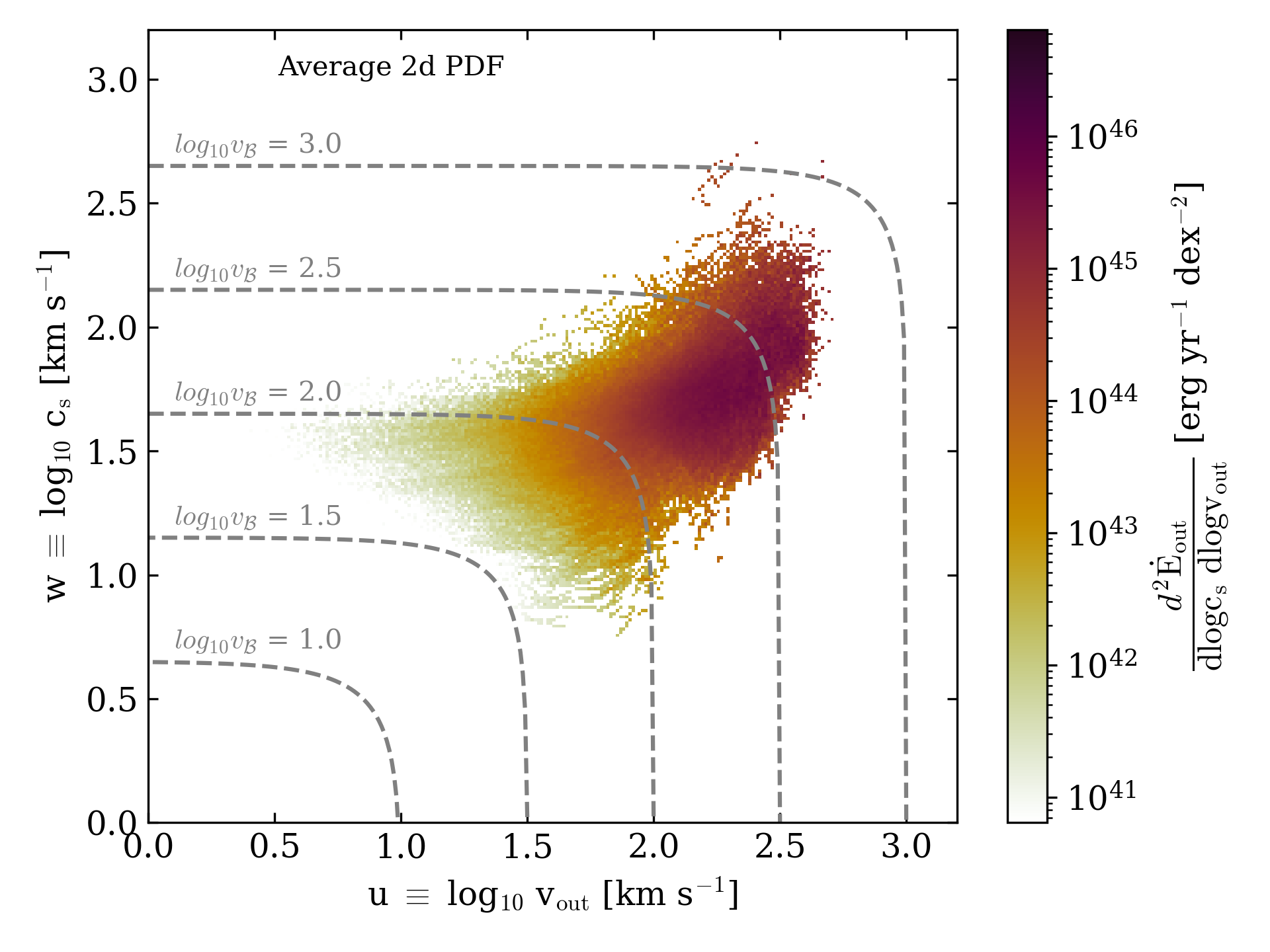}
    \caption{Compilation of the 2d joint PDFs for the energy outflow rate for the same radial and vertical regions as in Fig.~\ref{fig:2dpdfs_mass_comp}.}
    \label{fig:2dpdfs_energy_comp}
\end{figure*}

In Fig.~\ref{fig:phase} we show the time-averaged, mass-weighted phase diagrams in density-pressure phase space (left) and in density-temperature phase space (right). The phase diagrams are both evaluated within a slab of 100 pc around the midplane and within a radius of 5 kpc and generally reveal the expected multiphase structure of our simulated ISM. 

The warm and cold ISM dominates the total mass budget and covers a wide range of pressure, representing radial variation of the mean heating rate (lower pressure with lower photoelectric heating at outer regions and vice versa). The hot phase is established by SNe. Additionally, we can identify the HII-regions as a horizontal line at $T=10^{4}$ K. We note that there is a significantly higher amount of mass in the diffuse hot phase for our LMC run compared to the smaller WLM-like system of previous studies \citep[e.g.][]{Hu2016, Hu2017, Hu2019, Steinwandel2020, Smith2021, Hislop2022, Steinwandel2022}. Thus it is particularly interesting to qualitatively compare to the obtained ISM structure of the results of WLM-like systems that also include photo-ionising radiation, but which consistently show less hot and more warm diffuse gas in phase space \citep[e.g.][]{Hu2017, Smith2021, Steinwandel2022}.

\subsection{Definitions for outflow rates and loading factors  \label{sec:defs}}

We will measure the flow rates at different heights $|z|$: 0.5 kpc, 1 kpc, 3 kpc, 5 kpc and 10 kpc. Additionally, in some plots we will adopt radial cuts $0.0 < R < 5.0$ kpc,  $0.0 < R < 1.5$ kpc,  $1.5 < R < 3.0$ kpc and $3.0 < R < 5.0$ kpc, where $R$ denotes galactocentric distance $R = \sqrt{x^2 + y^z}$. Furthermore, to illustrate the complex structure of the wind it is useful to also measure flow rates in spherical shells, specifically at higher altitude above the midplane. Whenever wind profiles are measured in spherical shells they will be indicated with $r$, which is defined as $r = \sqrt{x^2 + y^z + z^2}$. Vertical profiles (perpendicular to the midplane) will be indicated as the third component of the gas position vector ($z$-axis).

We will adopt the following definitions for radial inflow and outflow rates, with the net total flow rates defined via $\dot{M} = \dot{M}_\mathrm{out} - \dot{M}_\mathrm{in}$ and $\dot{E} = \dot{E}_\mathrm{out} - \dot{E}_\mathrm{in}$. Outflow is defined by a positive value of the radial velocity given $\mathbf{v \cdot \hat{r}} = v_{r} > 0$. We can write down the discrete outflow rates for mass and energy:
\begin{equation}
    \dot{M}_\mathrm{out} = \sum_{i, v_{i,r} > 0} \frac{m_{i}v_{i,r}}{dr},
\end{equation}
\begin{equation}
    \dot{E}_\mathrm{out} = \sum_{i, v_{i,r} > 0} \frac{m_{i} [v_{i,r}^2 + \gamma u_{i}]v_{i,r}}{dr},
\end{equation}
with $\gamma = 5/3$ and $dr$ as the shell thickness in which we  measure the outflow rates. We note that for vertical measurement of outflow rates (in plane parallel cylinders at the given heights $|z$ above) we will adopt $dr = 0.1 \times |z|$. For the spherical measurement at $r_\mathrm{vir}$, we adopt a spherical shell at 0.1 $r_\mathrm{vir}$ with a thickness of 0.01 $r_\mathrm{vir}$.

It can be useful to investigate these quantities when normalized to a set of reference quantities, defined as the so-called loading factors for mass ($\eta_\mathrm{m}$), and energy ($\eta_\mathrm{e}$) that we define via:
\begin{enumerate}
    \item outflow mass loading factor: \\
    $\eta_\mathrm{m}^\mathrm{out} = \dot{M}_\mathrm{out} / \overline{\mathrm{SFR}}$,
    \item energy outflow loading factor: \\
    $\eta_\mathrm{e}^\mathrm{out} = \dot{E}_\mathrm{out}/(E_\mathrm{SN} \mathcal{R}_\mathrm{SN})$,
    \label{eq:loadings}
\end{enumerate}
In our work we adopt $E_\mathrm{SN} = 10^{51}$ erg, the supernova rate $\mathcal{R}_\mathrm{SN} = \overline{\mathrm{SFR}}/(100 M_{\odot})$ which are adopted for best comparison and stellar population synthesis modeling with earlier dwarf galaxy studies who adopt these values for normalization \citep[e.g.][]{Hu2019}.
We compute the mean star formation rate $\overline{\mathrm{SFR}}$ over a time span of 300 Myr between 200 and 500 Myr of evolution when we investigate the time evolution of loading factors. However, we note that the definition of loading factors in galactocentric, radial or vertical profiles is more difficult to define due to the complicated structure of the wind. Hence, when we investigate the spatial properties of the wind, we will do this mostly in ``outflow rate space'' but note that one can always compute a loading factor using the mean star formation rate of $\overline{\mathrm{SFR}}\sim 0.61$ M$_{\odot}$ yr$^{-1}$ and the SN-energy
that we provided for the computation of ${\cal R}_\mathrm{SN}$. However, the problem is that specifically at larger heights this approach is not very physical and the correct way to obtain loading factors would be to trace wind streamlines\footnote{It is worthwhile noting that this is a picture perfect use case of our fully Lagrangian method MFM.}, something we will postpone to future work. Thus the only time that we adopt explicit loading factors in this paper is in Sec.~\ref{sec:sfr_density} where we compute the loading factors in bins close to the disc at 0.5 and 1 kpc respectively, where there is still a spatial correlation possible with the star formation activity that is measured ``down the barrel'' 

Lastly, we define the Bernoulli velocity as:
\begin{equation}
    v_\mathrm{\mathcal{B}} = \left(v_\mathrm{out}^2 + \frac{2 \gamma}{\gamma-1} c_\mathrm{s}^{2}\right)^{\frac{1}{2}},
    \label{eq:Bernoulli}
\end{equation}
where $c_s$ is the isothermal sound speed, $v_\mathrm{out}$ is the velocity of the wind, and $\gamma=5/3$ everywhere in the simulation domain. The Bernoulli-velocity represents the terminal velocity that an adiabatic wind would reach in the limit of v$_\mathcal{B} \gg v_\mathrm{esc}$ 

\subsection{Time evolution of the outflows}

In Fig.~\ref{fig:sfr} we show the time evolution of the star formation rate (top), the mass loading factor $\eta_\mathrm{M}$ (center) and the energy loading factor $\eta_\mathrm{E}$ (bottom). We compute the star formation rate averaged over four different time intervals by summing up the masses of all stars with ages less than 20, 10, 5 and 1 Myr respectively. While this is merely a sanity check it indicates that the ``burstiness'' of the star formation history depends on the time interval over which it is averaged. Initially, the star formation rate peaks at a value (first peak) of $\sim 0.1$ M$_{\odot}$/yr, which is a remnant from the initial transient phase in this type of isolated galaxy simulation that leads to a small starburst. However, after this transient phase the star formation rate drops by a factor of around 2 and shows a slight increase around 150 Myr followed by a drop towards 250 Myr before a steeper increase around 310 Myrs, which is actually a result of the spatial overlap of several ``star clusters'' in the lower left quadrant of the galaxy, causing a second (smaller) starburst event. We note that the system has a period of around 150 Myr between peaks in star formation. A similar cycle can be seen in the WLM-like simulations \citep[e.g.][]{Hu2017, Hislop2022, Steinwandel2022} as well as in the \textsc{Tigress} ``tall box'' simulations \citep[e.g.][]{Kim2018, Kim2020_smaug}, carried out with the grid code \textsc{athena} \citep[][]{Stone2008}, for their low gas surface density model.

The mass loading and energy loading factors are computed at four different heights: 1 kpc, 3 kpc, 5 kpc, and 10 kpc in slabs above and below the midplane based on the definitions for outflow rates and mass loadings summarized in Section \ref{sec:defs}. Additionally, in Fig.~\ref{fig:sfr} we show results that correspond to a global measurement of the loading factors at 0.1 R$_\mathrm{vir} \sim 10.0$ kpc, similar to the scale often used in studies of gas flows in cosmological simulations \citep[e.g.][]{Pandya2021}. 
The thickness of the slices at $z=1$ kpc, $z=3$ kpc, $z=5$ kpc and $z=10$ kpc correspond to 10 per cent of the height $dr=0.1z$, which results in slightly thicker slabs at higher altitude to ensure that the number of particle in the respective slabs is high enough to obtain a meaningful average. Broadly, the peaks in star formation are followed by an increase in the loading factors with a delay time of roughly $\sim25$ Myr at 1kpc, $\sim30$ Myr at 3 kpc, $\sim35$ Myr at 5 kpc and $\sim50$ Myr at 10 kpc. These numbers are consistent between the two peaks in star formation at 50 Myr and 300 Myr that we can observe over the 500 Myrs of evolution of the system. The mass loading is around $1-5$ at 1 kpc and a factor of a few lower at higher altitude, while the energy loading is generally low with values around a few times of 0.01. 
 
\subsection{Phase structure of the outflows}

A key aspect of this outflow study is to investigate the multiphase nature of galactic outflows. In Fig. \ref{fig:mass_energy_outflow} we show the joint PDFs of outflow velocity and sound speed weighted by mass outflow rate (left) and energy outflow rate (right) for a radial region between $0.0 < R/\mathrm{kpc} < 5.0$ at a height of 1 kpc above and below the midplane. Generally, we find that the bulk of the mass is transported by the warm phase of the wind with velocities between 30 km s$^{-1}$ and 50 km s$^{-1}$ while for the energy we find a fast hot component that carries at least 50 percent of the energy of the wind. While this is similar to previous results as shown for example by \citet{Fielding2018}, \citet{Kim2020}, or \citet{Pandya2021}, we find a significant contribution to the energy outflow rate from the warm phase of the wind at least at lower altitude.  

In Fig.~\ref{fig:2dpdfs_mass_comp} we show a compilation of the mass outflow rate weighted joint PDFs for different regions above and below the midplane. Different columns indicate different radial regions within the slabs, while different rows indicate different heights above and below the midplane. The PDFs in the first column are measured between $0.0 < R < 1.5$ kpc at $|z|=$0.5 kpc, 1 kpc, 3 kpc, 5 kpc and 10 kpc above and below the midplane in slabs with thickness $dr=0.1z$. The other columns show the same respective heights, but are restricted to $1.5 < R < 3.0$ kpc (second column), $3.0 < R < 5.0$ kpc (third column). Breaking down the distribution shown in Fig.~\ref{fig:mass_energy_outflow} into different regions allows for a more detailed understanding of the complex structure of the outflows in global galaxy models. We highlight some important features here. At a height of 0.5 and 1 kpc we find that the outflow structure has only a very weak radial trend. However, the highest velocities and temperatures in the outflows are reached between $1.5 < R < 3.0$ kpc, which is likely related to the fact that the outflowing gas from the center is leaving the galaxy with some finite opening angle. Moreover, the wind carries more mass in the central regions than in the outskirts at lower altitude. However, the effect is modest at 0.5 and 1 kpc and is subsequently inverted at higher altitude, again indicating that the gas accelerated from the center is not ejected perfectly perpendicular to the midplane, most likely due to a combination of the asymmetry of large scale SN-bubbles and the large scale shear due to rotation of the galaxy. Considering the distribution in the third column one can observe a quite drastic increase in the high velocity hot component, which can be explained as the hot gas launched from the center that is then adiabatically cooled due to the velocity streamline opening at higher altitude. We note that the area of the rings that we average over is increasing from the center to the outskirts, which might be able to partially explain the inverted radial trend at high altitudes. Using mass fluxes (outflow rate normalized by area) instead of mass outflow rates, we checked that the trend persists.

Furthermore, we observe that there is a gradual loss of the warm component of the wind as we reach higher altitudes and we find that at a height of 10 kpc, only the (cooled) hot component of the wind is left, with most of the mass located in the outskirts of the wind. We find that the low specific energy part of the wind has dropped out. The slow material falls back to the disc. However, it is interesting to note that there is a part of the warm wind that is moving almost as fast as parts of the hot wind, that could potentially be entrained in the hot wind. This would be consistent with the lower sound speed of the wind material at higher altitude, indicating that there is either cooling due to adiabatic expansion of the wind material or because a significant amount of the warm fast part is mixing with the hot wind \citep[see][and references therein for the mixing of cold gas in a hot wind]{Fielding2022}. The exact mixing and/or adiabatic expansion of the wind is an interesting question in itself but would need a more detailed study of the distribution of specific entropy as a function of height.

In Fig.~\ref{fig:2dpdfs_energy_comp} we show the same cuts in height and radius as in Fig.~\ref{fig:2dpdfs_mass_comp} but this time for the energy outflow rate. Similar to the case for mass, we  find two distinct regions that transport the energy, a low temperature, high velocity part of the wind at a sound-speed of $c_\mathrm{s} \sim 10-15 {\rm km\, s}^{-1}$ with a temperature of around $\sim 10^4$K and a high temperature, high velocity part of the wind that reaches temperatures of up to $\sim 10^{7}$ K and outflow velocities up to a few 100 km s$^{-1}$. Qualitatively, we observe similar trends as for the mass outflow rate. At low altitude, just above the galactic midplane, we detect the highest energy flux, while the energy flux drops towards the galactic outskirts at low altitude. In contrast, we observe that the outskirts transport a significantly larger amount of energy than the central part of the galaxy at high altitudes (specifically at 10 kpc height). Furthermore, we again point out that the outflow velocity of the hot increased from 1 kpc to 3 kpc in the outermost ring ranging from $3.0 < R< 5.0$ kpc. This is a clear indication that the streamlines of the wind show an open topology as the outflow reaches higher altitudes above and below the midplane (see Sec.~\ref{sec:structure_outflow}). We note that it is not necessarily surprising that the streamlines of the wind are opening up, but it will have important consequences for the computation of loading factors, which is the main reason why we will focus in the remainder of the paper either on outflow rates directly or loading factors close to the midplane, and postpone a more detailed analysis of loading factors following streamline topology to future work.

\subsection{Radial and vertical structure of the outflows}

\label{sec:structure_outflow}

\begin{figure*}
    \centering
    \includegraphics[width=0.49\textwidth]{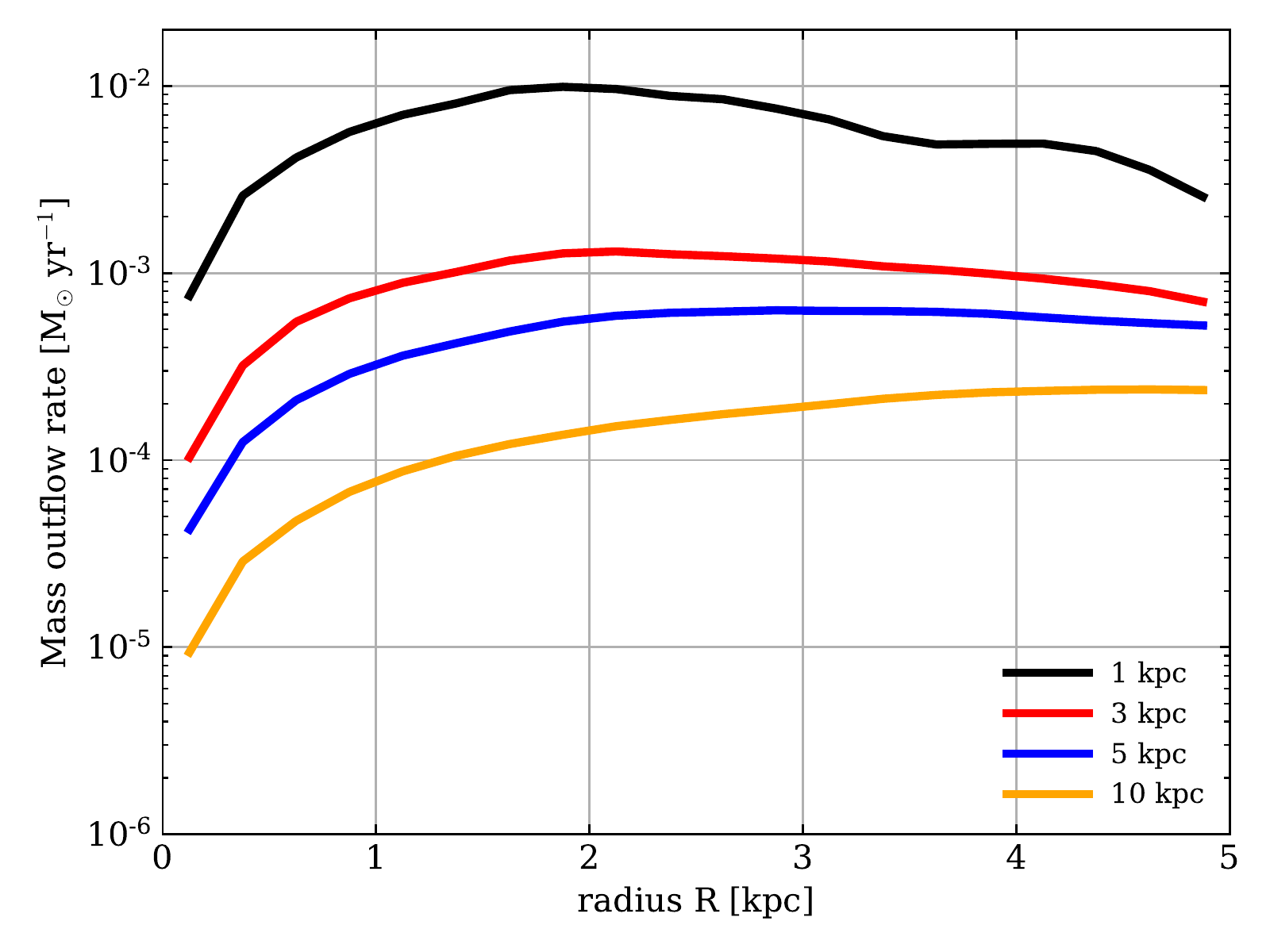}
    \includegraphics[width=0.49\textwidth]{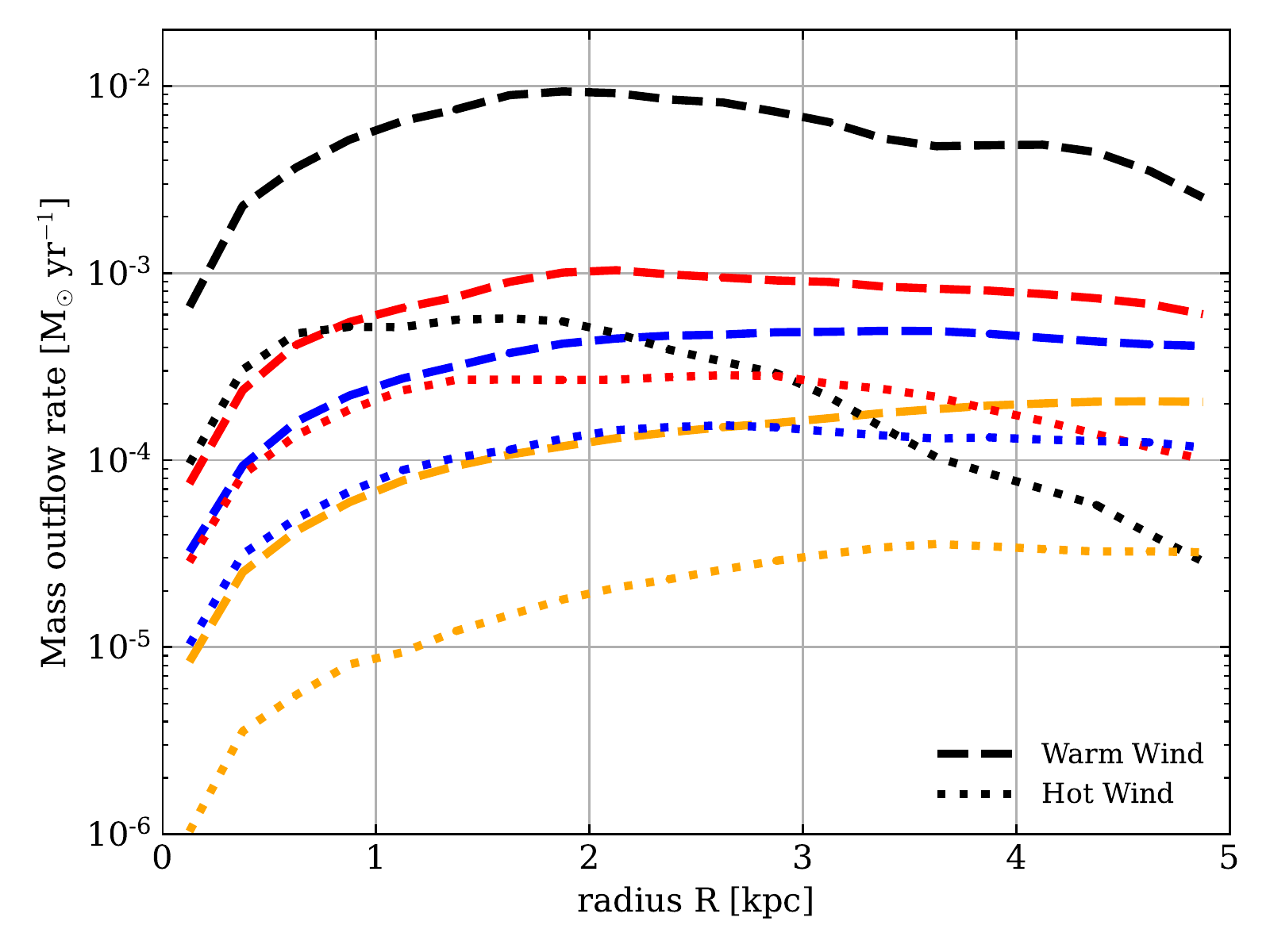}
    \caption{Time averaged (200 Myrs; 200-400 Myr) radial profiles for the total mass outflow rate at different heights of 1 kpc (black), 3 kpc (red), 5 kpc (blue) and 10 kpc (orange) on the left and the mass outflow rate split in warm ($T < 5 \times 10^5$ K, dashed lines) and hot ($T > 5 \times 10^5$ K, dotted lines) on the right. Generally, the mass outflow rate is higher at lower altitude and exhibits a peak at around $R=$1.5 to 2 kpc at low altitude. However at 10 kpc we find a gradually increasing trend outwards. If we look a the split in the warm and the hot component on the right it becomes clear that at all radii and at the four different heights at which we computed the radial profiles, we find that the mass transport is dominated by the warm component, where we find that at low altitude (1kpc) roughly 10 per cent of the total outflow rate is in the hot phase, while it appears that there is slightly more mass in the hot at the higher altitudes up to a maximum of around 20 per cent.}
    \label{fig:radial_loadings_mass}
\end{figure*}

\begin{figure*}
    \centering
    \includegraphics[width=0.49\textwidth]{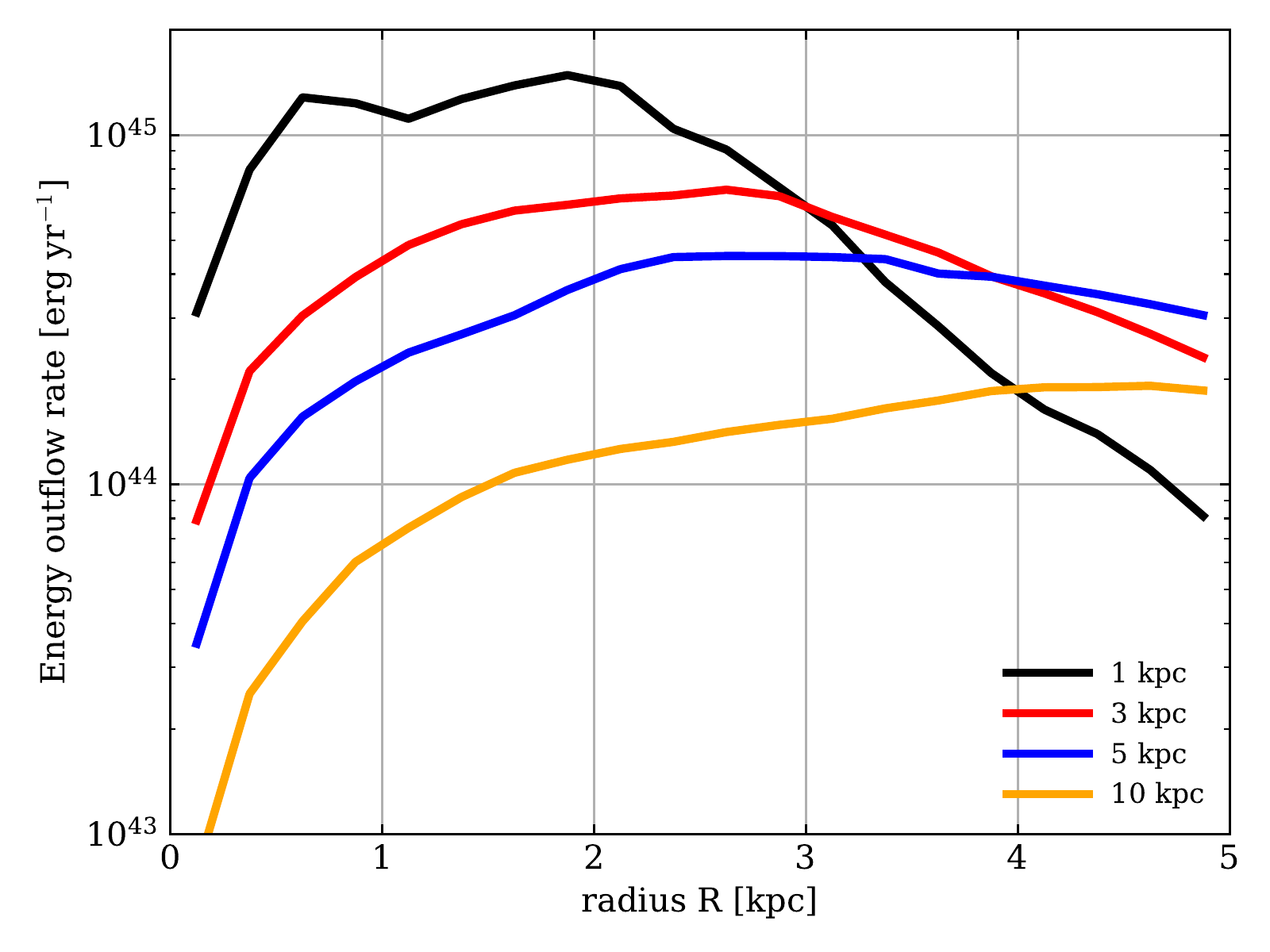}
    \includegraphics[width=0.49\textwidth]{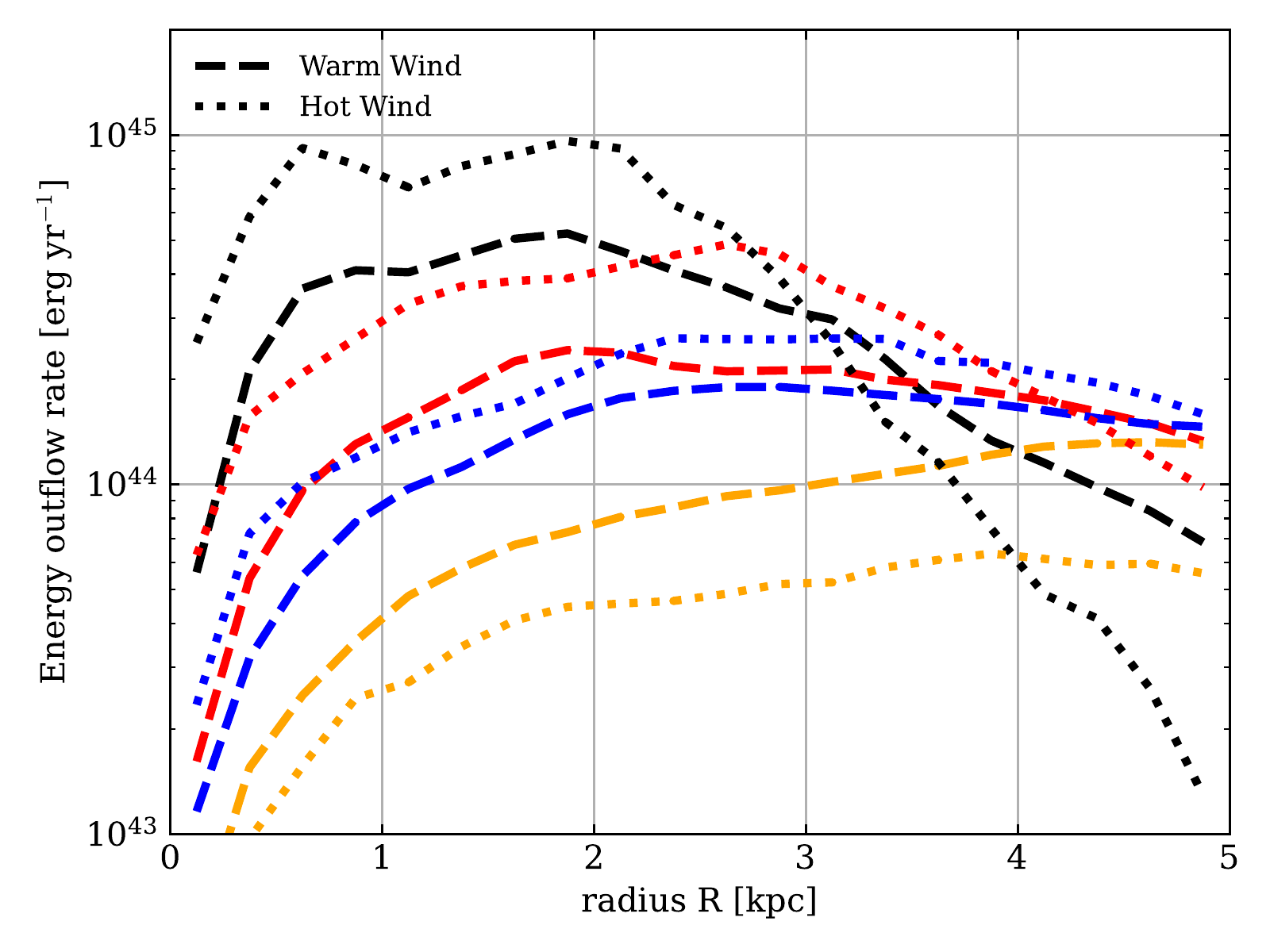}
    \caption{Time averaged (200 Myrs; 200-400 Myr) radial profiles for the total energy outflow rate at different heights of 1 kpc (black), 3 kpc (red), 5 kpc (blue) and 10 kpc (orange) on the left and the energy outflow rate split in warm ($T < 5 \times 10^5$ K, dashed lines) and hot ($T > 5 \times 10^5$ K, dotted lines) on the right. The energy outflow rate behaves similarly apart from the fact that there is a sharp drop off beyond a radius of 2 kpc at a height of 1 kpc. In combination with the radial trends of the mass outflow rate presented in Fig~\ref{fig:radial_loadings_mass}, these results indicate that the wind in the low altitude, inner regions has moved outward as it propagates upward (especially, energy-carrying hot component); i.e., the wind has an open streamline structure. The split in warm and hot wind on the right reveals that the energy transport is dominated by the hot phase for 1 kpc, 3 kpc and 5 kpc. At an altitude of 10 kpc the warm component is dominating the energetics of the wind. However, this is likely due to the fact that the formerly hot component of the wind is cooled adiabatically expanding as it is expanding.}
    \label{fig:radial_loadings_energy}
\end{figure*}

\begin{figure*}
    \centering
    \includegraphics[width=0.49\textwidth]{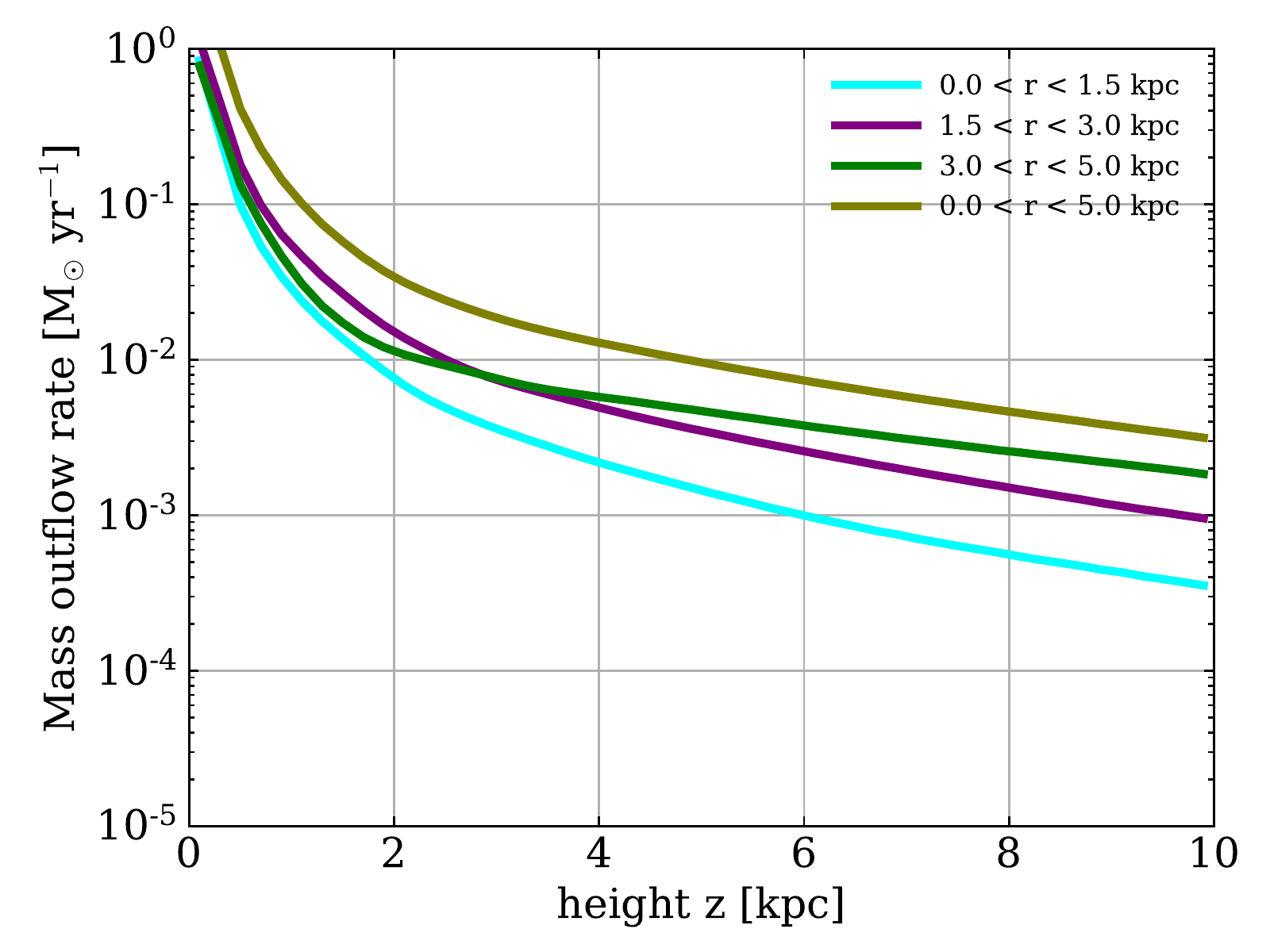}
    \includegraphics[width=0.49\textwidth]{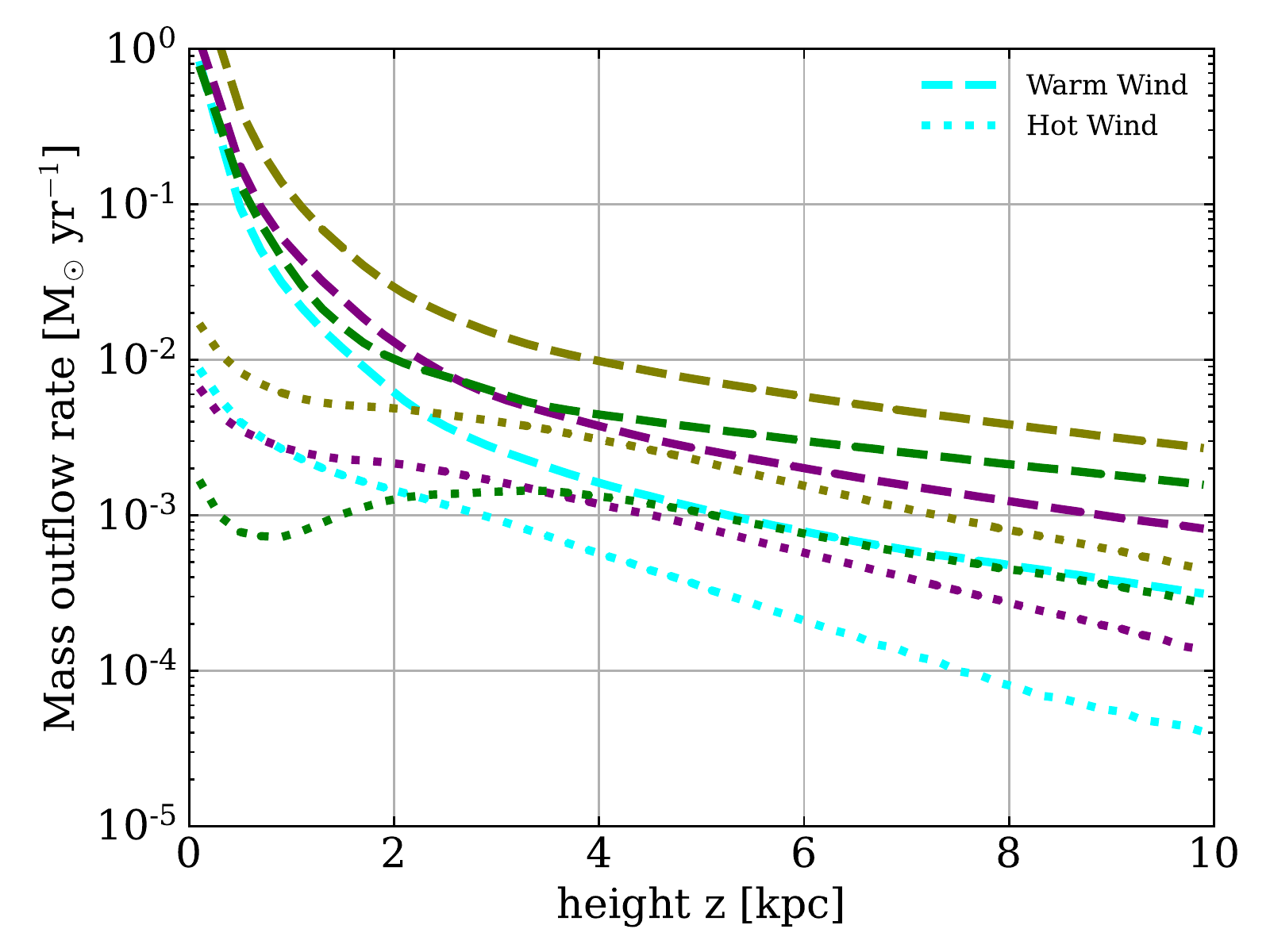}
    \caption{Vertical profiles for the total mass outflow rate (left), measured in different radial regions given via $0.0 < R < 1.5$ kpc (cyan), $1.5 < R < 3.0$ kpc (purple), $3.0 < R < 5.0$ kpc (green) and $0.0 < R < 5.0$ kpc (olive). All of these profiles are averaged over 200 Myrs from 200 to 400 Myrs. At low altitude out to around 2 kpc height the wind is transporting a similar amount of mass in the inner and outer parts of the galaxy. Beyond 2 kpc we find that most of the mass is transported in the outer parts of the wind, indicating an opening velocity streamline structure. If this is split into warm ($T < 5 \times 10^5$ K, dashed lines) and hot ($T > 5 \times 10^5$ K, dotted lines, right hand side), it becomes clear the warm wind is completely dominating the mass transport out to a height of 10 kpc.}
    \label{fig:vertical_loadings_mass}
\end{figure*}

\begin{figure*}
    \centering
    \includegraphics[width=0.49\textwidth]{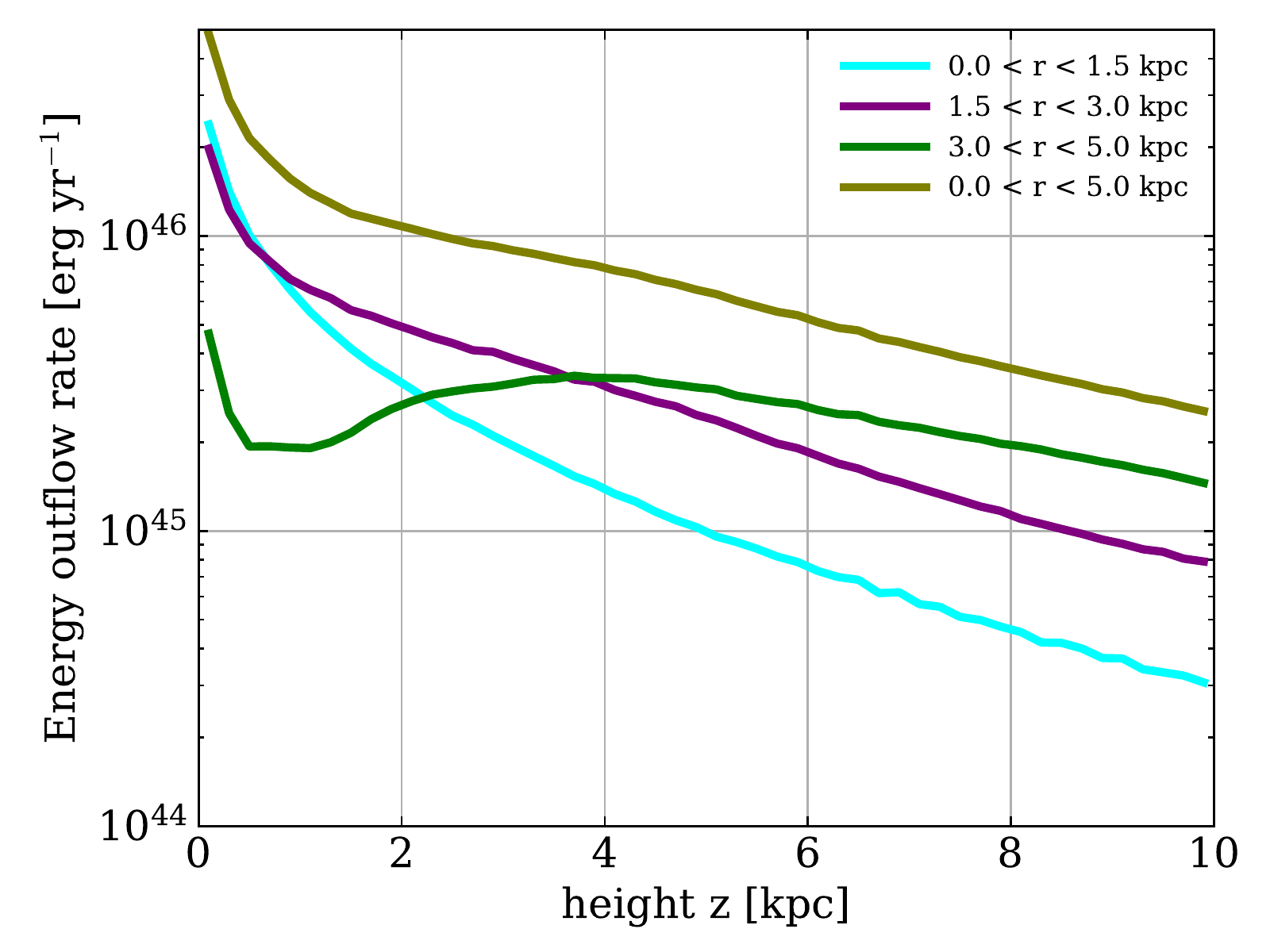}
    \includegraphics[width=0.49\textwidth]{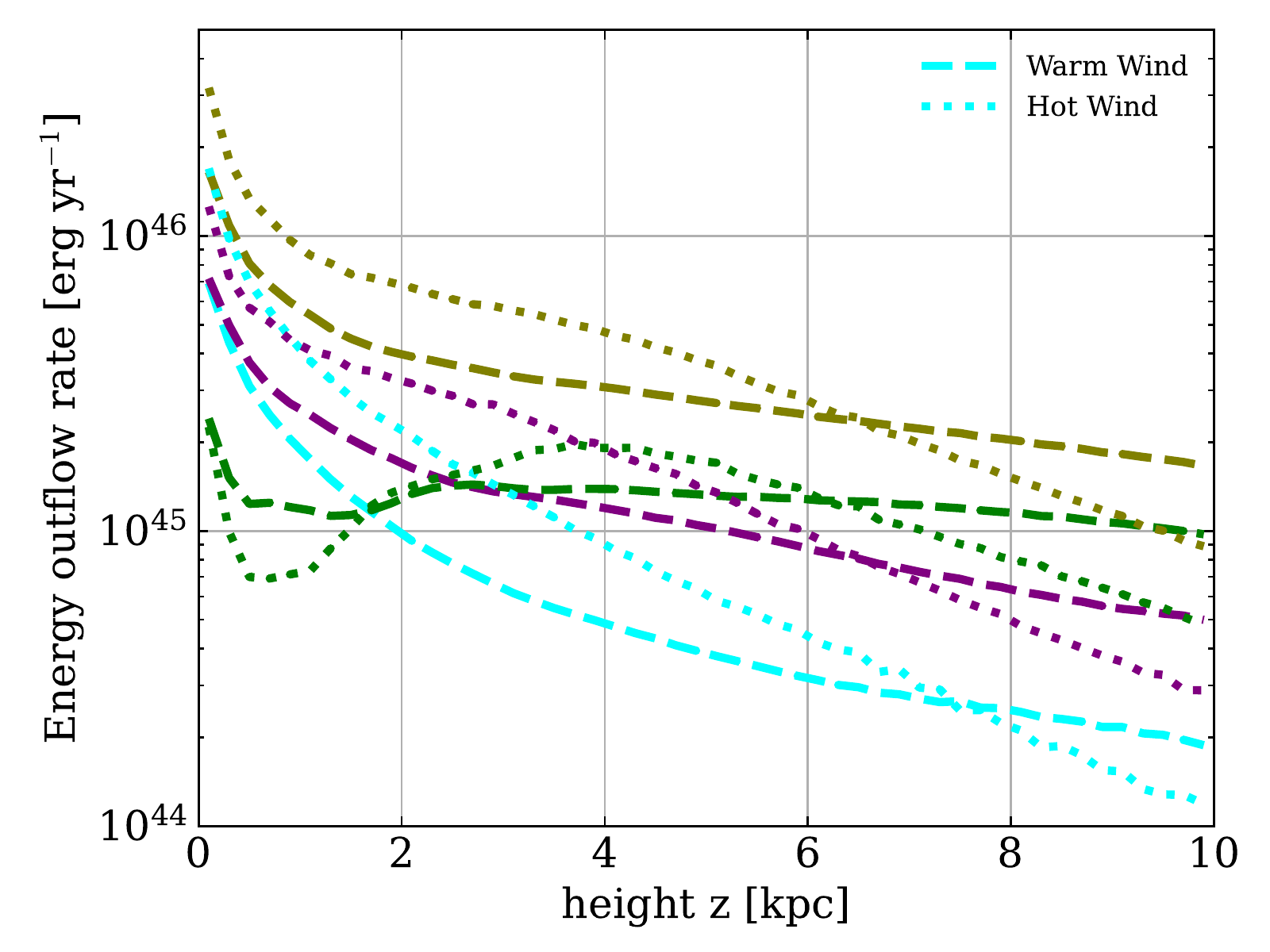}
    \caption{Vertical profiles for the total energy outflow rate (left), measured in different radial regions given via $0.0 < R < 1.5$ kpc (cyan), $1.5 < R < 3.0$ kpc (purple), $3.0 < R < 5.0$ kpc (green) and $0.0 < R < 5.0$ kpc (olive). All of these profiles are averaged over 200 Myrs, between 200 and 400 Myrs of evolution. At low altitude (1 kpc) most of the energy is transported in the central part of the wind, while at higher altitude we find that there is more energy in the outer parts of the wind. Separating this by warm ($T < 5 \times 10^5$ K, dashed lines) and hot ($T > 5 \times 10^5$ K  (right) gas indicates that at low altitude most of the energy is transported by the hot wind out to 2 kpc, in the innermost 2 rings we defined, out to a height of around 7 kpc when the warm wind starts to dominate the energy transport. In the outermost ring the warm wind dominates out to 2 kpc when the hot wind starts to dominate out to 7 kpc after which the warm dominates again. This turnaround in the outermost ring is likely the contribution from the high angle (hot) material from the central wind.}
    \label{fig:vertical_loadings_energy}
\end{figure*}

\begin{figure*}
    \centering
    \includegraphics[width=0.49\textwidth]{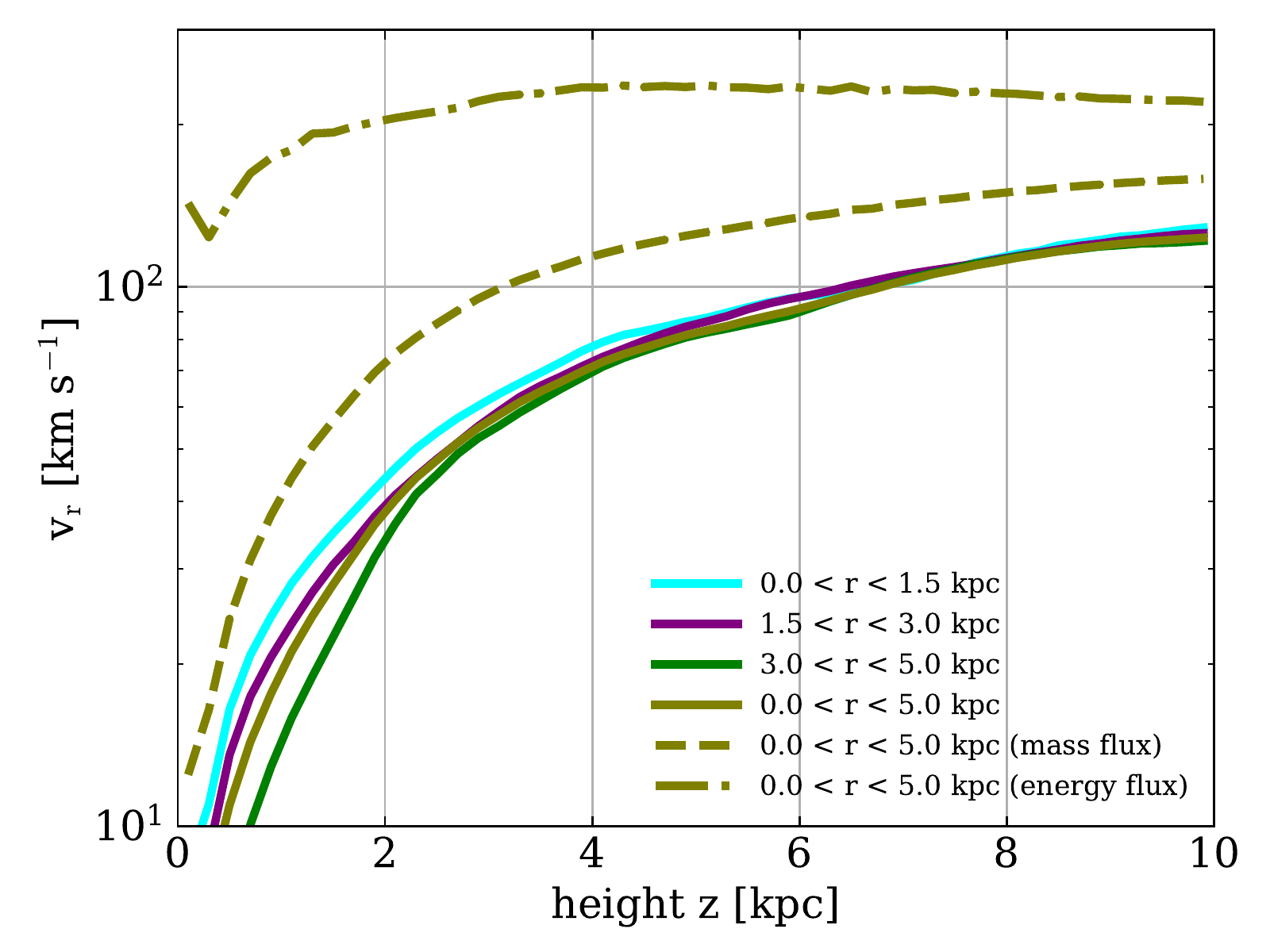}
    \includegraphics[width=0.49\textwidth]{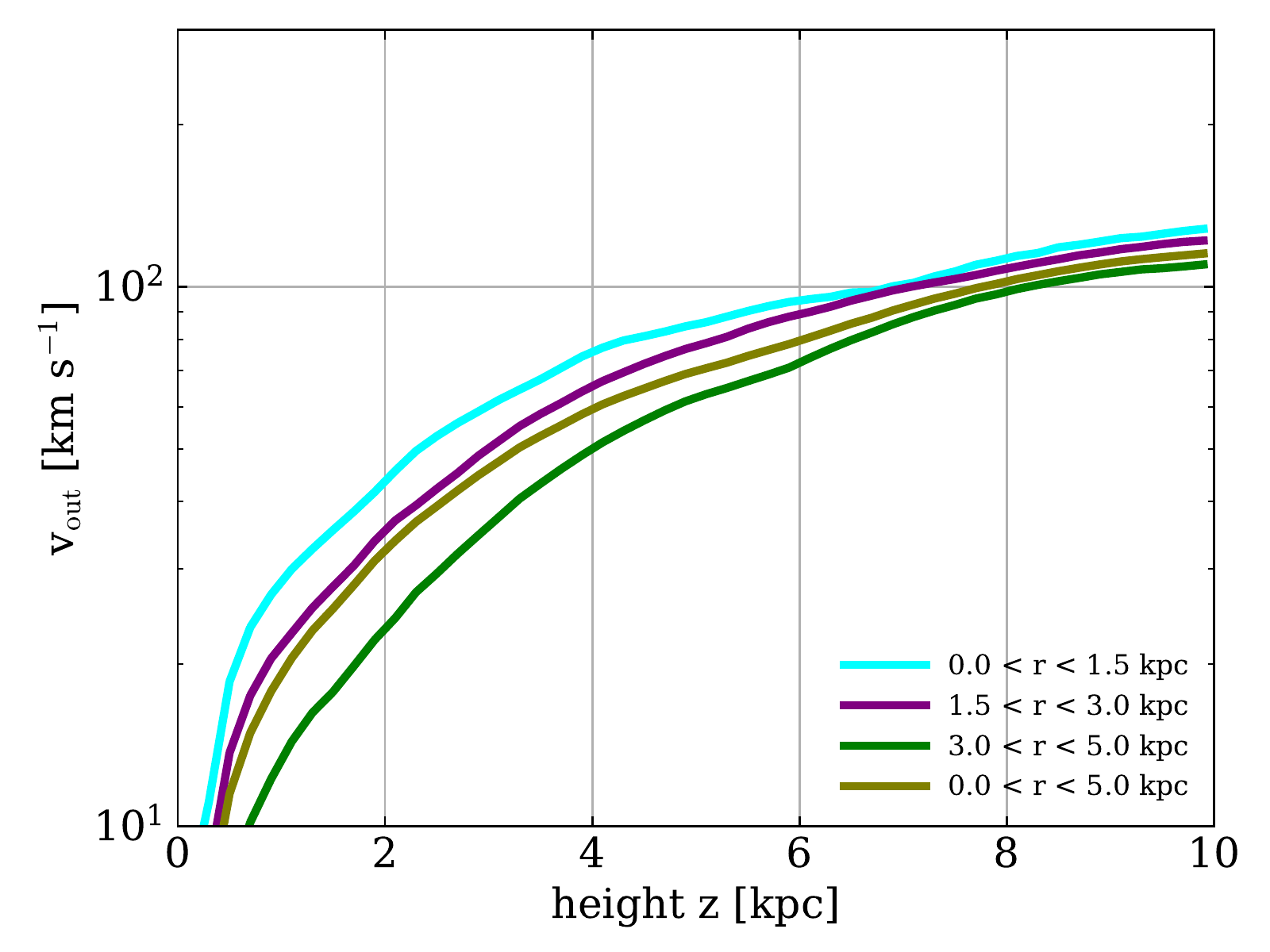}
    \includegraphics[width=0.49\textwidth]{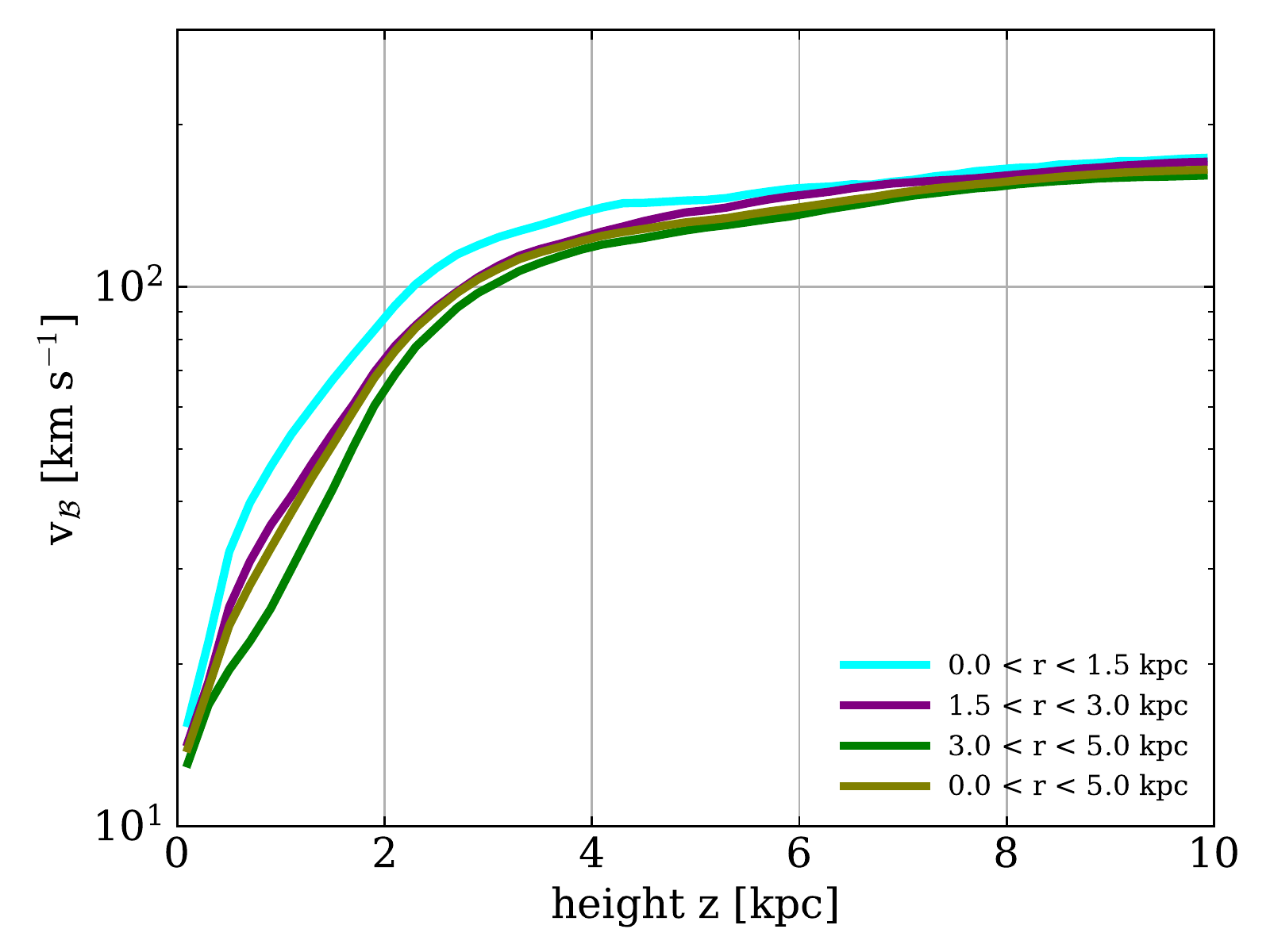}
    \includegraphics[width=0.49\textwidth]{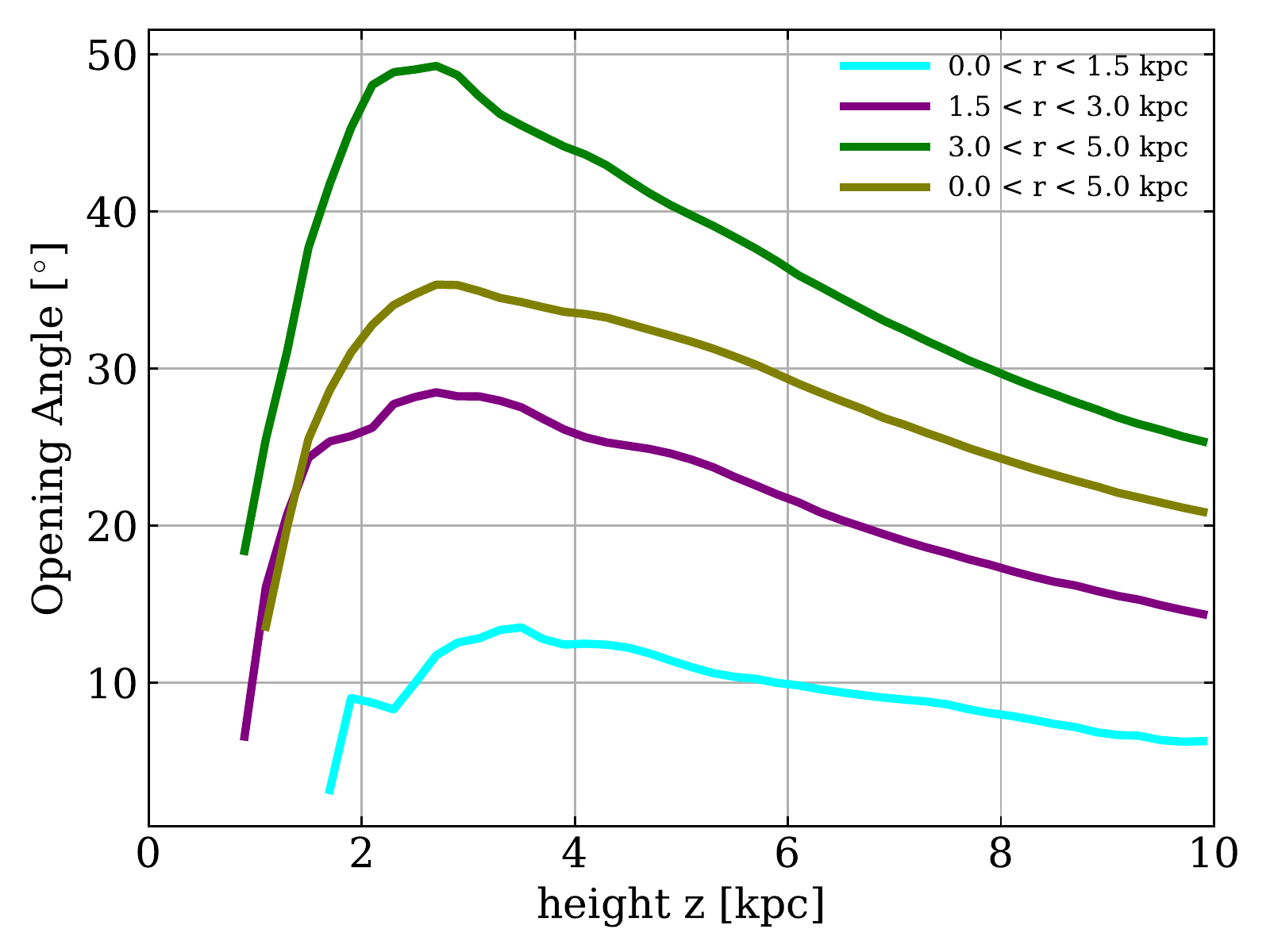}
    \caption{Mean values for the radial velocity (top left, v$_\mathrm{r} > 0$), vertical outflow velocity (top right), the Bernoulli velocity (bottom left), and the opening angle of the wind (defined as $360/2\pi \times \arccos(v_\mathrm{z}/ v_\mathrm{r})$) within different radial regions between $0.0 < R < 1.5$ kpc (black), $1.5 < R < 3.0$ kpc (red), $3.0 < R < 5.0$ kpc (blue) and $0.0 < R < 5.0$ kpc (orange). We note that these results are mass weighted, while our PDFs in Fig.~ \ref{fig:2dpdfs_mass_comp} and Fig.~\ref{fig:2dpdfs_energy_comp} are mass outflow rate or energy outflow rate weighted respectively. To demonstrate the effect of weighting either by mass or energy outflow rate we add the mass and energy flux weighted velocities for $0.0 < R < 5.0$ kpc as dashed orange (mass outflow rate) and dotted orange (energy outflow rate) lines to the top left panel.}
    \label{fig:vout}
\end{figure*}

\begin{figure*}
    \centering
    \includegraphics[width=0.87\textwidth]{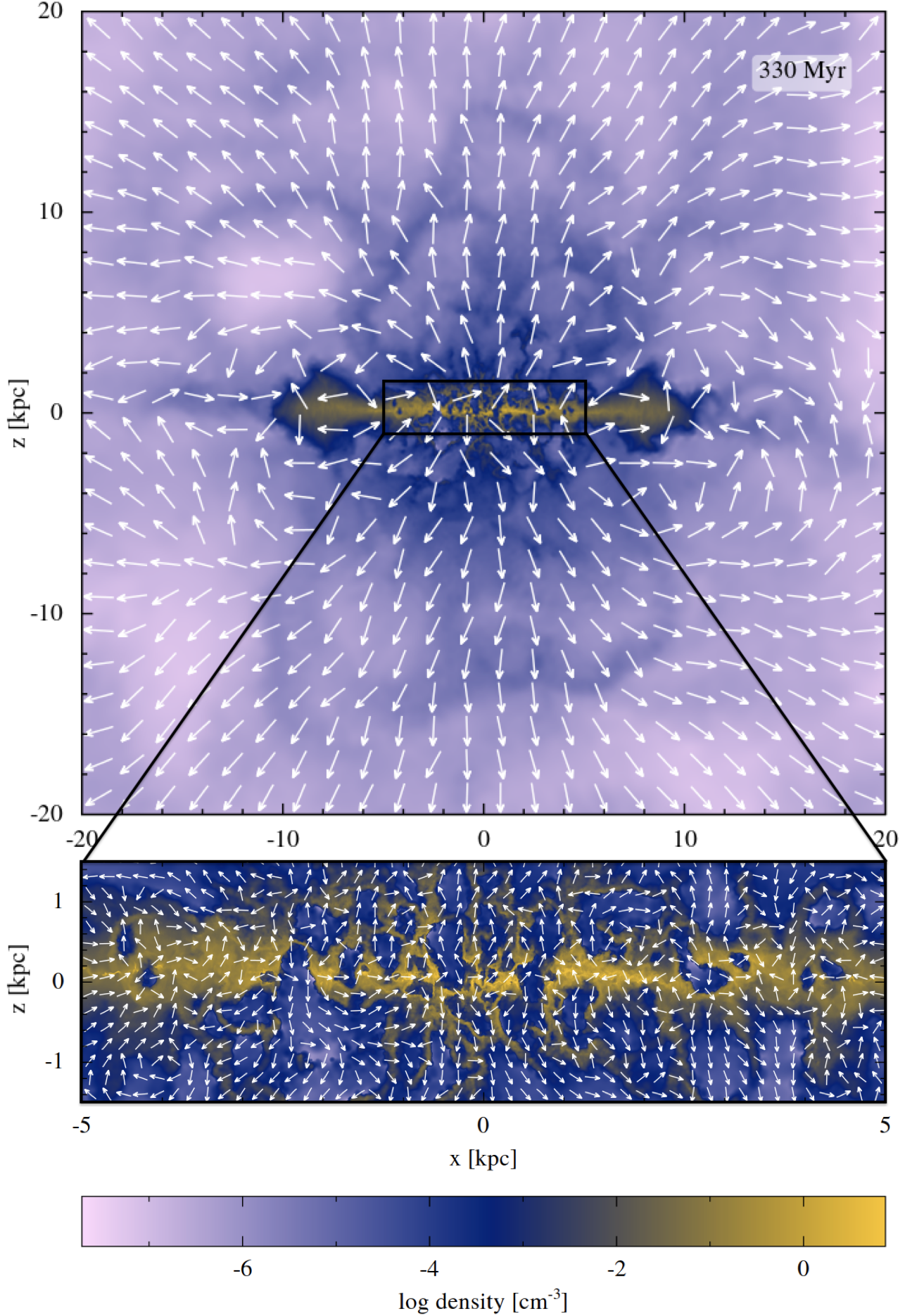}
    \caption{Top: Edge-on slice at $y=0$ with a thickness of 50 pc of hydrogen number density at $t=330$ Myr, demonstrating the large scale outflow from -20 kpc to 20 kpc in x and z direction indicating an opening streamline geometry. Bottom: Zoom on the innermost disc (-5 to 5 kpc width and -1.5 kpc to 1.5 kpc height) to capture parts of the smaller scale (turbulent) fountain flow. The vectors show the directionality (not magnitude) of the velocity field and clearly identify that the velocity field is opening up as a function of height.}
    \label{fig:streamlines}
\end{figure*}

In the left panels of Fig.~\ref{fig:radial_loadings_mass} and Fig.~\ref{fig:radial_loadings_energy} we show the radial trends of the mass outflow rate and the energy outflow rate respectively. 
Again, we plot these at four different heights (1 (black), 3 (red), 5 (blue), and 10 (orange) kpc). We find that at low altitude the mass outflow rate is higher than at high altitude as a function of cylindrical distance $R$ and peaks around a radius of 2 kpc. The energy outflow rate at low altitude is highest in the innermost part ($R < 2 kpc$) of the outflow, after which it sharply declines. However, at higher altitude the outskirts carry more of the total energy in the outflow. This trend indicates that the wind (especially, the hot component) delivers significant energy from inner, lower regions to outer, upper regions.

In the left panels of Fig.~\ref{fig:vertical_loadings_mass} and Fig.~\ref{fig:vertical_loadings_energy} we show the vertical dependence of the total respective outflow rates for mass and energy for the global disc ($0<R<5$ kpc, olive) as well as three different radial annuli between 0.0-1.5 kpc (cyan), 1.5-3.0 (purple) kpc and 3.0-5.0 (green) kpc. All of these represent the averaged vertical profiles between 200 and 400 Myr of evolution. We do this to illustrate the different outflow regimes in the center and in the outskirts as a function of height above the midplane, which we could already clearly identify in phase space in Fig.~\ref{fig:2dpdfs_mass_comp} and Fig.~\ref{fig:2dpdfs_energy_comp}. We find profiles that decrease with increasing radius at all heights. We note that the sum of the cyan, purple and green lines is the olive line. The variation of the outflow rates and the loading factors gets flatter (but not actually flat) above a height of 5 kpc. The instantaneous profiles are similar to the averaged profiles, modulo small amplitude perturbations that arise from the fluctuations in the star formation rate. This set of figures provides more evidence that the outflows are initially driven from the center and lose mass and energy as a function of increasing height. Most of the mass and the energy are transported to the outskirts of the outflow at higher altitude, while at lower altitude above the midplane the mass budget is comparable in the center and the outskirts. However, the bulk of the energy is initially transported outwards through the center until a height of around 4 kpc, after which the outskirts transport most of the energy. These findings are in good agreement with the detailed phase-space analysis of the outflow in Fig.~\ref{fig:2dpdfs_mass_comp} and Fig.~\ref{fig:2dpdfs_energy_comp}.

In the right panels  of Fig.~\ref{fig:radial_loadings_mass} and Fig.~\ref{fig:radial_loadings_energy} we show the radial trends of the mass outflow rate and the energy outflow rate respectively separated by the warm ($T < 5 \times 10^{5}$, dashed lines) and hot phase respectively ($T > 5 \times 10^{5}$, dotted line) for four different height above the midplane at 1 (black), 3 (red), 5 (blue) and 10 kpc (orange). The mass outflow rate as function of the radius R is dominated at all heights that we investigated by the warm wind and most of the mass ($\sim 90$ per cent) is transported by this component while the hot component transport $\sim 10$ per cent of the total mass. However, at low altitude most of the mass in the hot wind is transported  through the central ``chimney'', while at higher altitude most of the hot gas in the wind is located in the outskirts of the galaxy, indicating that the hot wind is leaving trough an angle and is mostly generated in the central starburst region, while the warm wind shows almost constant outflow rates as function of radius.

In the right panels  of Fig.~\ref{fig:vertical_loadings_mass} and Fig.~\ref{fig:vertical_loadings_energy} we show the vertical trends of the mass outflow rate and the energy outflow rate respectively separated by the warm ($T < 5 \times 10^{5}$, dashed lines) and hot phase respectively ($T > 5 \times 10^{5}$, dotted line) for four different radial annuli along the midplane given via $0.0<R<1.5$ (cyan), $1.5<R<3.0$ (purple), $3.0<R<5.0$ (green) and $0.0<R<5.0$ kpc (olive). In all radial bins and at all height the mass outflow is dominated by the warm wind, which is especially true at low altitude out to 2 kpc after which the mass fraction between hot and warm remains close to 10 per cent. At low altitude out to 500 pc the sharp increase in mass outflow rate in the warm is likely related to the fact that we pick up a lot of warm material that belongs to the disc and classify it as outflow. The energy outflow rate on the other hand is dominated by the warm wind out to a height of 7 kpc in all radial bins. This is likely due to the fact that the hot wind is adiabatically cooled with increase in height and the formerly hot wind is classified as warm wind above this specific height. In future studies this has to be investigated more closely by the profiles for specific entropy to disentangle the origin of the reduction of temperature of the hot out to higher heights which is beyond the scope of this paper which has the main focus on the phase separation of outflows. 

In Fig.~\ref{fig:vout} we show the time averaged (over 200 Myr) radial outflow velocity (top left), the time averaged vertical outflow velocity (top right), the time averaged Bernoulli velocity (bottom left) and the time averaged opening angle (bottom right) of the wind as a function of height above the midplane. We find that all averaged velocity profiles increase steeply until a height of around 2.5 to 3.5 kpc up to values of around $\sim$100 km s$^{-1}$, where the Bernoulli velocity reaches slightly higher values than the outflow velocity (see the definition of $v_\mathbf{\mathcal{B}}$). Generally, these results are very comparable to those of previous studies such as \citet[][]{Kim2020}. The solid lines in the panels in Fig.\ref{fig:vout} for radial, vertical and Bernoulli velocities are mass weighted. We state this specifically to avoid confusion with Fig.~\ref{fig:2dpdfs_mass_comp} and Fig.~\ref{fig:2dpdfs_energy_comp} which are the mass outflow rate and energy outflow rate weighted distributions (they are shifted by v$_{r}$). Hence, if we had computed the mean outflow velocity using the weighting from Fig.~\ref{fig:2dpdfs_mass_comp} and Fig.~\ref{fig:2dpdfs_energy_comp} respectively we would obtain a velocity profile with a higher mean velocity. We demonstrate that effect on the time averaged radial outflow velocity for $0 < R < 5.0$ kpc with the dashed (mass flux weighted) and the dashed-dotted line (energy flux weighted).

Finally, we note that the shape of the opening angle in the bottom right panel of Fig.~\ref{fig:vout} is increasing as a function of height, peaks around 2.5 to 3.0 kpc and declines outwards to a height of 10.0 kpc. We compute the opening angle $\alpha$ via:
\begin{equation}
    \alpha = \arccos{(v_{z} / v_{r})} \frac{360^\circ}{2 \pi}.
\end{equation}
Qualitatively, this shape can be understood as the different components of the wind. To visualize this we plot the streamlines of the wind in Fig. \ref{fig:streamlines} at $t=330$ Myrs corresponding to the peak in outflow rate following the peak in star formation at $t=300$ Myrs. In the top panel we show the large scale gas outflow out to 20 kpc height above and below the mid plane. The white arrows show the direction of the gas itself. We generally can say that, in the central part, the opening angles are small because the wind is accelerated almost vertically. At larger radii, the opening angle gets larger overall because the bulk of the wind is launched from the innermost 3.0 kpc, the region of the largest star formation rate. 
The streamlines  are then bent as the outflow widens and the opening angle increases at larger radii. The turnover towards decreasing opening angles at larger heights occurs because material with large opening angles will simply not reach these heights at the given radial bin and subsequently ``rain down'' on the outer galactic disc. In the bottom panel of Fig~\ref{fig:streamlines} we visualize the small scale turbulent flow closer to the mid plane indicating the fountain flow on smaller scales.

\subsection{Loading factor dependence}

\label{sec:sfr_density}
\begin{figure*}
    \centering
    \includegraphics[width=0.49\textwidth]{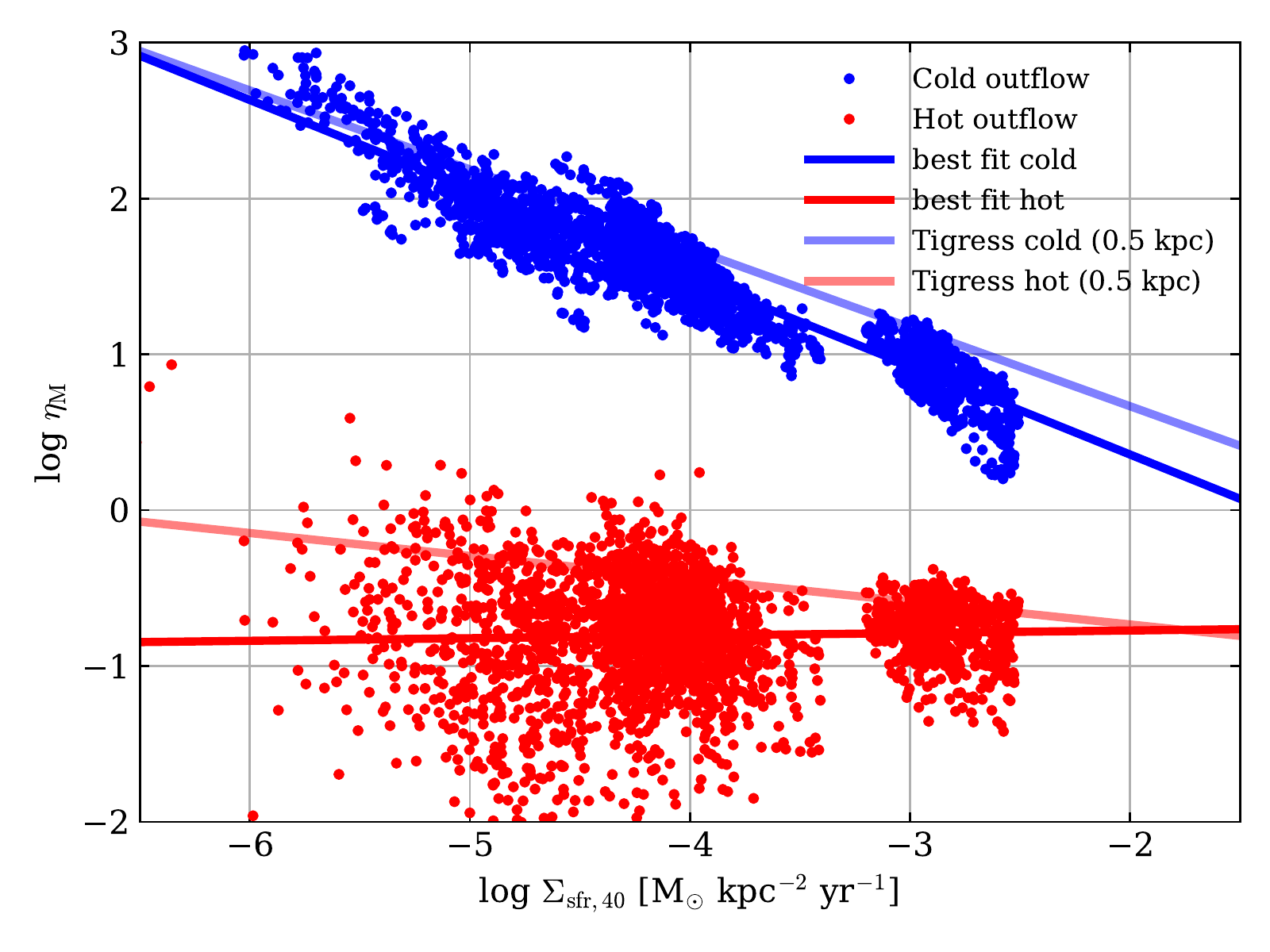}
    \includegraphics[width=0.49\textwidth]{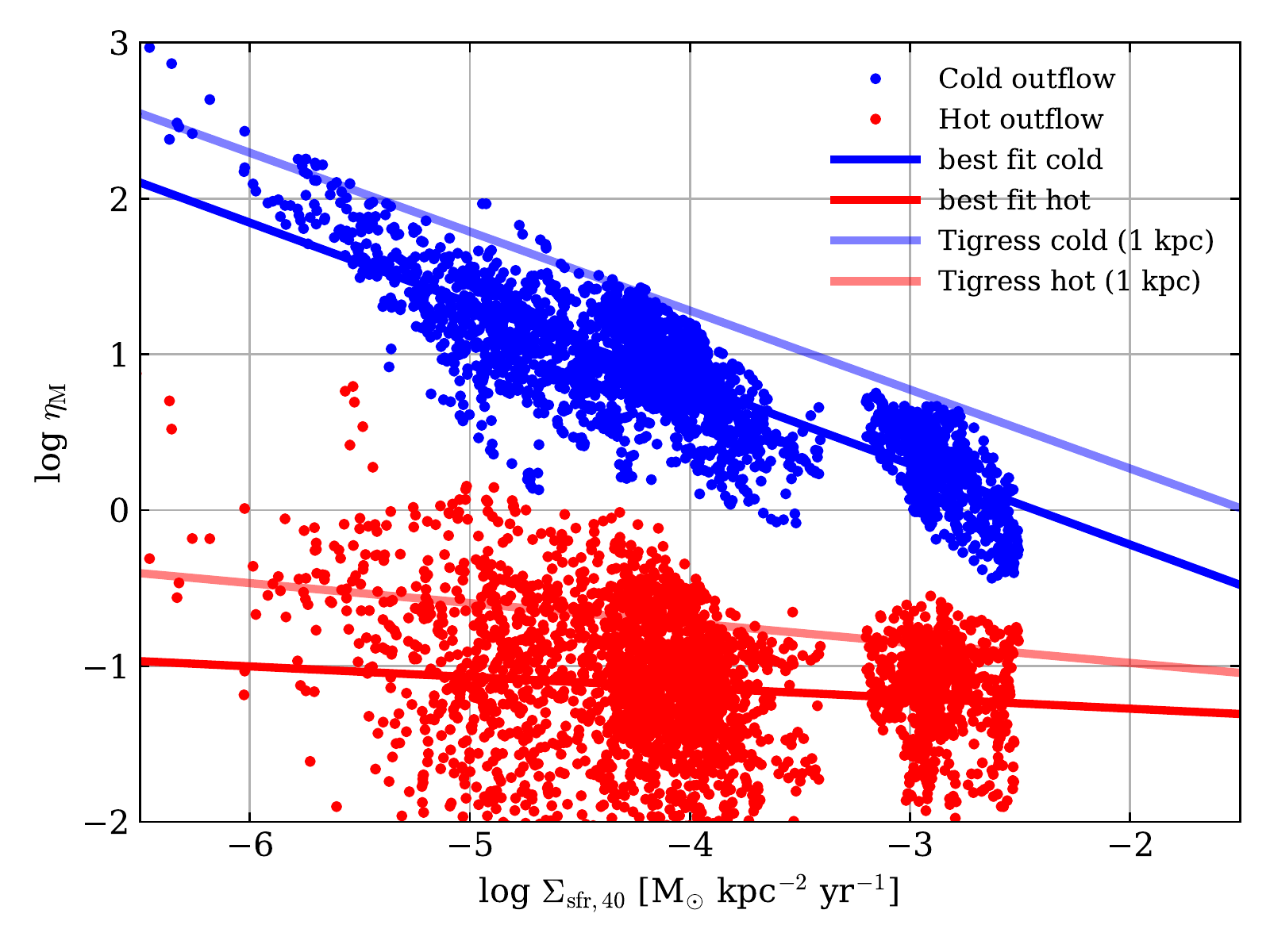}
    \includegraphics[width=0.49\textwidth]{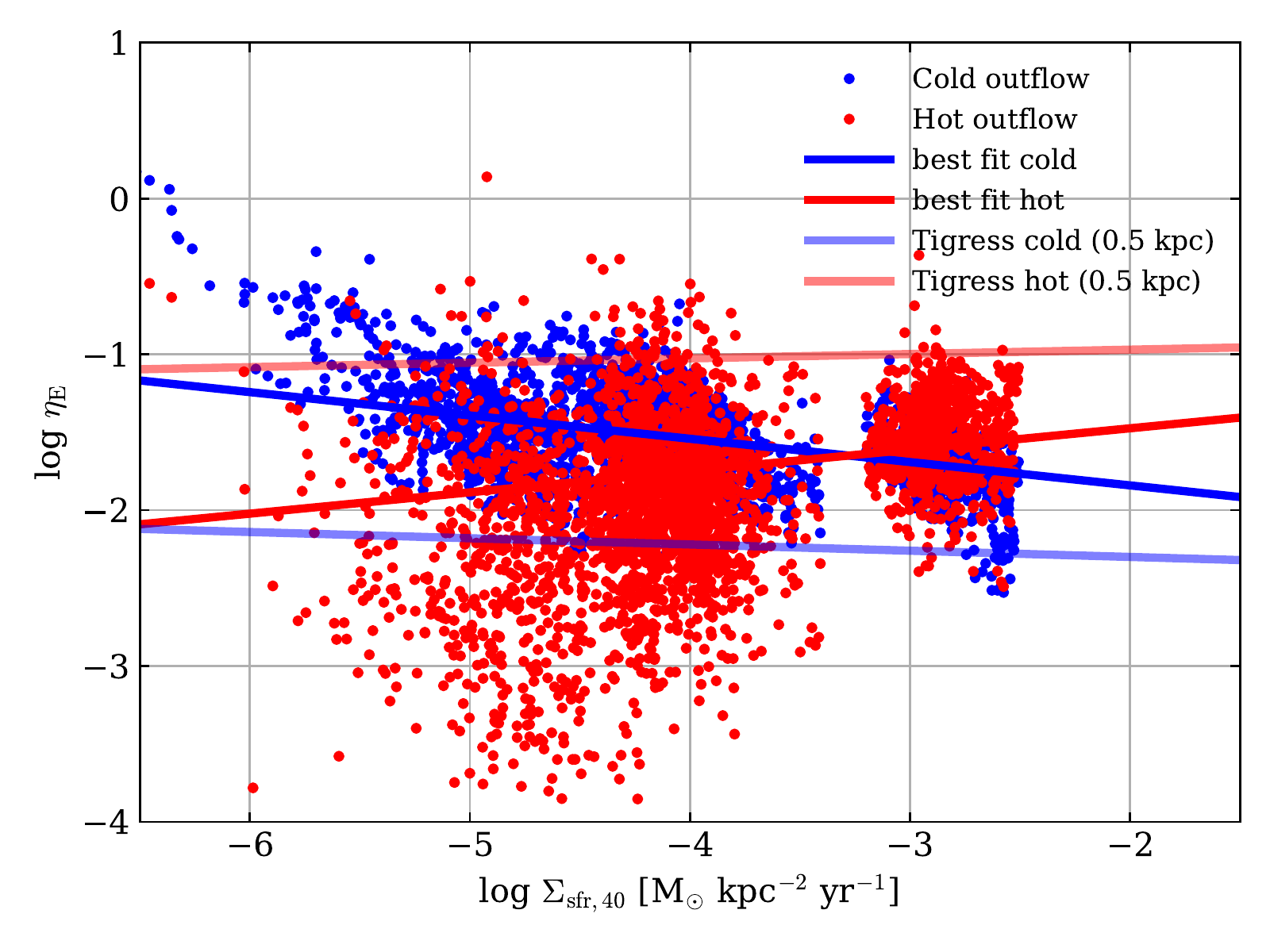}
    \includegraphics[width=0.49\textwidth]{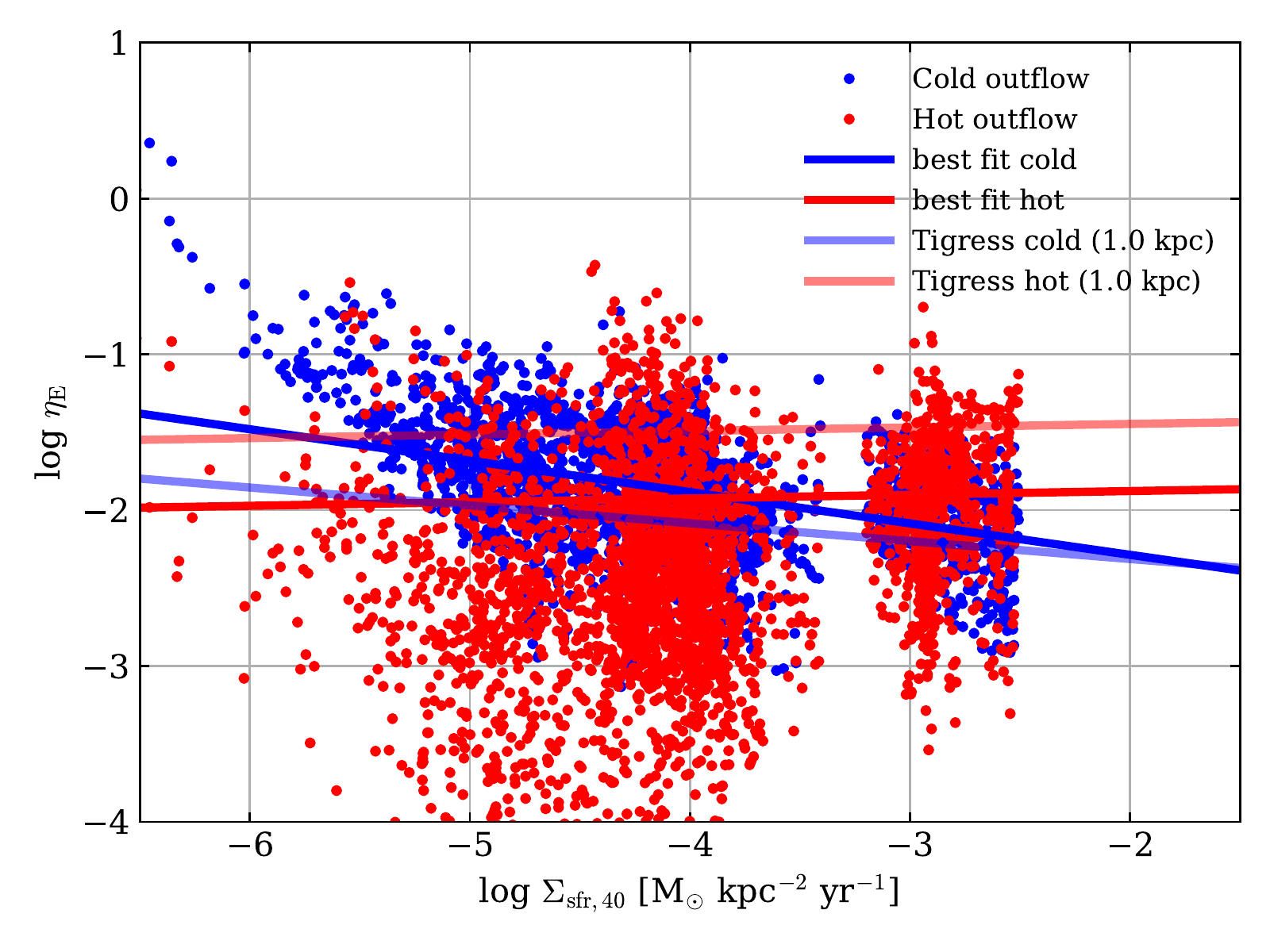}
    \caption{Mass loading (top) and energy loading (bottom) as a function of star formation rate surface density split into hot gas with $T > 5 \times 10^{5}$ K (red dots) and warm gas with $T < 5 \times 10^{5}$ K (blue dots) at two heights at 0.5 kpc (left) and 1 kpc (right). The red and blue lines represent a best fit in log-log space. 
    }
    \label{fig:sigma_sfr_phase}
\end{figure*}

In Fig.~\ref{fig:sigma_sfr_phase} we show the dependence of the mass loading factor $\eta_\mathrm{M}$ (top) and the energy loading factor $\eta_\mathrm{E}$ (center) for a height of 0.5 kpc (left) and a height of 1.0 kpc (right), on the star formation surface rate density. The data in these plots are computed on a grid with a cell size of $2.5 \times 2.5$ kpc$^2$ resulting in 16 cells across the innermost 5 kpc of the galaxy. The different colors represent the warm (blue) and hot (red) loading factors, where the warm and hot material are separated at $5 \times 10^{5}$ K. The data in these plots is collected over 20 snapshots from 200 Myrs to 400 Myrs and the star formation rate surface density is computed as a 40 Myr time average. For the mass outflow rate we find that at low altitude there is a weak correlation between the star formation rate surface density and the outflow rate, while there is almost no correlation with the hot mass loading. We provide the fits for the loading factors obtained at heights of 0.5 and 1.0 kpc, for the mass and energy in the warm and hot wind respectively. The scalings we find are as follows, where we give $\Sigma_\mathrm{sfr, 40Myr}$ in the following equations in units of M$_{\odot}$ yr$^{-1}$ kpc$^{-2}$
\begin{equation}
    \log \eta_\mathrm{M, warm} (0.5 \mathrm{kpc}) = - 0.56 \log \left(\frac{\Sigma_\mathrm{sfr, 40Myr}}{\mathrm{M}_{\odot} \mathrm{yr}^{-1} \mathrm{kpc}^{-2}}\right) -0.78 
\end{equation}
\begin{equation}
    \log \eta_\mathrm{M, hot} (0.5 \mathrm{kpc}) = + 0.016 \log \left(\frac{\Sigma_\mathrm{sfr, 40Myr}}{\mathrm{M}_{\odot} \mathrm{yr}^{-1} \mathrm{kpc}^{-2}}\right) - 0.73
\end{equation}
\begin{equation}
    \log \eta_\mathrm{M, warm} (1 \mathrm{kpc}) = - 0.516 \log \left(\frac{\Sigma_\mathrm{sfr, 40Myr}}{\mathrm{M}_{\odot} \mathrm{yr}^{-1} \mathrm{kpc}^{-2}}\right) -1.25
\end{equation}
\begin{equation}
    \log \eta_\mathrm{M, hot} (1 \mathrm{kpc}) = - 0.06 \log \left(\frac{\Sigma_\mathrm{sfr, 40Myr}}{\mathrm{M}_{\odot} \mathrm{yr}^{-1} \mathrm{kpc}^{-2}}\right) -1.40
\end{equation}
\begin{equation}
    \log \eta_\mathrm{E, warm} (0.5 \mathrm{kpc}) = - 0.14 \log \left(\frac{\Sigma_\mathrm{sfr, 40Myr}}{\mathrm{M}_{\odot} \mathrm{yr}^{-1} \mathrm{kpc}^{-2}}\right) -2.14
\end{equation}
\begin{equation}
    \log \eta_\mathrm{E, hot} (0.5 \mathrm{kpc}) = + 0.13 \log \left(\frac{\Sigma_\mathrm{sfr, 40Myr}}{\mathrm{M}_{\odot} \mathrm{yr}^{-1} \mathrm{kpc}^{-2}}\right) -1.20
\end{equation}
\begin{equation}
    \log \eta_\mathrm{E, warm} (1 \mathrm{kpc}) = - 0.20 \log \left(\frac{\Sigma_\mathrm{sfr, 40Myr}}{\mathrm{M}_{\odot} \mathrm{yr}^{-1} \mathrm{kpc}^{-2}}\right) - 2.68 
\end{equation}
\begin{equation}
    \log \eta_\mathrm{E, hot} (1 \mathrm{kpc}) = + 0.023 \log \left(\frac{\Sigma_\mathrm{sfr, 40Myr}}{\mathrm{M}_{\odot} \mathrm{yr}^{-1} \mathrm{kpc}^{-2}}\right) -1.82
\end{equation}
Furthermore, we directly compare to the fits found in the \textsc{tigress} simulation suite based on the results of \citet{Kim2020_smaug}. We note that we generally find good agreement with the slopes of the scalings from \textsc{tigress} for the warm wind, while the normalization in our simulation is lower by 0.5-1 dex. Additionally, there are some notable differences for the hot wind in our simulations. The reasons for this will be discussed below.

\section{Discussion} \label{sec:discussion}

In this section we discuss the implications of the results presented in Section \ref{sec:results}, compare to previous work, and describe limitations and pathways for future and follow up work. 

\subsection{The multiphase ISM}

Fig.~\ref{fig:galaxy} shows a turbulent galactic ISM with large SN-driven bubbles all over the midplane. The strong stellar feedback is generating large holes in the ISM with diameters of up to one kpc. The observed structure of the multiphase ISM is similar to that of previous studies of  WLM-like dwarfs such as presented in \citet[][]{Hu2016, Hu2017, Hu2019}, \citet{Lahen2020a, Lahen2020b, Lahen2022}, \citet[][]{Smith2021}, \citet{Gutcke2021} and \citet{Hislop2022}. From these models the simulations of \citet{Hu2017} labeled as \textit{SN-PI-PE} have the most comparable physical setup. However, our dwarf is an LMC-like analogue that is 10 times more massive (gas mass and halo mass)than the WLM-like simulation of \citet{Hu2017}. Generally, we find that the phase structure of the ISM in density-temperature phase space as presented in the bottom panel of Fig.~\ref{fig:phase} looks very similar to the results of \citet{Hu2017} with the difference that we find more mass in the hot phase in our LMC-run. Hence, our density-temperature phase diagram shows more similarities to the phase diagram of \citet{Lahen2020a} who simulated a merger of two WLM-like dwarfs that shows a peak of star formation with similar values to those we find in our LMC-like run of around 0.1 M$_{\odot}$ yr$^{-1}$ based on our Fig.~\ref{fig:sfr}. The same can be said when comparing their density-pressure phase space diagram with ours, which shows a similar thermally unstable region around a density of 0.1 cm$^{-3}$ and pressure of 3000 K cm$^{-3}$ as in the merger simulations of \citet{Lahen2020a} and \citet{Lahen2020b}. The studies of \citet{Smith2021} and \citet{Hislop2022} put the focus on a more compact version of the WLM-like dwarf galaxy with a factor of 20 lower average star formation rate than our system. While \citet{Hislop2022} did not directly look into the ISM-phase structure, the initial conditions and the code version used are the same as in the compact WLM version of \citet{Hu2017} and similar results are expected. The phase-structure of \citet{Smith2021} looks slightly different from the one we find, but they also use a different cooling and chemistry network based on \textsc{grackle} \citep{Smith2017}. The same can be said for the simulations published by \citet{Gutcke2021}, who use a tabulated cooling function and a density threshold for star formation. 
We note that a more comprehensive comparison between different hydrodynamical methods can be found in \citet{Hu2023} for the WLM regime.

\subsection{The structure of the outflow}

The central result of this study is the morphology and phase structure of the outflow driven by SN-feedback which we report for the first time in a simulation of an LMC-like analogue at a mass resolution of 4 M$_{\odot}$, which allows us to self-consistently resolve SN-feedback within the multiphase ISM. Some of the first simulations of global galaxy simulations with improved treatment of the ISM-physics were carried out by \citet{Hopkins2012}, who simulated four galaxies in the mass range of a Milky Way (MW), an SMC, a gas rich Sbc and a thicker disc that represented higher redshift conditions. The run we present here is in between the runs SMC and Sbc of \citet{Hopkins2012}, where we note that the SMC model of \citet{Hopkins2012} is a factor of 1.5 more massive (in gas) than our LMC run based on the gas mass adopted, while our LMC is a scaled up version of our previous WLM runs \citep[][]{Hu2017, Steinwandel2022}. Additionally, \citet{Hopkins2012} measured mass loading factors based on a positive Bernoulli parameter which they define as b = (v$_{out}^2 + 3 c_\mathrm{s} - v_\mathrm{esc}$) and compared to a definition where the outflow is comprised out of gas with v$_{z} > 100$ km s$^{-1}$ and $|z| > 500$ pc, where they compute a mean by $\int{M_\mathrm{out}}/\int{M_{\star}}$, and report similar number for these procedures.  Our definition is based on fluxes in plane parallel slabs at different height. We find a similar value of around 1 to 5 for eta$_\mathrm{eta}$ at a height of 500 pc for the time evolution of the mass loading factors as shown in the center panel of Fig.~\ref{fig:sfr} when directly compared to Fig.~6 of \citet{Hopkins2012} for their SMC and Sbc runs. Our mass loading factors decrease by around 1.5 dex at higher altitude.

More recently, \citet{Pandya2021} performed a detailed study of mass, metal, momentum and energy loading in the FIRE-2 cosmological zoom-in simulations. The most comparable halos to our simulation are the FIRE-2 m11 halos which show higher mass loading factors compared to the ones that we report based on our results in the center panel of Fig.~\ref{fig:sfr} within 0.1 R$_\mathrm{vir}$. The energy loading reported in \citet{Pandya2021} is higher than our time averaged global values that we find based on the results of the bottom panel of Fig.~\ref{fig:sfr}, where we find an energy loading of 0.01 to 0.07 at 1 kpc height, but if we measure at R$_\mathrm{vir}$ where \citet{Pandya2021} measured their energy loadings we find a value that is consistently around 0.01 while \citet{Pandya2021} report values from 0.05 to 0.3 for the systems that come closest to our simulations (m11a, m11b, m11q, m11c, m11v, m11f).

One of the central parts of our study is the detailed analysis of the 2d joint PDFs of outflow velocity and sound speed for the mass flux (Fig.~\ref{fig:2dpdfs_mass_comp}) and the energy flux (Fig.~\ref{fig:2dpdfs_energy_comp}). These have been investigated previously in a number of studies including \citet{Fielding2018}, \citet{Kim2020}, \citet{Schneider2020} and \citet{Steinwandel2022} with the general conclusion that the bulk of the mass is transported by the warm phase while most of the energy is transported in the hot phase of the wind. However, \citet{Pandya2021} pointed out that the hot outflow loading dominates for M$_{\star} > 10^8M_{\odot}$ \citet{Fielding2018} and \citet{Kim2020} studied these PDFs for vertically stratified galactic discs and we note that we generally find good agreement between  the shape of their 2d PDFs and ours, when compared at the appropriate  heights (200pc and 540 pc for \citet{Fielding2018} and at the scale height H, and 2H as well as at 500 pc and 1000 pc for \citet{Kim2020}). These heights are in the range of the first two rows of our Fig.~\ref{fig:2dpdfs_mass_comp} and Fig.~\ref{fig:2dpdfs_energy_comp} at 500 pc and 1000 pc, for which we find quite good agreement on the mass outflow rate where we can clearly see that the bulk of the mass is transported by the warm wind. However, the energy flux shows a more bimodal distribution, with the warm component gradually vanishing at larger heights. Ultimately, this indicates two concepts. First, the slower part of the warm wind component is falling back to the disc (the left part of the phase diagram vanishes at larger heights). Second, the faster part of the warm wind component ($v_\mathrm{out} > 50$ km s$^{-1}$) is lost at higher heights, most likely due to mixing with the hot wind or its adiabatic expansion which would both be consistent with the lower temperature of the hot wind at larger height. In addition, our simulation allows for a more detailed investigation of the shape of the wind by quantifying the radial structure as we described in Section \ref{sec:results}. 
We note that the comparison with the results of \citet{Pandya2021} based on the FIRE-2 simulations \citep[][]{Hopkins2018} is more complex since they measured the outflow rates in shells between 0.1 and 0.2 of R$_\mathrm{vir}$. Qualitatively, the simulations indicate a similar phase structure.

Finally, it is worth discussing the complex structure of the wind in greater detail, for which we find multiple indicators that there has to be an opening velocity streamline trajectory. The first indicator for this can be seen based on the radial outflow rate profiles of Fig.~\ref{fig:radial_loadings_mass} and Fig.~\ref{fig:radial_loadings_energy} for different heights above the midplane, which shows that the energy outflow rate is decreasing at low height as a function of radius and slightly increasing at the largest height of 10 kpc. This is evidence that there is a redistribution of energy from the center at low heights to the outskirts at large heights. If we consider vertical profiles of the outflow velocities that are of significance for this study we can try to obtain a first order estimate of the opening angle of the wind based on the misalignment of the radial velocity and its vertical component for which we show radial profiles in Fig.~\ref{fig:vout}.
These can be used to obtain the opening angle which we measure at different radii as a function of height above the mid plane, where we find the largest opening angle in the outermost regions above the disc and the lowest opening angles in the central parts, as expected. 

The shape of the opening angle is quite complex, but one can straightforwardly show that a similar shape can be obtained by assuming a ballistic wind with uniform streamlines and constant angle dependent injection velocity. The idea is then that, in the inner parts of the galaxy at low altitude one is simply measuring a wind component perpendicular to the disc. Since the wind is opening up as a function of height we measure first a larger opening angle until around a height of 3 kpc, after which we find a decline again, simply because most of the bent field lines will not reach that altitude, at least not within the innermost radial region above the disc. This then explains the increase in the opening angle in the different radial regions as we measure more and more of the material that is driven out of the center. 

We also find that the wind shows similar scalings for the dependence of the loading factors as a function of the star formation rate surface density in comparison to previous work. The derived scalings for the warm wind are in good agreement with the results of \citet{Kim2020_smaug}. For the hot wind we find similar slopes for the mass and the energy loading but the loading factors are lower than in the simulations of \citet{Kim2020_smaug}. That being said, the mass loading factors at low altitude are comparable for the warm wind and only slightly lower for the hot wind than the ones reported by \citet{Kim2020_smaug}. The energy loading at low altitude is more dominated for our set of simulations by the warm component when compared to \citet{Kim2020_smaug} in the sense that we have similar energy loadings in warm and hot phase with higher laodings in the warm than \citet{Kim2020_smaug} and lower loadings in the hot. At higher altitude we find an offset by half a dex for the mass loading (warm and hot), while the energy loading is consistently lower in the hot than in \citet{Kim2020_smaug} but the warm loadings are comparable. One possible explanation for this could be the suppression of clustering of the stars and subsequently the feedback \citep[e.g.,][]{Fielding2017,Fielding2018}, because our simulations include photo-ionizing feedback that has been shown in previous simulations to decrease the clustering of the SNe \citep[][]{Hu2017, Smith2021} and is a component that was not active in the simulations of \citet{Kim2020_smaug}

Additionally, we note that the lower energy loading factors are generally also found in the lower mass systems we simulated such as WLM and LeoP which will be presented with a detailed discussion on the wind phase-structure in a future study. Specifically, when considering the results of \citet{Steinwandel2022}, it is quite clear that there is almost no hot wind in the WLM-fiducial run in the classical sense and a hot wind was only able to be established by the inclusion of runaway stars. Alternatively, one can increase the gas surface density of the simulations. 

We note that \citet{Andersson2022} find a weaker impact of runaways stars on the phase structure of galactic outflows in WLM-like simulations, but they also simulated a higher column density (shorter disc scale length) version of the dwarf that is presented in \citet{Steinwandel2022} with a different physics implementation. 

Recently, \citet{Rathjen2022} investigated the phase structure of the winds in the SILCC simulations. When only SNe, stellar winds and photo-ionizing radiation are included they find similar results for the PDFs for mass outflow rate and energy outflow rate in terms of outflow velocities (bulk mass outflow between 10 and 100 km s$^{-1}$) and thermal state of the gas (warm gas with c$_\mathrm{s} \sim $ 10 km s$^{-1}$) comparable to what we find. While we do not include a proper treatment for stellar winds, we adopt a treatment for the photo-ionizing radiation, which makes our runs quite comparable to theirs considering the underlying gas cooling is almost identical between SILCC and our simulations, although they adopt higher metallicity. Its unlikely that stellar winds would change this picture strongly given recent work on stellar wind bubbles \citep[e.g.][]{Lancaster2021a, Lancaster2021b, Lancaster2021c}. Additionally, a Lagrangian code such as ours is not necessarily best suited for a stellar wind feedback implementation due to the lack of resolution elements with which one can resolve the bubble. Resolving the thermal mixing layer of the bubble which is ultimately driving its evolution requires much higher resolution. Hence, any stellar wind treatment at our resolution has to be labeled as ``sub-grid''. 
Furthermore, \citet{Rathjen2022} investigate the impact of cosmic ray (CR) diffusion driven winds on the phase structure, which leads to an additional cold phase wind launched by CRs above the midplane. Since, we do not adopt this mechanism it is obviously absent from our phase structure. That being said it is  not true for our simulations that we do not obtain any cold material in our phase structure below a sound speed of 10 km s$^{-1}$, but it is difficult to properly compare since they exclusively select the snapshots which have mass loading higher than one in their analysis, while we average over a full wind driving cycle. Finally, we note that they simulated much higher column densities in their main paper than we did although they show some comparisons to their lower column density runs in the appendix (which are still higher column density than our LMC analogue by at least half dex). 

\section{Conclusions} \label{sec:conclusions}

Finally, we summarize the main findings of this work. We present initial results from an LMC-like simulation with solar mass and sub-parsec resolution with a special focus on the supernova-driven galactic outflows. This is the largest galaxy simulated with a resolved ISM to date.

\begin{enumerate}
    \item Our system shows a star formation rate that varies relatively little over the evolution time, with a mean star formation rate of around $\sim 0.05$ M$_{\odot}$ yr$^{-1}$ and a cycle of roughly 150 Myrs. The values for gas surface density and star formation rate surface density we put forward in Fig.~\ref{fig:surface_dens} indicate that the galaxy is forming stars in agreement with the observations of \citet{Bigiel2010} for the dwarf galaxy regime, corresponding to a gas depletion time scale of 10 Gyr.
    
    \item We measured the time evolution of loading factors for mass and energy. At low altitude (1 kpc) we find a mass loading around $\sim$ 1.0 while at higher altitudes we find a roughly constant global value of around 0.1. The global averages for the energy loading are roughly constant with a slight decrease from $2 \times 10^{-2}$ to $9 \times 10^{-3}$ as we move outwards from 1 kpc to 10 kpc. This indicated that the wind at higher altitude is dominated by the hot wind. 
    
    \item We discussed the mass and energy flux weighted joint PDFs of outflow velocity and sound speed in different radial regions and different heights. We see a bimodal wind structure, where both the warm and hot wind transport significant amounts of mass and energy. These results are qualitatively similar to the findings of \citet{Kim2020} as well as the simulations of \citet{Fielding2018} but differ in terms of the total mass and energy loading budget as in our simulations hot and warm mass and energy loading are more similar to one another than in \citet{Kim2020} and \citet{Fielding2018}.

    \item The wind shows an opening streamline structure with an average opening angle of around 30 degrees that has a peak at a height of around 3 kpc. The structure of the streamlines is in rough agreement with a (uniform) ballistic wind. However, a more detailed analytical model for the streamline structure needs to be developed to better understand the complex structure of the opening angle.  
    
    \item The mass loading of the warm wind is decreasing with the star formation rate surface density $\Sigma_{\rm SFR}$ with a slope of around -0.5 at 500 pc and 1000 pc height. The hot wind shows a scaling with $\Sigma_{\rm SFR}$ with a slope of around +0.01 at 500 pc and -0.06 at 1000 pc. The slopes of the warm and hot wind scaling are in good agreement with the results of \citet{Kim2020} but both wind components show lower loading factors by roughly half a dex. The energy loading shows asymmetric behavior for the hot and the warm wind with slopes of -0.14 for the warm and +0.13 for the hot wind at 500 pc height with similar trends at 1000pc height for the warm wind, but shallower for the hot wind (slope of +0.023). The hot wind has lower energy loading and the warm wind has higher energy loading than in \citet{Kim2020_smaug} at similar star formation rate surface density. There are a number of reasons for why exactly that could be the case, including the different gravitational potential between our runs and \textsc{tigress} or the different geometry adopted. Hence, this is not necessarily to be interpreted as an apples to apples comparison to \textsc{tigress} and could be explained by testing against pressure regulated feedback modulated theory of \citet{Ostriker2022} which we postpone to a future study.
\end{enumerate}
.
\section*{Acknowledgements}
UPS thanks Matt Orr, Viraj Pandya, Eliot Quataert and Romain Teyssier for helpful discussions. UPS thanks Daniel J. Price for making the software \textsc{splash} publicly available \citep{Price2007}. UPS acknowledges the computing time provided on SuperMuc-NG in Garching under the project number pn72bu. RSS is supported by the Simons Foundation. GLB acknowledges support from the NSF (AST-2108470, XSEDE grant MCA06N030), NASA TCAN award 80NSSC21K1053, and the Simons Foundation (grant 822237). ECO and CGK acknowledge support from the Simons Foundation (888968). CGK acknowledges support from NASA ATP award 80NSSC22K0717. This work was supported by the Simons Collaboration on ``Learning the Universe'' 

\vspace{5mm}
\facilities{SuperMUC-NG (LRZ Garching), Rusty (Flatiron Institute)}

\software{P-Gadget3 \citep[][]{springel05, Hu2014, Steinwandel2020}, astropy \citep{2013A&A...558A..33A,2018AJ....156..123A}, CMasher \citep{cmasher}, matplotlib \citep{Matplotlib}, scipy \citep[][]{scipy}, numpy \citep[][]{numpy}}

\bibliography{sample631}{}

\begin{thebibliography}{}
\expandafter\ifx\csname natexlab\endcsname\relax\def\natexlab#1{#1}\fi
\providecommand{\url}[1]{\href{#1}{#1}}
\providecommand{\dodoi}[1]{doi:~\href{http://doi.org/#1}{\nolinkurl{#1}}}
\providecommand{\doeprint}[1]{\href{http://ascl.net/#1}{\nolinkurl{http://ascl.net/#1}}}
\providecommand{\doarXiv}[1]{\href{https://arxiv.org/abs/#1}{\nolinkurl{https://arxiv.org/abs/#1}}}

\bibitem[{{Andersson} {et~al.}(2022){Andersson}, {Agertz}, {Renaud}, \&
  {Teyssier}}]{Andersson2022}
{Andersson}, E.~P., {Agertz}, O., {Renaud}, F., \& {Teyssier}, R. 2022, arXiv
  e-prints, arXiv:2209.06218.
\newblock \doarXiv{2209.06218}

\bibitem[{{Ashley} {et~al.}(2020){Ashley}, {Fox}, {Jenkins}, {Wakker},
  {Bordoloi}, {Lockman}, {Savage}, \& {Karim}}]{Ashley2020}
{Ashley}, T., {Fox}, A.~J., {Jenkins}, E.~B., {et~al.} 2020, \apj, 898, 128,
  \dodoi{10.3847/1538-4357/ab9ff8}

\bibitem[{{Astropy Collaboration} {et~al.}(2013){Astropy Collaboration},
  {Robitaille}, {Tollerud}, {Greenfield}, {Droettboom}, {Bray}, {Aldcroft},
  {Davis}, {Ginsburg}, {Price-Whelan}, {Kerzendorf}, {Conley}, {Crighton},
  {Barbary}, {Muna}, {Ferguson}, {Grollier}, {Parikh}, {Nair}, {Unther},
  {Deil}, {Woillez}, {Conseil}, {Kramer}, {Turner}, {Singer}, {Fox}, {Weaver},
  {Zabalza}, {Edwards}, {Azalee Bostroem}, {Burke}, {Casey}, {Crawford},
  {Dencheva}, {Ely}, {Jenness}, {Labrie}, {Lim}, {Pierfederici}, {Pontzen},
  {Ptak}, {Refsdal}, {Servillat}, \& {Streicher}}]{2013A&A...558A..33A}
{Astropy Collaboration}, {Robitaille}, T.~P., {Tollerud}, E.~J., {et~al.} 2013,
  \aap, 558, A33, \dodoi{10.1051/0004-6361/201322068}

\bibitem[{{Astropy Collaboration} {et~al.}(2018){Astropy Collaboration},
  {Price-Whelan}, {Sip{\H{o}}cz}, {G{\"u}nther}, {Lim}, {Crawford}, {Conseil},
  {Shupe}, {Craig}, {Dencheva}, {Ginsburg}, {VanderPlas}, {Bradley},
  {P{\'e}rez-Su{\'a}rez}, {de Val-Borro}, {Aldcroft}, {Cruz}, {Robitaille},
  {Tollerud}, {Ardelean}, {Babej}, {Bach}, {Bachetti}, {Bakanov}, {Bamford},
  {Barentsen}, {Barmby}, {Baumbach}, {Berry}, {Biscani}, {Boquien}, {Bostroem},
  {Bouma}, {Brammer}, {Bray}, {Breytenbach}, {Buddelmeijer}, {Burke},
  {Calderone}, {Cano Rodr{\'\i}guez}, {Cara}, {Cardoso}, {Cheedella}, {Copin},
  {Corrales}, {Crichton}, {D'Avella}, {Deil}, {Depagne}, {Dietrich}, {Donath},
  {Droettboom}, {Earl}, {Erben}, {Fabbro}, {Ferreira}, {Finethy}, {Fox},
  {Garrison}, {Gibbons}, {Goldstein}, {Gommers}, {Greco}, {Greenfield},
  {Groener}, {Grollier}, {Hagen}, {Hirst}, {Homeier}, {Horton}, {Hosseinzadeh},
  {Hu}, {Hunkeler}, {Ivezi{\'c}}, {Jain}, {Jenness}, {Kanarek}, {Kendrew},
  {Kern}, {Kerzendorf}, {Khvalko}, {King}, {Kirkby}, {Kulkarni}, {Kumar},
  {Lee}, {Lenz}, {Littlefair}, {Ma}, {Macleod}, {Mastropietro}, {McCully},
  {Montagnac}, {Morris}, {Mueller}, {Mumford}, {Muna}, {Murphy}, {Nelson},
  {Nguyen}, {Ninan}, {N{\"o}the}, {Ogaz}, {Oh}, {Parejko}, {Parley}, {Pascual},
  {Patil}, {Patil}, {Plunkett}, {Prochaska}, {Rastogi}, {Reddy Janga},
  {Sabater}, {Sakurikar}, {Seifert}, {Sherbert}, {Sherwood-Taylor}, {Shih},
  {Sick}, {Silbiger}, {Singanamalla}, {Singer}, {Sladen}, {Sooley},
  {Sornarajah}, {Streicher}, {Teuben}, {Thomas}, {Tremblay}, {Turner},
  {Terr{\'o}n}, {van Kerkwijk}, {de la Vega}, {Watkins}, {Weaver}, {Whitmore},
  {Woillez}, {Zabalza}, \& {Astropy Contributors}}]{2018AJ....156..123A}
{Astropy Collaboration}, {Price-Whelan}, A.~M., {Sip{\H{o}}cz}, B.~M., {et~al.}
  2018, \aj, 156, 123, \dodoi{10.3847/1538-3881/aabc4f}

\bibitem[{{Aumer} \& {White}(2013)}]{Aumer2013}
{Aumer}, M., \& {White}, S.~D.~M. 2013, \mnras, 428, 1055,
  \dodoi{10.1093/mnras/sts083}

\bibitem[{{Bakes} \& {Tielens}(1994)}]{Bakes1994}
{Bakes}, E.~L.~O., \& {Tielens}, A.~G.~G.~M. 1994, \apj, 427, 822,
  \dodoi{10.1086/174188}

\bibitem[{{Baldry} {et~al.}(2008){Baldry}, {Glazebrook}, \&
  {Driver}}]{Baldry2008}
{Baldry}, I.~K., {Glazebrook}, K., \& {Driver}, S.~P. 2008, \mnras, 388, 945,
  \dodoi{10.1111/j.1365-2966.2008.13348.x}

\bibitem[{{Baldry} {et~al.}(2012){Baldry}, {Driver}, {Loveday}, {Taylor},
  {Kelvin}, {Liske}, {Norberg}, {Robotham}, {Brough}, {Hopkins}, {Bamford},
  {Peacock}, {Bland-Hawthorn}, {Conselice}, {Croom}, {Jones}, {Parkinson},
  {Popescu}, {Prescott}, {Sharp}, \& {Tuffs}}]{Baldry2012}
{Baldry}, I.~K., {Driver}, S.~P., {Loveday}, J., {et~al.} 2012, \mnras, 421,
  621, \dodoi{10.1111/j.1365-2966.2012.20340.x}

\bibitem[{{Barnes} \& {Hut}(1986)}]{Barnes1986}
{Barnes}, J., \& {Hut}, P. 1986, \nat, 324, 446, \dodoi{10.1038/324446a0}

\bibitem[{{Barnes} {et~al.}(2011){Barnes}, {Haehnelt}, {Tescari}, \&
  {Viel}}]{Barnes2011}
{Barnes}, L.~A., {Haehnelt}, M.~G., {Tescari}, E., \& {Viel}, M. 2011, \mnras,
  416, 1723, \dodoi{10.1111/j.1365-2966.2011.18789.x}

\bibitem[{{Bergin} {et~al.}(2004){Bergin}, {Hartmann}, {Raymond}, \&
  {Ballesteros-Paredes}}]{Bergin2004}
{Bergin}, E.~A., {Hartmann}, L.~W., {Raymond}, J.~C., \& {Ballesteros-Paredes},
  J. 2004, \apj, 612, 921, \dodoi{10.1086/422578}

\bibitem[{{Bernardi} {et~al.}(2013){Bernardi}, {Meert}, {Sheth}, {Vikram},
  {Huertas-Company}, {Mei}, \& {Shankar}}]{Bernardi2013}
{Bernardi}, M., {Meert}, A., {Sheth}, R.~K., {et~al.} 2013, \mnras, 436, 697,
  \dodoi{10.1093/mnras/stt1607}

\bibitem[{{Bieri} {et~al.}(2022){Bieri}, {Naab}, {Geen}, {Coles}, {Pakmor}, \&
  {Walch}}]{Bieri2022}
{Bieri}, R., {Naab}, T., {Geen}, S., {et~al.} 2022, arXiv e-prints,
  arXiv:2209.06842.
\newblock \doarXiv{2209.06842}

\bibitem[{{Bigiel} {et~al.}(2010){Bigiel}, {Leroy}, {Walter}, {Blitz},
  {Brinks}, {de Blok}, \& {Madore}}]{Bigiel2010}
{Bigiel}, F., {Leroy}, A., {Walter}, F., {et~al.} 2010, \aj, 140, 1194,
  \dodoi{10.1088/0004-6256/140/5/1194}

\bibitem[{{Bolatto} {et~al.}(2013){Bolatto}, {Warren}, {Leroy}, {Walter},
  {Veilleux}, {Ostriker}, {Ott}, {Zwaan}, {Fisher}, {Weiss}, {Rosolowsky}, \&
  {Hodge}}]{Bolatto2013}
{Bolatto}, A.~D., {Warren}, S.~R., {Leroy}, A.~K., {et~al.} 2013, \nat, 499,
  450, \dodoi{10.1038/nature12351}

\bibitem[{{Chen} {et~al.}(2010){Chen}, {Tremonti}, {Heckman}, {Kauffmann},
  {Weiner}, {Brinchmann}, \& {Wang}}]{Chen2010}
{Chen}, Y.-M., {Tremonti}, C.~A., {Heckman}, T.~M., {et~al.} 2010, \aj, 140,
  445, \dodoi{10.1088/0004-6256/140/2/445}

\bibitem[{{Chieffi} \& {Limongi}(2004)}]{Chieffi2004}
{Chieffi}, A., \& {Limongi}, M. 2004, \apj, 608, 405, \dodoi{10.1086/392523}

\bibitem[{{Chisholm} {et~al.}(2018){Chisholm}, {Bordoloi}, {Rigby}, \&
  {Bayliss}}]{Chisholm2018}
{Chisholm}, J., {Bordoloi}, R., {Rigby}, J.~R., \& {Bayliss}, M. 2018, \mnras,
  474, 1688, \dodoi{10.1093/mnras/stx2848}

\bibitem[{{Chisholm} {et~al.}(2017){Chisholm}, {Tremonti}, {Leitherer}, \&
  {Chen}}]{Chisolm2017}
{Chisholm}, J., {Tremonti}, C.~A., {Leitherer}, C., \& {Chen}, Y. 2017, \mnras,
  469, 4831, \dodoi{10.1093/mnras/stx1164}

\bibitem[{{Chisholm} {et~al.}(2015){Chisholm}, {Tremonti}, {Leitherer}, {Chen},
  {Wofford}, \& {Lundgren}}]{Chisholm2015}
{Chisholm}, J., {Tremonti}, C.~A., {Leitherer}, C., {et~al.} 2015, \apj, 811,
  149, \dodoi{10.1088/0004-637X/811/2/149}

\bibitem[{{Cicone} {et~al.}(2014){Cicone}, {Maiolino}, {Sturm},
  {Graci{\'a}-Carpio}, {Feruglio}, {Neri}, {Aalto}, {Davies}, {Fiore},
  {Fischer}, {Garc{\'\i}a-Burillo}, {Gonz{\'a}lez-Alfonso}, {Hailey-Dunsheath},
  {Piconcelli}, \& {Veilleux}}]{Cicone2014}
{Cicone}, C., {Maiolino}, R., {Sturm}, E., {et~al.} 2014, \aap, 562, A21,
  \dodoi{10.1051/0004-6361/201322464}

\bibitem[{{Cioffi} {et~al.}(1988){Cioffi}, {McKee}, \&
  {Bertschinger}}]{Cioffi1988}
{Cioffi}, D.~F., {McKee}, C.~F., \& {Bertschinger}, E. 1988, \apj, 334, 252,
  \dodoi{10.1086/166834}

\bibitem[{{Creasey} {et~al.}(2013){Creasey}, {Theuns}, \&
  {Bower}}]{Creasey2013}
{Creasey}, P., {Theuns}, T., \& {Bower}, R.~G. 2013, \mnras, 429, 1922,
  \dodoi{10.1093/mnras/sts439}

\bibitem[{{Dav{\'e}} {et~al.}(2019){Dav{\'e}}, {Angl{\'e}s-Alc{\'a}zar},
  {Narayanan}, {Li}, {Rafieferantsoa}, \& {Appleby}}]{Dave2019}
{Dav{\'e}}, R., {Angl{\'e}s-Alc{\'a}zar}, D., {Narayanan}, D., {et~al.} 2019,
  \mnras, 486, 2827, \dodoi{10.1093/mnras/stz937}

\bibitem[{{Davies} {et~al.}(2019){Davies}, {F{\"o}rster Schreiber},
  {{\"U}bler}, {Genzel}, {Lutz}, {Renzini}, {Tacchella}, {Tacconi}, {Belli},
  {Burkert}, {Carollo}, {Davies}, {Herrera-Camus}, {Lilly}, {Mancini}, {Naab},
  {Nelson}, {Price}, {Shimizu}, {Sternberg}, {Wisnioski}, \&
  {Wuyts}}]{Davis2019}
{Davies}, R.~L., {F{\"o}rster Schreiber}, N.~M., {{\"U}bler}, H., {et~al.}
  2019, \apj, 873, 122, \dodoi{10.3847/1538-4357/ab06f1}

\bibitem[{{de Avillez} \& {Berry}(2001)}]{deAvillez2001}
{de Avillez}, M.~A., \& {Berry}, D.~L. 2001, \mnras, 328, 708,
  \dodoi{10.1046/j.1365-8711.2001.04865.x}

\bibitem[{{D'Souza} {et~al.}(2015){D'Souza}, {Vegetti}, \&
  {Kauffmann}}]{Dsouza2015}
{D'Souza}, R., {Vegetti}, S., \& {Kauffmann}, G. 2015, \mnras, 454, 4027,
  \dodoi{10.1093/mnras/stv2234}

\bibitem[{{Emerick} {et~al.}(2019){Emerick}, {Bryan}, \& {Mac
  Low}}]{Emerick2019}
{Emerick}, A., {Bryan}, G.~L., \& {Mac Low}, M.-M. 2019, \mnras, 482, 1304,
  \dodoi{10.1093/mnras/sty2689}

\bibitem[{{Erkal} {et~al.}(2019){Erkal}, {Belokurov}, {Laporte}, {Koposov},
  {Li}, {Grillmair}, {Kallivayalil}, {Price-Whelan}, {Evans}, {Hawkins},
  {Hendel}, {Mateu}, {Navarro}, {del Pino}, {Slater}, {Sohn}, \& {Orphan Aspen
  Treasury Collaboration}}]{Erkal2019}
{Erkal}, D., {Belokurov}, V., {Laporte}, C.~F.~P., {et~al.} 2019, \mnras, 487,
  2685, \dodoi{10.1093/mnras/stz1371}

\bibitem[{{Fielding} {et~al.}(2018){Fielding}, {Quataert}, \&
  {Martizzi}}]{Fielding2018}
{Fielding}, D., {Quataert}, E., \& {Martizzi}, D. 2018, \mnras, 481, 3325,
  \dodoi{10.1093/mnras/sty2466}

\bibitem[{{Fielding} {et~al.}(2017){Fielding}, {Quataert}, {Martizzi}, \&
  {Faucher-Gigu{\`e}re}}]{Fielding2017}
{Fielding}, D., {Quataert}, E., {Martizzi}, D., \& {Faucher-Gigu{\`e}re}, C.-A.
  2017, \mnras, 470, L39, \dodoi{10.1093/mnrasl/slx072}

\bibitem[{{Fielding} \& {Bryan}(2022)}]{Fielding2022}
{Fielding}, D.~B., \& {Bryan}, G.~L. 2022, \apj, 924, 82,
  \dodoi{10.3847/1538-4357/ac2f41}

\bibitem[{{F{\"o}rster Schreiber} {et~al.}(2014){F{\"o}rster Schreiber},
  {Genzel}, {Newman}, {Kurk}, {Lutz}, {Tacconi}, {Wuyts}, {Bandara}, {Burkert},
  {Buschkamp}, {Carollo}, {Cresci}, {Daddi}, {Davies}, {Eisenhauer}, {Hicks},
  {Lang}, {Lilly}, {Mainieri}, {Mancini}, {Naab}, {Peng}, {Renzini}, {Rosario},
  {Shapiro Griffin}, {Shapley}, {Sternberg}, {Tacchella}, {Vergani},
  {Wisnioski}, {Wuyts}, \& {Zamorani}}]{FS2014}
{F{\"o}rster Schreiber}, N.~M., {Genzel}, R., {Newman}, S.~F., {et~al.} 2014,
  \apj, 787, 38, \dodoi{10.1088/0004-637X/787/1/38}

\bibitem[{{F{\"o}rster Schreiber} {et~al.}(2019){F{\"o}rster Schreiber},
  {{\"U}bler}, {Davies}, {Genzel}, {Wisnioski}, {Belli}, {Shimizu}, {Lutz},
  {Fossati}, {Herrera-Camus}, {Mendel}, {Tacconi}, {Wilman}, {Beifiori},
  {Brammer}, {Burkert}, {Carollo}, {Davies}, {Eisenhauer}, {Fabricius},
  {Lilly}, {Momcheva}, {Naab}, {Nelson}, {Price}, {Renzini}, {Saglia},
  {Sternberg}, {van Dokkum}, \& {Wuyts}}]{FS2019}
{F{\"o}rster Schreiber}, N.~M., {{\"U}bler}, H., {Davies}, R.~L., {et~al.}
  2019, \apj, 875, 21, \dodoi{10.3847/1538-4357/ab0ca2}

\bibitem[{{Fujii} {et~al.}(2021){Fujii}, {Saitoh}, {Hirai}, \&
  {Wang}}]{Fujii2021}
{Fujii}, M.~S., {Saitoh}, T.~R., {Hirai}, Y., \& {Wang}, L. 2021, \pasj, 73,
  1074, \dodoi{10.1093/pasj/psab061}

\bibitem[{{Gaburov} \& {Nitadori}(2011)}]{Gaburov2011}
{Gaburov}, E., \& {Nitadori}, K. 2011, \mnras, 414, 129,
  \dodoi{10.1111/j.1365-2966.2011.18313.x}

\bibitem[{{Gatto} {et~al.}(2017){Gatto}, {Walch}, {Naab}, {Girichidis},
  {W{\"u}nsch}, {Glover}, {Klessen}, {Clark}, {Peters}, {Derigs}, {Baczynski},
  \& {Puls}}]{Gatto2017}
{Gatto}, A., {Walch}, S., {Naab}, T., {et~al.} 2017, \mnras, 466, 1903,
  \dodoi{10.1093/mnras/stw3209}

\bibitem[{{Geen} {et~al.}(2016){Geen}, {Hennebelle}, {Tremblin}, \&
  {Rosdahl}}]{Geen2016}
{Geen}, S., {Hennebelle}, P., {Tremblin}, P., \& {Rosdahl}, J. 2016, \mnras,
  463, 3129, \dodoi{10.1093/mnras/stw2235}

\bibitem[{{Georgy} {et~al.}(2013){Georgy}, {Ekstr{\"o}m}, {Eggenberger},
  {Meynet}, {Haemmerl{\'e}}, {Maeder}, {Granada}, {Groh}, {Hirschi}, {Mowlavi},
  {Yusof}, {Charbonnel}, {Decressin}, \& {Barblan}}]{Georgy2013}
{Georgy}, C., {Ekstr{\"o}m}, S., {Eggenberger}, P., {et~al.} 2013, \aap, 558,
  A103, \dodoi{10.1051/0004-6361/201322178}

\bibitem[{{Girichidis} {et~al.}(2018){Girichidis}, {Seifried}, {Naab},
  {Peters}, {Walch}, {W{\"u}nsch}, {Glover}, \& {Klessen}}]{Girichidis2018}
{Girichidis}, P., {Seifried}, D., {Naab}, T., {et~al.} 2018, \mnras, 480, 3511,
  \dodoi{10.1093/mnras/sty2016}

\bibitem[{{Girichidis} {et~al.}(2016){Girichidis}, {Walch}, {Naab}, {Gatto},
  {W{\"u}nsch}, {Glover}, {Klessen}, {Clark}, {Peters}, {Derigs}, \&
  {Baczynski}}]{Girichidis2016}
{Girichidis}, P., {Walch}, S., {Naab}, T., {et~al.} 2016, \mnras, 456, 3432,
  \dodoi{10.1093/mnras/stv2742}

\bibitem[{{Glover} \& {Clark}(2012)}]{Glover2012}
{Glover}, S.~C.~O., \& {Clark}, P.~C. 2012, \mnras, 421, 9,
  \dodoi{10.1111/j.1365-2966.2011.19648.x}

\bibitem[{{Glover} \& {Mac Low}(2007{\natexlab{a}})}]{Glover2007a}
{Glover}, S.~C.~O., \& {Mac Low}, M.-M. 2007{\natexlab{a}}, \apjs, 169, 239,
  \dodoi{10.1086/512238}

\bibitem[{{Glover} \& {Mac Low}(2007{\natexlab{b}})}]{Glover2007b}
---. 2007{\natexlab{b}}, \apj, 659, 1317, \dodoi{10.1086/512227}

\bibitem[{{Gong} {et~al.}(2017){Gong}, {Ostriker}, \& {Wolfire}}]{Gong2016}
{Gong}, M., {Ostriker}, E.~C., \& {Wolfire}, M.~G. 2017, \apj, 843, 38,
  \dodoi{10.3847/1538-4357/aa7561}

\bibitem[{{Grudi{\'c}} {et~al.}(2021){Grudi{\'c}}, {Guszejnov}, {Hopkins},
  {Offner}, \& {Faucher-Gigu{\`e}re}}]{Grudic2021}
{Grudi{\'c}}, M.~Y., {Guszejnov}, D., {Hopkins}, P.~F., {Offner}, S. S.~R., \&
  {Faucher-Gigu{\`e}re}, C.-A. 2021, \mnras, 506, 2199,
  \dodoi{10.1093/mnras/stab1347}

\bibitem[{{Gutcke} {et~al.}(2021){Gutcke}, {Pakmor}, {Naab}, \&
  {Springel}}]{Gutcke2021}
{Gutcke}, T.~A., {Pakmor}, R., {Naab}, T., \& {Springel}, V. 2021, \mnras, 501,
  5597, \dodoi{10.1093/mnras/staa3875}

\bibitem[{Harris {et~al.}(2020)Harris, Millman, van~der Walt, Gommers,
  Virtanen, Cournapeau, Wieser, Taylor, Berg, Smith, Kern, Picus, Hoyer, van
  Kerkwijk, Brett, Haldane, del R{\'{i}}o, Wiebe, Peterson,
  G{\'{e}}rard-Marchant, Sheppard, Reddy, Weckesser, Abbasi, Gohlke, \&
  Oliphant}]{numpy}
Harris, C.~R., Millman, K.~J., van~der Walt, S.~J., {et~al.} 2020, Nature, 585,
  357, \dodoi{10.1038/s41586-020-2649-2}

\bibitem[{{Hartwell} {et~al.}(2004){Hartwell}, {Stevens}, {Strickland},
  {Heckman}, \& {Summers}}]{Hartwell2004}
{Hartwell}, J.~M., {Stevens}, I.~R., {Strickland}, D.~K., {Heckman}, T.~M., \&
  {Summers}, L.~K. 2004, \mnras, 348, 406,
  \dodoi{10.1111/j.1365-2966.2004.07375.x}

\bibitem[{{Heckman} {et~al.}(2015){Heckman}, {Alexandroff}, {Borthakur},
  {Overzier}, \& {Leitherer}}]{Heckman2015}
{Heckman}, T.~M., {Alexandroff}, R.~M., {Borthakur}, S., {Overzier}, R., \&
  {Leitherer}, C. 2015, \apj, 809, 147, \dodoi{10.1088/0004-637X/809/2/147}

\bibitem[{{Heckman} {et~al.}(1995){Heckman}, {Dahlem}, {Lehnert}, {Fabbiano},
  {Gilmore}, \& {Waller}}]{Heckman1995}
{Heckman}, T.~M., {Dahlem}, M., {Lehnert}, M.~D., {et~al.} 1995, \apj, 448, 98,
  \dodoi{10.1086/175944}

\bibitem[{{Heckman} {et~al.}(2000){Heckman}, {Lehnert}, {Strickland}, \&
  {Armus}}]{Heckman2000}
{Heckman}, T.~M., {Lehnert}, M.~D., {Strickland}, D.~K., \& {Armus}, L. 2000,
  \apjs, 129, 493, \dodoi{10.1086/313421}

\bibitem[{{Hill} {et~al.}(2012){Hill}, {Joung}, {Mac Low}, {Benjamin},
  {Haffner}, {Klingenberg}, \& {Waagan}}]{Hill2012}
{Hill}, A.~S., {Joung}, M.~R., {Mac Low}, M.-M., {et~al.} 2012, \apj, 750, 104,
  \dodoi{10.1088/0004-637X/750/2/104}

\bibitem[{{Hirai} {et~al.}(2021){Hirai}, {Fujii}, \& {Saitoh}}]{Hirai2021}
{Hirai}, Y., {Fujii}, M.~S., \& {Saitoh}, T.~R. 2021, \pasj, 73, 1036,
  \dodoi{10.1093/pasj/psab038}

\bibitem[{{Hirschmann} {et~al.}(2014){Hirschmann}, {Dolag}, {Saro}, {Bachmann},
  {Borgani}, \& {Burkert}}]{Hirschmann2014}
{Hirschmann}, M., {Dolag}, K., {Saro}, A., {et~al.} 2014, \mnras, 442, 2304,
  \dodoi{10.1093/mnras/stu1023}

\bibitem[{{Hislop} {et~al.}(2022){Hislop}, {Naab}, {Steinwandel}, {Lah{\'e}n},
  {Irodotou}, {Johansson}, \& {Walch}}]{Hislop2022}
{Hislop}, J.~M., {Naab}, T., {Steinwandel}, U.~P., {et~al.} 2022, \mnras, 509,
  5938, \dodoi{10.1093/mnras/stab3347}

\bibitem[{{Hodges-Kluck} {et~al.}(2020){Hodges-Kluck}, {Yukita}, {Tanner},
  {Ptak}, {Bregman}, \& {Li}}]{Hodges2020}
{Hodges-Kluck}, E.~J., {Yukita}, M., {Tanner}, R., {et~al.} 2020, \apj, 903,
  35, \dodoi{10.3847/1538-4357/abb884}

\bibitem[{{Hopkins}(2013)}]{Hopkins2013}
{Hopkins}, P.~F. 2013, \mnras, 428, 2840, \dodoi{10.1093/mnras/sts210}

\bibitem[{{Hopkins}(2015)}]{Hopkins2015}
---. 2015, \mnras, 450, 53, \dodoi{10.1093/mnras/stv195}

\bibitem[{{Hopkins} {et~al.}(2014){Hopkins}, {Kere{\v s}}, {O{\~n}orbe},
  {Faucher-Gigu{\`e}re}, {Quataert}, {Murray}, \& {Bullock}}]{Hopkins2014}
{Hopkins}, P.~F., {Kere{\v s}}, D., {O{\~n}orbe}, J., {et~al.} 2014, \mnras,
  445, 581, \dodoi{10.1093/mnras/stu1738}

\bibitem[{{Hopkins} {et~al.}(2012){Hopkins}, {Quataert}, \&
  {Murray}}]{Hopkins2012}
{Hopkins}, P.~F., {Quataert}, E., \& {Murray}, N. 2012, \mnras, 421, 3522,
  \dodoi{10.1111/j.1365-2966.2012.20593.x}

\bibitem[{{Hopkins} {et~al.}(2018){Hopkins}, {Wetzel}, {Kere{\v s}},
  {Faucher-Gigu{\`e}re}, {Quataert}, {Boylan-Kolchin}, {Murray}, {Hayward},
  {Garrison-Kimmel}, {Hummels}, {Feldmann}, {Torrey}, {Ma},
  {Angl{\'e}s-Alc{\'a}zar}, {Su}, {Orr}, {Schmitz}, {Escala}, {Sanderson},
  {Grudi{\'c}}, {Hafen}, {Kim}, {Fitts}, {Bullock}, {Wheeler}, {Chan},
  {Elbert}, \& {Narayanan}}]{Hopkins2018}
{Hopkins}, P.~F., {Wetzel}, A., {Kere{\v s}}, D., {et~al.} 2018, \mnras, 480,
  800, \dodoi{10.1093/mnras/sty1690}

\bibitem[{{Hopkins} {et~al.}(2022){Hopkins}, {Wetzel}, {Wheeler}, {Sanderson},
  {Grudic}, {Sameie}, {Boylan-Kolchin}, {Orr}, {Ma}, {Faucher-Giguere},
  {Keres}, {Quataert}, {Su}, {Moreno}, {Feldmann}, {Bullock}, {Loebman},
  {Angles-Alcazar}, {Stern}, {Necib}, \& {Hayward}}]{Hopkins_fire3}
{Hopkins}, P.~F., {Wetzel}, A., {Wheeler}, C., {et~al.} 2022, arXiv e-prints,
  arXiv:2203.00040.
\newblock \doarXiv{2203.00040}

\bibitem[{{Hu}(2019)}]{Hu2019}
{Hu}, C.-Y. 2019, \mnras, 483, 3363, \dodoi{10.1093/mnras/sty3252}

\bibitem[{{Hu} {et~al.}(2017){Hu}, {Naab}, {Glover}, {Walch}, \&
  {Clark}}]{Hu2017}
{Hu}, C.-Y., {Naab}, T., {Glover}, S.~C.~O., {Walch}, S., \& {Clark}, P.~C.
  2017, \mnras, 471, 2151, \dodoi{10.1093/mnras/stx1773}

\bibitem[{{Hu} {et~al.}(2016){Hu}, {Naab}, {Walch}, {Glover}, \&
  {Clark}}]{Hu2016}
{Hu}, C.-Y., {Naab}, T., {Walch}, S., {Glover}, S.~C.~O., \& {Clark}, P.~C.
  2016, \mnras, 458, 3528, \dodoi{10.1093/mnras/stw544}

\bibitem[{{Hu} {et~al.}(2014){Hu}, {Naab}, {Walch}, {Moster}, \&
  {Oser}}]{Hu2014}
{Hu}, C.-Y., {Naab}, T., {Walch}, S., {Moster}, B.~P., \& {Oser}, L. 2014,
  ArXiv e-prints.
\newblock \doarXiv{1402.1788}

\bibitem[{{Hu} {et~al.}(2022{\natexlab{a}}){Hu}, {Schruba}, {Sternberg}, \&
  {van Dishoeck}}]{Hu2022}
{Hu}, C.-Y., {Schruba}, A., {Sternberg}, A., \& {van Dishoeck}, E.~F.
  2022{\natexlab{a}}, arXiv e-prints, arXiv:2201.03885.
\newblock \doarXiv{2201.03885}

\bibitem[{{Hu} {et~al.}(2021){Hu}, {Sternberg}, \& {van Dishoeck}}]{Hu2021}
{Hu}, C.-Y., {Sternberg}, A., \& {van Dishoeck}, E.~F. 2021, \apj, 920, 44,
  \dodoi{10.3847/1538-4357/ac0dbd}

\bibitem[{{Hu} {et~al.}(2022{\natexlab{b}}){Hu}, {Smith}, {Teyssier}, {Bryan},
  {Verbeke}, {Emerick}, {Somerville}, {Burkhart}, {Li}, {Forbes}, \&
  {Starkenburg}}]{Hu2023}
{Hu}, C.-Y., {Smith}, M.~C., {Teyssier}, R., {et~al.} 2022{\natexlab{b}}, arXiv
  e-prints, arXiv:2208.10528.
\newblock \doarXiv{2208.10528}

\bibitem[{Hunter(2007)}]{Matplotlib}
Hunter, J.~D. 2007, Computing in Science \& Engineering, 9, 90,
  \dodoi{10.1109/MCSE.2007.55}

\bibitem[{{Joung} {et~al.}(2009){Joung}, {Mac Low}, \& {Bryan}}]{Joung2009}
{Joung}, M.~R., {Mac Low}, M.-M., \& {Bryan}, G.~L. 2009, \apj, 704, 137,
  \dodoi{10.1088/0004-637X/704/1/137}

\bibitem[{{Kacprzak} {et~al.}(2015){Kacprzak}, {Muzahid}, {Churchill},
  {Nielsen}, \& {Charlton}}]{Kacprzak2015}
{Kacprzak}, G.~G., {Muzahid}, S., {Churchill}, C.~W., {Nielsen}, N.~M., \&
  {Charlton}, J.~C. 2015, \apj, 815, 22, \dodoi{10.1088/0004-637X/815/1/22}

\bibitem[{{Kannan} {et~al.}(2019){Kannan}, {Vogelsberger}, {Marinacci},
  {McKinnon}, {Pakmor}, \& {Springel}}]{Kannan2019}
{Kannan}, R., {Vogelsberger}, M., {Marinacci}, F., {et~al.} 2019, \mnras, 485,
  117, \dodoi{10.1093/mnras/stz287}

\bibitem[{{Kim} {et~al.}(2022){Kim}, {Kim}, {Gong}, \& {Ostriker}}]{Kim2022}
{Kim}, C.-G., {Kim}, J.-G., {Gong}, M., \& {Ostriker}, E.~C. 2022, arXiv
  e-prints, arXiv:2211.13293.
\newblock \doarXiv{2211.13293}

\bibitem[{{Kim} \& {Ostriker}(2015)}]{Kim2015}
{Kim}, C.-G., \& {Ostriker}, E.~C. 2015, \apj, 802, 99,
  \dodoi{10.1088/0004-637X/802/2/99}

\bibitem[{{Kim} \& {Ostriker}(2018)}]{Kim2018}
---. 2018, \apj, 853, 173, \dodoi{10.3847/1538-4357/aaa5ff}

\bibitem[{{Kim} {et~al.}(2020{\natexlab{a}}){Kim}, {Ostriker}, {Fielding},
  {Smith}, {Bryan}, {Somerville}, {Forbes}, {Genel}, \& {Hernquist}}]{Kim2020}
{Kim}, C.-G., {Ostriker}, E.~C., {Fielding}, D.~B., {et~al.}
  2020{\natexlab{a}}, \apjl, 903, L34, \dodoi{10.3847/2041-8213/abc252}

\bibitem[{{Kim} {et~al.}(2020{\natexlab{b}}){Kim}, {Ostriker}, {Somerville},
  {Bryan}, {Fielding}, {Forbes}, {Hayward}, {Hernquist}, \&
  {Pandya}}]{Kim2020_smaug}
{Kim}, C.-G., {Ostriker}, E.~C., {Somerville}, R.~S., {et~al.}
  2020{\natexlab{b}}, \apj, 900, 61, \dodoi{10.3847/1538-4357/aba962}

\bibitem[{{Kimm} {et~al.}(2015){Kimm}, {Cen}, {Devriendt}, {Dubois}, \&
  {Slyz}}]{Kimm2015}
{Kimm}, T., {Cen}, R., {Devriendt}, J., {Dubois}, Y., \& {Slyz}, A. 2015,
  \mnras, 451, 2900, \dodoi{10.1093/mnras/stv1211}

\bibitem[{{Kobulnicky} \& {Skillman}(2008)}]{Kobulnicky2008}
{Kobulnicky}, H.~A., \& {Skillman}, E.~D. 2008, \aj, 135, 527,
  \dodoi{10.1088/0004-6256/135/2/527}

\bibitem[{{Lah{\'e}n} {et~al.}(2019{\natexlab{a}}){Lah{\'e}n}, {Naab},
  {Johansson}, {Elmegreen}, {Hu}, \& {Walch}}]{Lahen2020a}
{Lah{\'e}n}, N., {Naab}, T., {Johansson}, P.~H., {et~al.} 2019{\natexlab{a}},
  \apjl, 879, L18, \dodoi{10.3847/2041-8213/ab2a13}

\bibitem[{{Lah{\'e}n} {et~al.}(2019{\natexlab{b}}){Lah{\'e}n}, {Naab},
  {Johansson}, {Elmegreen}, {Hu}, {Walch}, {Steinwand el}, \&
  {Moster}}]{Lahen2020b}
---. 2019{\natexlab{b}}, arXiv e-prints, arXiv:1911.05093.
\newblock \doarXiv{1911.05093}

\bibitem[{{Lah{\'e}n} {et~al.}(2022){Lah{\'e}n}, {Naab}, {Kauffmann},
  {Sz{\'e}csi}, {Hislop}, {Rantala}, {Kozyreva}, {Walch}, \& {Hu}}]{Lahen2022}
{Lah{\'e}n}, N., {Naab}, T., {Kauffmann}, G., {et~al.} 2022, arXiv e-prints,
  arXiv:2211.15705.
\newblock \doarXiv{2211.15705}

\bibitem[{{Lancaster} {et~al.}(2021{\natexlab{a}}){Lancaster}, {Ostriker},
  {Kim}, \& {Kim}}]{Lancaster2021a}
{Lancaster}, L., {Ostriker}, E.~C., {Kim}, J.-G., \& {Kim}, C.-G.
  2021{\natexlab{a}}, \apj, 914, 89, \dodoi{10.3847/1538-4357/abf8ab}

\bibitem[{{Lancaster} {et~al.}(2021{\natexlab{b}}){Lancaster}, {Ostriker},
  {Kim}, \& {Kim}}]{Lancaster2021b}
---. 2021{\natexlab{b}}, \apj, 914, 89, \dodoi{10.3847/1538-4357/abf8ab}

\bibitem[{{Lancaster} {et~al.}(2021{\natexlab{c}}){Lancaster}, {Ostriker},
  {Kim}, \& {Kim}}]{Lancaster2021c}
---. 2021{\natexlab{c}}, \apjl, 922, L3, \dodoi{10.3847/2041-8213/ac3333}

\bibitem[{{Lanson} \& {Vila}(2008)}]{Lanson2008}
{Lanson}, N., \& {Vila}, J. 2008, SIAM J. Numer. Anal, 46, 1935

\bibitem[{{Lehnert} \& {Heckman}(1996)}]{Lehnert1996}
{Lehnert}, M.~D., \& {Heckman}, T.~M. 1996, \apj, 462, 651,
  \dodoi{10.1086/177180}

\bibitem[{{Lejeune} {et~al.}(1997){Lejeune}, {Cuisinier}, \&
  {Buser}}]{Lejeune1997}
{Lejeune}, T., {Cuisinier}, F., \& {Buser}, R. 1997, \aaps, 125, 229,
  \dodoi{10.1051/aas:1997373}

\bibitem[{{Lejeune} {et~al.}(1998){Lejeune}, {Cuisinier}, \&
  {Buser}}]{Lejeune1998}
---. 1998, \aaps, 130, 65, \dodoi{10.1051/aas:1998405}

\bibitem[{{Lelli} {et~al.}(2014){Lelli}, {Verheijen}, \&
  {Fraternali}}]{Lelli2014}
{Lelli}, F., {Verheijen}, M., \& {Fraternali}, F. 2014, \aap, 566, A71,
  \dodoi{10.1051/0004-6361/201322657}

\bibitem[{{Leroy} {et~al.}(2015){Leroy}, {Walter}, {Martini}, {Roussel},
  {Sandstrom}, {Ott}, {Weiss}, {Bolatto}, {Schuster}, \&
  {Dessauges-Zavadsky}}]{Leroy2015}
{Leroy}, A.~K., {Walter}, F., {Martini}, P., {et~al.} 2015, \apj, 814, 83,
  \dodoi{10.1088/0004-637X/814/2/83}

\bibitem[{{Li} {et~al.}(2020){Li}, {Li}, {Bryan}, {Ostriker}, \&
  {Quataert}}]{Li2020}
{Li}, M., {Li}, Y., {Bryan}, G.~L., {Ostriker}, E.~C., \& {Quataert}, E. 2020,
  \apj, 898, 23, \dodoi{10.3847/1538-4357/ab9c22}

\bibitem[{{Lopez} {et~al.}(2020){Lopez}, {Mathur}, {Nguyen}, {Thompson}, \&
  {Olivier}}]{Lopez2020}
{Lopez}, L.~A., {Mathur}, S., {Nguyen}, D.~D., {Thompson}, T.~A., \& {Olivier},
  G.~M. 2020, \apj, 904, 152, \dodoi{10.3847/1538-4357/abc010}

\bibitem[{{Marasco} {et~al.}(2022){Marasco}, {Belfiore}, {Cresci}, {Lelli},
  {Venturi}, {Hunt}, {Concas}, {Marconi}, {Mannucci}, {Mingozzi}, {McLeod},
  {Kumari}, {Carniani}, {Vanzi}, \& {Ginolfi}}]{Marasco2022}
{Marasco}, A., {Belfiore}, F., {Cresci}, G., {et~al.} 2022, arXiv e-prints,
  arXiv:2209.02726.
\newblock \doarXiv{2209.02726}

\bibitem[{{Marinacci} {et~al.}(2019){Marinacci}, {Sales}, {Vogelsberger},
  {Torrey}, \& {Springel}}]{Marinacci2019}
{Marinacci}, F., {Sales}, L.~V., {Vogelsberger}, M., {Torrey}, P., \&
  {Springel}, V. 2019, \mnras, 489, 4233, \dodoi{10.1093/mnras/stz2391}

\bibitem[{{Martin}(1999)}]{Martin1999}
{Martin}, C.~L. 1999, \apj, 513, 156, \dodoi{10.1086/306863}

\bibitem[{{Martin}(2005)}]{Martin2005}
---. 2005, \apj, 621, 227, \dodoi{10.1086/427277}

\bibitem[{{Martin} \& {Bouch{\'e}}(2009)}]{Martin2009}
{Martin}, C.~L., \& {Bouch{\'e}}, N. 2009, \apj, 703, 1394,
  \dodoi{10.1088/0004-637X/703/2/1394}

\bibitem[{{Martini} {et~al.}(2018){Martini}, {Leroy}, {Mangum}, {Bolatto},
  {Keating}, {Sandstrom}, \& {Walter}}]{Martini2018}
{Martini}, P., {Leroy}, A.~K., {Mangum}, J.~G., {et~al.} 2018, \apj, 856, 61,
  \dodoi{10.3847/1538-4357/aab08e}

\bibitem[{{Martizzi} {et~al.}(2015){Martizzi}, {Faucher-Gigu{\`e}re}, \&
  {Quataert}}]{Martizzi2015}
{Martizzi}, D., {Faucher-Gigu{\`e}re}, C.-A., \& {Quataert}, E. 2015, \mnras,
  450, 504, \dodoi{10.1093/mnras/stv562}

\bibitem[{{McKeith} {et~al.}(1995){McKeith}, {Greve}, {Downes}, \&
  {Prada}}]{Keith1995}
{McKeith}, C.~D., {Greve}, A., {Downes}, D., \& {Prada}, F. 1995, \aap, 293,
  703

\bibitem[{{McQuinn} {et~al.}(2018){McQuinn}, {Skillman}, {Heilman}, {Mitchell},
  \& {Kelley}}]{McQuinn2018}
{McQuinn}, K. B.~W., {Skillman}, E.~D., {Heilman}, T.~N., {Mitchell}, N.~P., \&
  {Kelley}, T. 2018, \mnras, 477, 3164, \dodoi{10.1093/mnras/sty839}

\bibitem[{{McQuinn} {et~al.}(2019){McQuinn}, {van Zee}, \&
  {Skillman}}]{McQuinn2019}
{McQuinn}, K. B.~W., {van Zee}, L., \& {Skillman}, E.~D. 2019, \apj, 886, 74,
  \dodoi{10.3847/1538-4357/ab4c37}

\bibitem[{{Meurer}(2004)}]{Meurer2004}
{Meurer}, G.~R. 2004, in Recycling Intergalactic and Interstellar Matter, ed.
  P.-A. {Duc}, J.~{Braine}, \& E.~{Brinks}, Vol. 217, 287.
\newblock \doarXiv{astro-ph/0311184}

\bibitem[{{Meurer} {et~al.}(2006){Meurer}, {Hanish}, {Ferguson}, {Knezek},
  {Kilborn}, {Putman}, {Smith}, {Koribalski}, {Meyer}, {Oey}, {Ryan-Weber},
  {Zwaan}, {Heckman}, {Kennicutt}, {Lee}, {Webster}, {Bland-Hawthorn},
  {Dopita}, {Freeman}, {Doyle}, {Drinkwater}, {Staveley-Smith}, \&
  {Werk}}]{Meurer2006}
{Meurer}, G.~R., {Hanish}, D.~J., {Ferguson}, H.~C., {et~al.} 2006, \apjs, 165,
  307, \dodoi{10.1086/504685}

\bibitem[{{Micic} {et~al.}(2012){Micic}, {Glover}, {Federrath}, \&
  {Klessen}}]{Micic2012}
{Micic}, M., {Glover}, S.~C.~O., {Federrath}, C., \& {Klessen}, R.~S. 2012,
  \mnras, 421, 2531, \dodoi{10.1111/j.1365-2966.2012.20477.x}

\bibitem[{{Naab} \& {Ostriker}(2017)}]{Naab2017}
{Naab}, T., \& {Ostriker}, J.~P. 2017, \araa, 55, 59,
  \dodoi{10.1146/annurev-astro-081913-040019}

\bibitem[{{Nelson} \& {Langer}(1997)}]{Nelson1997}
{Nelson}, R.~P., \& {Langer}, W.~D. 1997, \apj, 482, 796,
  \dodoi{10.1086/304167}

\bibitem[{{Newman} {et~al.}(2012){Newman}, {Genzel}, {F{\"o}rster-Schreiber},
  {Shapiro Griffin}, {Mancini}, {Lilly}, {Renzini}, {Bouch{\'e}}, {Burkert},
  {Buschkamp}, {Carollo}, {Cresci}, {Davies}, {Eisenhauer}, {Genel}, {Hicks},
  {Kurk}, {Lutz}, {Naab}, {Peng}, {Sternberg}, {Tacconi}, {Vergani}, {Wuyts},
  \& {Zamorani}}]{Newmann2012}
{Newman}, S.~F., {Genzel}, R., {F{\"o}rster-Schreiber}, N.~M., {et~al.} 2012,
  \apj, 761, 43, \dodoi{10.1088/0004-637X/761/1/43}

\bibitem[{{Nielsen} {et~al.}(2015){Nielsen}, {Churchill}, {Kacprzak}, {Murphy},
  \& {Evans}}]{Nielsen2015}
{Nielsen}, N.~M., {Churchill}, C.~W., {Kacprzak}, G.~G., {Murphy}, M.~T., \&
  {Evans}, J.~L. 2015, \apj, 812, 83, \dodoi{10.1088/0004-637X/812/1/83}

\bibitem[{{Ohlin} {et~al.}(2019){Ohlin}, {Renaud}, \& {Agertz}}]{Ohlin2019}
{Ohlin}, L., {Renaud}, F., \& {Agertz}, O. 2019, arXiv e-prints.
\newblock \doarXiv{1902.00028}

\bibitem[{{Ostriker} \& {Kim}(2022)}]{Ostriker2022}
{Ostriker}, E.~C., \& {Kim}, C.-G. 2022, \apj, 936, 137,
  \dodoi{10.3847/1538-4357/ac7de2}

\bibitem[{{Ott} {et~al.}(2005){Ott}, {Walter}, \& {Brinks}}]{Ott2005}
{Ott}, J., {Walter}, F., \& {Brinks}, E. 2005, \mnras, 358, 1453,
  \dodoi{10.1111/j.1365-2966.2005.08863.x}

\bibitem[{{Pandya} {et~al.}(2021){Pandya}, {Fielding},
  {Angl{\'e}s-Alc{\'a}zar}, {Somerville}, {Bryan}, {Hayward}, {Stern}, {Kim},
  {Quataert}, {Forbes}, {Faucher-Gigu{\`e}re}, {Feldmann}, {Hafen}, {Hopkins},
  {Kere{\v{s}}}, {Murray}, \& {Wetzel}}]{Pandya2021}
{Pandya}, V., {Fielding}, D.~B., {Angl{\'e}s-Alc{\'a}zar}, D., {et~al.} 2021,
  \mnras, 508, 2979, \dodoi{10.1093/mnras/stab2714}

\bibitem[{{Peters} {et~al.}(2017){Peters}, {Naab}, {Walch}, {Glover},
  {Girichidis}, {Pellegrini}, {Klessen}, {W{\"u}nsch}, {Gatto}, \&
  {Baczynski}}]{Peters2017}
{Peters}, T., {Naab}, T., {Walch}, S., {et~al.} 2017, \mnras, 466, 3293,
  \dodoi{10.1093/mnras/stw3216}

\bibitem[{{Pettini} {et~al.}(2001){Pettini}, {Shapley}, {Steidel}, {Cuby},
  {Dickinson}, {Moorwood}, {Adelberger}, \& {Giavalisco}}]{Pettini2001}
{Pettini}, M., {Shapley}, A.~E., {Steidel}, C.~C., {et~al.} 2001, \apj, 554,
  981, \dodoi{10.1086/321403}

\bibitem[{{Pillepich} {et~al.}(2018){Pillepich}, {Springel}, {Nelson}, {Genel},
  {Naiman}, {Pakmor}, {Hernquist}, {Torrey}, {Vogelsberger}, {Weinberger}, \&
  {Marinacci}}]{Pillepich2018}
{Pillepich}, A., {Springel}, V., {Nelson}, D., {et~al.} 2018, \mnras, 473,
  4077, \dodoi{10.1093/mnras/stx2656}

\bibitem[{{Price}(2007)}]{Price2007}
{Price}, D.~J. 2007, \pasa, 24, 159, \dodoi{10.1071/AS07022}

\bibitem[{{Ptak} {et~al.}(1997){Ptak}, {Serlemitsos}, {Yaqoob}, {Mushotzky}, \&
  {Tsuru}}]{Ptak1997}
{Ptak}, A., {Serlemitsos}, P., {Yaqoob}, T., {Mushotzky}, R., \& {Tsuru}, T.
  1997, \aj, 113, 1286, \dodoi{10.1086/118342}

\bibitem[{{Rathjen} {et~al.}(2022){Rathjen}, {Naab}, {Walch}, {Seifried},
  {Girichidis}, \& {W{\"u}nsch}}]{Rathjen2022}
{Rathjen}, T.-E., {Naab}, T., {Walch}, S., {et~al.} 2022, arXiv e-prints,
  arXiv:2211.15419.
\newblock \doarXiv{2211.15419}

\bibitem[{{Rathjen} {et~al.}(2021){Rathjen}, {Naab}, {Girichidis}, {Walch},
  {W{\"u}nsch}, {Dinnbier}, {Seifried}, {Klessen}, \& {Glover}}]{Rathjen2021}
{Rathjen}, T.-E., {Naab}, T., {Girichidis}, P., {et~al.} 2021, \mnras, 504,
  1039, \dodoi{10.1093/mnras/stab900}

\bibitem[{{Rubin} {et~al.}(2011){Rubin}, {Prochaska}, {M{\'e}nard}, {Murray},
  {Kasen}, {Koo}, \& {Phillips}}]{Rubin2011}
{Rubin}, K. H.~R., {Prochaska}, J.~X., {M{\'e}nard}, B., {et~al.} 2011, \apj,
  728, 55, \dodoi{10.1088/0004-637X/728/1/55}

\bibitem[{{Rupke} {et~al.}(2005){Rupke}, {Veilleux}, \& {Sanders}}]{Rupke2005}
{Rupke}, D.~S., {Veilleux}, S., \& {Sanders}, D.~B. 2005, \apjs, 160, 115,
  \dodoi{10.1086/432889}

\bibitem[{{Schaye} {et~al.}(2015){Schaye}, {Crain}, {Bower}, {Furlong},
  {Schaller}, {Theuns}, {Dalla Vecchia}, {Frenk}, {McCarthy}, {Helly},
  {Jenkins}, {Rosas-Guevara}, {White}, {Baes}, {Booth}, {Camps}, {Navarro},
  {Qu}, {Rahmati}, {Sawala}, {Thomas}, \& {Trayford}}]{Schaye2015}
{Schaye}, J., {Crain}, R.~A., {Bower}, R.~G., {et~al.} 2015, \mnras, 446, 521,
  \dodoi{10.1093/mnras/stu2058}

\bibitem[{{Schneider} {et~al.}(2020){Schneider}, {Ostriker}, {Robertson}, \&
  {Thompson}}]{Schneider2020}
{Schneider}, E.~E., {Ostriker}, E.~C., {Robertson}, B.~E., \& {Thompson}, T.~A.
  2020, \apj, 895, 43, \dodoi{10.3847/1538-4357/ab8ae8}

\bibitem[{{Shapley} {et~al.}(2003){Shapley}, {Steidel}, {Pettini}, \&
  {Adelberger}}]{Shapely2003}
{Shapley}, A.~E., {Steidel}, C.~C., {Pettini}, M., \& {Adelberger}, K.~L. 2003,
  \apj, 588, 65, \dodoi{10.1086/373922}

\bibitem[{{Smith} {et~al.}(2017){Smith}, {Bryan}, {Glover}, {Goldbaum}, {Turk},
  {Regan}, {Wise}, {Schive}, {Abel}, {Emerick}, {O'Shea}, {Anninos}, {Hummels},
  \& {Khochfar}}]{Smith2017}
{Smith}, B.~D., {Bryan}, G.~L., {Glover}, S. C.~O., {et~al.} 2017, \mnras, 466,
  2217, \dodoi{10.1093/mnras/stw3291}

\bibitem[{{Smith} {et~al.}(2021){Smith}, {Bryan}, {Somerville}, {Hu},
  {Teyssier}, {Burkhart}, \& {Hernquist}}]{Smith2021}
{Smith}, M.~C., {Bryan}, G.~L., {Somerville}, R.~S., {et~al.} 2021, \mnras,
  506, 3882, \dodoi{10.1093/mnras/stab1896}

\bibitem[{{Somerville} \& {Dav{\'e}}(2015)}]{Somerville2015}
{Somerville}, R.~S., \& {Dav{\'e}}, R. 2015, \araa, 53, 51,
  \dodoi{10.1146/annurev-astro-082812-140951}

\bibitem[{{Springel}(2005)}]{springel05}
{Springel}, V. 2005, \mnras, 364, 1105,
  \dodoi{10.1111/j.1365-2966.2005.09655.x}

\bibitem[{{Springel} {et~al.}(2005){Springel}, {Di Matteo}, \&
  {Hernquist}}]{springel05b}
{Springel}, V., {Di Matteo}, T., \& {Hernquist}, L. 2005, \mnras, 361, 776,
  \dodoi{10.1111/j.1365-2966.2005.09238.x}

\bibitem[{{Steidel} {et~al.}(2010){Steidel}, {Erb}, {Shapley}, {Pettini},
  {Reddy}, {Bogosavljevi{\'c}}, {Rudie}, \& {Rakic}}]{Steidel2010}
{Steidel}, C.~C., {Erb}, D.~K., {Shapley}, A.~E., {et~al.} 2010, \apj, 717,
  289, \dodoi{10.1088/0004-637X/717/1/289}

\bibitem[{{Steinwandel} {et~al.}(2022){Steinwandel}, {Bryan}, {Somerville},
  {Hayward}, \& {Burkhart}}]{Steinwandel2022}
{Steinwandel}, U.~P., {Bryan}, G.~L., {Somerville}, R.~S., {Hayward}, C.~C., \&
  {Burkhart}, B. 2022, arXiv e-prints, arXiv:2205.09774.
\newblock \doarXiv{2205.09774}

\bibitem[{{Steinwandel} {et~al.}(2020){Steinwandel}, {Moster}, {Naab}, {Hu}, \&
  {Walch}}]{Steinwandel2020}
{Steinwandel}, U.~P., {Moster}, B.~P., {Naab}, T., {Hu}, C.-Y., \& {Walch}, S.
  2020, \mnras, \dodoi{10.1093/mnras/staa821}

\bibitem[{{Stone} {et~al.}(2008){Stone}, {Gardiner}, {Teuben}, {Hawley}, \&
  {Simon}}]{Stone2008}
{Stone}, J.~M., {Gardiner}, T.~A., {Teuben}, P., {Hawley}, J.~F., \& {Simon},
  J.~B. 2008, \apjs, 178, 137, \dodoi{10.1086/588755}

\bibitem[{{Strickland} \& {Heckman}(2009)}]{Stricklan2009}
{Strickland}, D.~K., \& {Heckman}, T.~M. 2009, \apj, 697, 2030,
  \dodoi{10.1088/0004-637X/697/2/2030}

\bibitem[{{Summers} {et~al.}(2003){Summers}, {Stevens}, {Strickland}, \&
  {Heckman}}]{Summers2003}
{Summers}, L.~K., {Stevens}, I.~R., {Strickland}, D.~K., \& {Heckman}, T.~M.
  2003, \mnras, 342, 690, \dodoi{10.1046/j.1365-8711.2003.06590.x}

\bibitem[{{Summers} {et~al.}(2004){Summers}, {Stevens}, {Strickland}, \&
  {Heckman}}]{Summers2004}
---. 2004, \mnras, 351, 1, \dodoi{10.1111/j.1365-2966.2004.07749.x}

\bibitem[{{Thornton} {et~al.}(1998){Thornton}, {Gaudlitz}, {Janka}, \&
  {Steinmetz}}]{Thornton1998}
{Thornton}, K., {Gaudlitz}, M., {Janka}, H.-T., \& {Steinmetz}, M. 1998, \apj,
  500, 95, \dodoi{10.1086/305704}

\bibitem[{{Tremonti} {et~al.}(2004){Tremonti}, {Heckman}, {Kauffmann},
  {Brinchmann}, {Charlot}, {White}, {Seibert}, {Peng}, {Schlegel}, {Uomoto},
  {Fukugita}, \& {Brinkmann}}]{Tremonti2004}
{Tremonti}, C.~A., {Heckman}, T.~M., {Kauffmann}, G., {et~al.} 2004, \apj, 613,
  898, \dodoi{10.1086/423264}

\bibitem[{{van der Velden}(2020)}]{cmasher}
{van der Velden}, E. 2020, The Journal of Open Source Software, 5, 2004,
  \dodoi{10.21105/joss.02004}

\bibitem[{{Vijayan} {et~al.}(2020){Vijayan}, {Kim}, {Armillotta}, {Ostriker},
  \& {Li}}]{Vijayan2020}
{Vijayan}, A., {Kim}, C.-G., {Armillotta}, L., {Ostriker}, E.~C., \& {Li}, M.
  2020, \apj, 894, 12, \dodoi{10.3847/1538-4357/ab8474}

\bibitem[{Virtanen {et~al.}(2020)Virtanen, Gommers, Oliphant, Haberland, Reddy,
  Cournapeau, Burovski, Peterson, Weckesser, Bright, {van der Walt}, Brett,
  Wilson, Millman, Mayorov, Nelson, Jones, Kern, Larson, Carey, Polat, Feng,
  Moore, {VanderPlas}, Laxalde, Perktold, Cimrman, Henriksen, Quintero, Harris,
  Archibald, Ribeiro, Pedregosa, {van Mulbregt}, \& {SciPy 1.0
  Contributors}}]{scipy}
Virtanen, P., Gommers, R., Oliphant, T.~E., {et~al.} 2020, Nature Methods, 17,
  261, \dodoi{10.1038/s41592-019-0686-2}

\bibitem[{{Vogelsberger} {et~al.}(2014){Vogelsberger}, {Genel}, {Springel},
  {Torrey}, {Sijacki}, {Xu}, {Snyder}, {Nelson}, \&
  {Hernquist}}]{Vogelsberger2014}
{Vogelsberger}, M., {Genel}, S., {Springel}, V., {et~al.} 2014, \mnras, 444,
  1518, \dodoi{10.1093/mnras/stu1536}

\bibitem[{{Walch} {et~al.}(2015){Walch}, {Girichidis}, {Naab}, {Gatto},
  {Glover}, {W{\"u}nsch}, {Klessen}, {Clark}, {Peters}, {Derigs}, \&
  {Baczynski}}]{Walch2015}
{Walch}, S., {Girichidis}, P., {Naab}, T., {et~al.} 2015, \mnras, 454, 238,
  \dodoi{10.1093/mnras/stv1975}

\bibitem[{{Walter} {et~al.}(2002){Walter}, {Weiss}, \& {Scoville}}]{Walter2002}
{Walter}, F., {Weiss}, A., \& {Scoville}, N. 2002, \apjl, 580, L21,
  \dodoi{10.1086/345287}

\bibitem[{{Westera} {et~al.}(2002){Westera}, {Lejeune}, {Buser}, {Cuisinier},
  \& {Bruzual}}]{Westera2002}
{Westera}, P., {Lejeune}, T., {Buser}, R., {Cuisinier}, F., \& {Bruzual}, G.
  2002, \aap, 381, 524, \dodoi{10.1051/0004-6361:20011493}

\bibitem[{{Westmoquette} {et~al.}(2009){Westmoquette}, {Smith}, {Gallagher},
  {Trancho}, {Bastian}, \& {Konstantopoulos}}]{Westermoquette2009}
{Westmoquette}, M.~S., {Smith}, L.~J., {Gallagher}, J.~S., I., {et~al.} 2009,
  \apj, 696, 192, \dodoi{10.1088/0004-637X/696/1/192}

\bibitem[{{Wiersma} {et~al.}(2009){Wiersma}, {Schaye}, \&
  {Smith}}]{Wiersma2009}
{Wiersma}, R.~P.~C., {Schaye}, J., \& {Smith}, B.~D. 2009, \mnras, 393, 99,
  \dodoi{10.1111/j.1365-2966.2008.14191.x}

\bibitem[{{Wolfire} {et~al.}(2003){Wolfire}, {McKee}, {Hollenbach}, \&
  {Tielens}}]{Wolfire2003}
{Wolfire}, M.~G., {McKee}, C.~F., {Hollenbach}, D., \& {Tielens}, A.~G.~G.~M.
  2003, \apj, 587, 278, \dodoi{10.1086/368016}

\bibitem[{{Xu} {et~al.}(2022){Xu}, {Heckman}, {Henry}, {Berg}, {Chisholm},
  {James}, {Martin}, {Stark}, {Aloisi}, {Amor{\'\i}n}, {Arellano-C{\'o}rdova},
  {Bordoloi}, {Charlot}, {Chen}, {Hayes}, {Mingozzi}, {Sugahara}, {Kewley},
  {Ouchi}, {Scarlata}, \& {Steidel}}]{Xu2022}
{Xu}, X., {Heckman}, T., {Henry}, A., {et~al.} 2022, \apj, 933, 222,
  \dodoi{10.3847/1538-4357/ac6d56}

\end{thebibliography}
\bibliographystyle{aasjournal}

\end{document}